\documentclass[12pt]{article}
\usepackage{latexsym,amssymb}

\begin{document}

\newcommand{\eqn}[1]{(\ref{#1})}
\newcommand{\eq}[1]{(\ref{#1})}

\newcommand{\bm}[1]{\mbox{\boldmath $#1$}}
\newcommand{\be}{\begin{equation}}
\newcommand{\ee}{\end{equation}}
\newcommand{\bea}{\begin{eqnarray}}
\newcommand{\eea}{\end{eqnarray}}
\newcommand{\nn}{\nonumber}
\newcommand{\ov}{\overline}
\newcommand{\ba}{\begin{eqnarray}}
\newcommand{\ea}{\end{eqnarray}}

\def\oddmarkC{\thepage}
\def\oddmarkB{}
\def\oddmarkD{}
\def\oddmarkE{}
\def\oddmarkF{}
\def\footmsgA{}
\def\sk{\vskip .4cm}
\def\ibar{{\bar \imath}}
\def\jbar{{\bar \jmath}}
\def\kbar{{\bar k}}
\def\lbar{{\bar \ell}}
\def\mbar{{\bar m}}
\def\Im{{\rm Im }}
\def\Re{{\rm Re }}
\def\IP{\relax{\rm I\kern-.18em P}}
\def\arccosh{{\rm arccosh ~}}
\def\Es{\bf E_{7(7)}}
\def\muun{\underline \mu}
\def\mun{\underline m}
\def\nuun{\underline \nu}
\def\nun{\underline n}
\def\buun{\underline \bullet}
\def\Eb{{\bf E}}
\def\bu{\bullet}
\def\we{\wedge}
\font\cmss=cmss10 \font\cmsss=cmss10 at 7pt
\def\twomat#1#2#3#4{\left(\matrix{#1 & #2 \cr #3 & #4}\right)}
\def\inbar{\vrule height1.5ex width.4pt depth0pt}
\def\IC{\relax\,\hbox{$\inbar\kern-.3em{\rm C}$}}
\def\IG{\relax\,\hbox{$\inbar\kern-.3em{\rm G}$}}
\def\IB{\relax{\rm I\kern-.18em B}}
\def\ID{\relax{\rm I\kern-.18em D}}
\def\IL{\relax{\rm I\kern-.18em L}}
\def\IF{\relax{\rm I\kern-.18em F}}
\def\IH{\relax{\rm I\kern-.18em H}}
\def\II{\relax{\rm I\kern-.17em I}}
\def\IN{\relax{\rm I\kern-.18em N}}
\def\IP{\relax{\rm I\kern-.18em P}}
\def\IQ{\relax\,\hbox{$\inbar\kern-.3em{\rm Q}$}}
\def\bfzero{\relax\,\hbox{$\inbar\kern-.3em{\rm 0}$}}
\def\IK{\relax{\rm I\kern-.18em K}}
\def\IG{\relax\,\hbox{$\inbar\kern-.3em{\rm G}$}}
 \font\cmss=cmss10 \font\cmsss=cmss10 at 7pt
\def\IR{\relax{\rm I\kern-.18em R}}
\def\ZZ{\relax\ifmmode\mathchoice
{\hbox{\cmss Z\kern-.4em Z}}{\hbox{\cmss Z\kern-.4em Z}}
{\lower.9pt\hbox{\cmsss Z\kern-.4em Z}} {\lower1.2pt\hbox{\cmsss
Z\kern-.4em Z}}\else{\cmss Z\kern-.4em Z}\fi}
\def\bfone{\relax{\rm 1\kern-.35em 1}}
\def\dop{{\rm d}\hskip -1pt}
\def\real{{\rm Re}\hskip 1pt}
\def\trace{{\rm Tr}\hskip 1pt}
\def\ii{{\rm i}}
\def\diag{{\rm diag}}
\def\sch#1#2{\{#1;#2\}}
\def\IU{\relax\,\hbox{$\inbar\kern-.3em{\rm U}$}}
\def\sk{\vskip .4cm}
\def\noi{\noindent}
\def\om{\omega}
\def\Om{\Omega}
\def\al{\alpha}
\def\la{\lambda}
\def\be{\beta}
\def\ga{\gamma}
\def\Ga{\Gamma}
\def\de{\delta}
\def\epsi{\varepsilon}
\def\we{\wedge}
\def\part{\partial}
\def\bu{\bullet}
\def\ci{\circ}
\def\square{{\,\lower0.9pt\vbox{\hrule \hbox{\vrule height 0.2 cm
\hskip 0.2 cm \vrule height 0.2 cm}\hrule}\,}}
\def\muun{\underline \mu}
\def\mun{\underline m}
\def\nuun{\underline \nu}
\def\nun{\underline n}
\def\buun{\underline \bullet}
\def\Rb{{\bf R}}
\def\Eb{{\bf E}}
\def\gb{{\bf g}}
\def\dt{{\tilde d}}
\def\Dt{{\tilde D}}
\def\Dcal{{\cal D}}
\def\R#1#2{ R^{#1}_{~~~#2} }
\def\ome#1#2{\om^{#1}_{~#2}}
\def\Rf#1#2{ R^{\underline #1}_{~~~{\underline #2}} }
\def\Rfu#1#2{ R^{{\underline #1}{\underline #2}} }
\def\Rfd#1#2{ R_{{\underline #1}{\underline #2}} }
\def\Rfb#1#2{ {\bf R}^{\underline #1}_{~~~{\underline #2}} }
\def\omef#1#2{\om^{\underline #1}_{~{\underline #2}}}
\def\omefb#1#2{{ \omb}^{\underline #1}_{~{\underline #2}}}
\def\omefu#1#2{\om^{{\underline #1} {\underline #2}}}
\def\omefub#1#2{{\omb}^{{\underline #1} {\underline #2}}}
\def\Ef#1{E^{\underline #1}}
\def\Efb#1{{\bf E}^{\underline #1}}
\def\omb{\bf \mbox{\boldmath $\om$}}
\def\bfone{\relax{\rm 1\kern-.35em 1}}
\font\cmss=cmss10 \font\cmsss=cmss10 at 7pt
\def\a{\alpha} \def\b{\beta} \def\d{\delta}
\def\e{\epsilon} \def\c{\gamma}
\def\G{\Gamma} \def\l{\lambda}
\def\L{\Lambda} \def\s{\sigma}
\def\cA{{\cal A}} \def\cB{{\cal B}}
\def\cS{{\cal S}}
\def\cC{{\cal C}} \def\cD{{\cal D}}
\def\cF{{\cal F}} \def\cG{{\cal G}}
\def\cH{{\cal H}} \def\cI{{\cal I}}
\def\cJ{{\cal J}} \def\cK{{\cal K}}
\def\cL{{\cal L}} \def\cM{{\cal M}}
\def\cN{{\cal N}} \def\cO{{\cal O}}
\def\cP{{\cal P}} \def\cQ{{\cal Q}}
\def\cR{{\cal R}} \def\cV{{\cal V}}\def\cW{{\cal W}}
%
%
%
\def\crr{\crcr\noalign{\vskip {8.3333pt}}}
\def\tilde{\widetilde}
\def\bar{\overline}
\def\us#1{\underline{#1}}
\let\shat=\hat
\def\hat{\widehat}
\def\hyp{\vrule height 2.3pt width 2.5pt depth -1.5pt}
\def\Coeff#1#2{\frac{#1}{ #2}}
\def\Coe#1.#2.{\frac{#1}{ #2}}
\def\coeff#1#2{\relax{\textstyle {#1 \over #2}}\displaystyle}
\def\coe#1.#2.{\relax{\textstyle {#1 \over #2}}\displaystyle}
\def\half{{1 \over 2}}
\def\shalf{\relax{\textstyle \frac{1}{ 2}}\displaystyle}
\def\dag#1{#1\!\!\!/\,\,\,}
\def\to{\rightarrow}
\def\notin{\hbox{{$\in$}\kern-.51em\hbox{/}}}
\def\shdot{\!\cdot\!}
\def\ket#1{\,\big|\,#1\,\big>\,}
\def\bra#1{\,\big<\,#1\,\big|\,}
\def\equaltop#1{\mathrel{\mathop=^{#1}}}
\def\Trbel#1{\mathop{{\rm Tr}}_{#1}}
\def\inserteq#1{\noalign{\vskip-.2truecm\hbox{#1\hfil}
\vskip-.2cm}}
\def\attac#1{\Bigl\vert
{\phantom{X}\atop{{\rm\scriptstyle #1}}\phantom{X}}}
\def\exx#1{e^{{\displaystyle #1}}}
\def\del{\partial}
\def\delbar{\bar\partial}
\def\nex#1{$N\!=\!#1$}
\def\dex#1{$d\!=\!#1$}
\def\cex#1{$c\!=\!#1$}
\def\eg{{\it e.g.}} \def\ie{{\it i.e.}}
\def\IE{\relax{{\rm I\kern-.18em E}}}
\def\cE{{\cal E}}
\def\rt{{\cR^{(3)}}}
\def\IGam{\relax{{\rm I}\kern-.18em \Gamma}}
\def\IGa{\IA}
\def\ii{{\rm i}}
\def\inbar{\vrule height1.5ex width.4pt depth0pt}
\def\bfzero{\relax{\rm I\kern-.18em 0}}
\def\bfone{\relax{\rm 1\kern-.35em 1}}
\def\twomat#1#2#3#4{\left(\begin{array}{cc}
\end{array}
\right)}
\def\twovec#1#2{\left(\begin{array}{c}
{#1}\\ {#2}\\
\end{array}
\right)}
\def\o#1#2{{{#1}\over{#2}}}
\newcommand{\La}{{\Lambda}}
\newcommand{\Si}{{\Sigma}}
\newcommand{\im}{{\rm Im\ }}
\def\cA{{\cal A}} \def\cB{{\cal B}}
\def\cC{{\cal C}} \def\cD{{\cal D}}
\def\cF{{\cal F}} \def\cG{{\cal G}}
\def\cH{{\cal H}} \def\cI{{\cal I}}
\def\cJ{{\cal J}} \def\cK{{\cal K}}
\def\cL{{\cal L}} \def\cM{{\cal M}}
\def\cN{{\cal N}} \def\cO{{\cal O}}
\def\cP{{\cal P}} \def\cQ{{\cal Q}}
\def\cR{{\cal R}} \def\cV{{\cal V}}\def\cW{{\cal W}}
\def\a{\alpha} \def\b{\beta} \def\d{\delta}
\def\e{\epsilon} \def\c{\gamma}
\def\G{\Gamma} \def\l{\lambda}
\def\L{\Lambda} \def\s{\sigma}
\def\T{T}

\begin{flushright}
CERN-PH-TH/2006-248
\end{flushright}
\vskip 1cm
  \begin{center}{\LARGE \bf  Extremal Black Holes in Supergravity}
\vskip 1.5cm
{L. Andrianopoli$^{1,2,3}$,  R. D'Auria$^2$, S. Ferrara$^{3,4}$ and M. Trigiante$^2
$}
\end{center}
 \vskip 3mm
\noindent
{\small
$^1$ Centro E. Fermi, Compendio Viminale,
I-00184 Rome, Italy \\
$^2$
Dipartimento di Fisica,
  Politecnico di Torino, Corso Duca degli Abruzzi 24, I-10129
  Turin, Italy and Istituto Nazionale di Fisica Nucleare (INFN)
  Sezione di Torino, Italy
  \\ \texttt{riccardo.dauria@polito.it}\\ \texttt{mario.trigiante@polito.it}\\
 $^3$
   CERN PH-TH Division, CH 1211 Geneva 23, Switzerland
  \\  \texttt{laura.andrianopoli@cern.ch}\\ \texttt{sergio.ferrara@cern.ch}
 \\
 $^4$ Istituto Nazionale di Fisica Nucleare (INFN)
  Laboratori Nazionali di Frascati, Italy }

\vfill
\vskip 1,5 cm
\begin{center}
{\it To appear in the book "String theory and fundamental interactions"
published in celebration of the 65th birthday of Gabriele Veneziano,
\\
Eds. M. Gasperini and J. Maharana \\(Lecture Notes in Physics,
Springer Berlin/Heidelberg, 2007)}, \texttt{ www.springerlink.com/content/1616-6361}
\end{center}

\vskip 1.5cm

\begin{abstract}
We present the main features of the physics of extremal black holes embedded in supersymmetric theories of gravitation, with a detailed analysis of the attractor mechanism for BPS and non-BPS black-hole solutions in four dimensions.
\end{abstract}

 \vfill\eject
\section{Introduction:
Extremal Black Holes from Classical General Relativity to String Theory} \label{intro1} The
physics of black holes \cite{wald}, with its theoretical and phenomenological implications, has a fertile
impact on many branches of natural science, such as astrophysics, cosmology, particle physics
and, more recently, mathematical physics \cite{moore} and quantum information theory
\cite{information}. This is not so astonishing in view of the fact that, owing to the
singularity theorems of Penrose and Hawking \cite{ph}, the existence of black holes seems to
be an unavoidable consequence of Einstein's theory of general relativity and of its modern
generalizations such as supergravity \cite{black}, superstrings and M-theory \cite{review}.

A fascinating aspect of black-hole physics is in their thermodynamic properties that seem to
encode fundamental insights  of a so far not  established final theory of quantum gravity.
In this context a central role is played by the Bekenstein--Hawking (in the following, B-H)
entropy formula \cite{entrop}:
\begin{equation}
S_{\mbox{\small B-H}} = \frac{k_B}{\ell_P^2} \, \frac{1}{4} \, \mbox{Area}_H\,, \label{bekhaw}
\end{equation}
where $k_B$ is the Boltzman constant, $\ell_P^2=G \hbar/c^3$ is the
squared Planck length  while $\mbox{Area}_H$ denotes the area of the
horizon surface (from now on we shall use the natural units
$\hbar=c=G=k_B=1$).
\par
This relation between a thermodynamic quantity ($S_{\mbox{\small B-H}}$) and a geometric quantity ($
\mbox{Area}_H$) is a puzzling aspect that motivated much theoretical work in the last decades.
In fact a microscopic statistical explanation of the area/entropy  formula, related to
microstate counting, has been  regarded as possible only within a consistent and satisfactory
formulation of quantum gravity. Superstring theory is the most serious candidate for a theory
of quantum gravity and, as such, should eventually provide such a microscopic explanation of the
area law  \cite{blackmicro}. Since black holes are a typical non-perturbative phenomenon,
perturbative string theory could say very little about their entropy: only non-perturbative
string theory could have a handle on it. Progress in this direction came after 1995
\cite{witten}, through the recognition of the role of string dualities. These dualities allow
one to relate the strong coupling regime of one superstring model to the weak coupling regime of
another. Interestingly enough, there is evidence that the (perturbative and
non-perturbative) string dualities are all encoded in the global symmetry group (the
$U$-duality group) of the low energy supergravity effective action \cite{huto}.
\par
Let us introduce a particular class of black-hole solutions, which
will be particularly relevant to our discussion: the {\it extremal
black holes}. The simplest instance of  these solutions may be found
within the class of the so-called Reissner--Nordstr\"om (R-N)
space-time \cite{rn}, whose metric describes a static, isotropic
black hole of mass $M$ and electric (or magnetic) charge $Q$:
\begin{equation}
ds^2 \, = \, dt^2 \, \left( 1- \frac{2 M}{\rho}+\frac{Q^2}{\rho^2}
\right) - d\rho^2 \, \left( 1- \frac{2 M}{\rho}+\frac{Q^2}{\rho^2}
\right)^{-1} - \rho^2 \, d\Omega^2\,, \label{reinor}
\end{equation}
where $d\Omega^2= (d\theta^2 + \sin^2\theta \, d\phi^2)$ is the
metric on a $2$-sphere. The metric \eqn{reinor} admits two Killing
horizons, where the norm of the Killing vector
$\frac{\partial}{\partial t}$  changes sign. The horizons are
located at the two roots of the quadratic polynomial $\Delta \equiv
-2  M  \rho + Q^2 + \rho^2$:
\begin{equation}
 \rho_\pm = M \pm \sqrt{M^2-Q^2}\,.
 \label{2hor}
\end{equation}
If $M < |Q|$ the two horizons disappear and we have a naked
singularity. In classical general relativity people have postulated
the so-called {\it cosmic censorship} conjecture \cite{cato,black}:
space-time singularities should always be hidden inside a horizon.
This conjecture implies, in the R-N case, the bound:
\begin{equation}
M \, \ge \, |Q|\,. \label{censur}
\end{equation}
 Of
particular interest are the states that saturate the bound
\eqn{censur}. If
\begin{eqnarray}M=|Q|\,, \label{rhor}
\end{eqnarray}
the two horizons coincide and, setting: $\rho=r+M$ (where  $r^2 =
{\vec x}\, \cdot \, {\vec x}$), the metric \eqn{reinor} can be rewritten as:
\begin{eqnarray}
ds^2 & =& dt^2 \, \left( 1+ \frac{Q}{r}\right)^{-2} - \left( 1+
\frac{Q}{r}\right)^{ 2} \, \left( dr^2 +r^2 \,
d\Omega^2\right)\nonumber\\
& =&  \, H^{-2}(r)\, dt^2  - H^{2}(r)\, d{\vec x} \, \cdot \, d{\vec
x} \label{guarda!}
\end{eqnarray}
in terms of the harmonic function
\begin{equation}
\label{given}
 H (r)= \left(1 + \frac{Q}{r}\right)\,.
\end{equation}
\par
As eq. \eq{guarda!} shows, the extremal R-N configuration may be
regarded as  a soliton of classical general relativity,
interpolating between two vacua of the theory: the flat Minkowski
space-time, asymptotically reached at spatial infinity $r\to
\infty$, and the Bertotti--Robinson (B-R) metric \cite{br},
describing the conformally flat geometry $AdS_2 \times S^2$ near the
horizon $r \to 0$ \cite{black}:
 \begin{equation}ds^2_{\mbox{\small B-R}} =  \frac{r^2}{M_{\mbox{\small B-R}}^2}  dt^2 -
\frac{M_{\mbox{\small B-R}}^2}{r^2}\left( dr^2+r^2 d\Omega \right)\,. \label{br}
\end{equation} Last, let us note that the condition $M= |Q|$ can be
regarded as a no-force condition between the gravitational
attraction $F_g=\frac{M}{r^2}$ and the electric repulsion
$F_q=-\frac{Q}{r^2}$ on a unit mass carrying a unit charge.
\par

Until now we have reviewed the concept of extremal black holes as it
arises in classical general relativity.  However, extremal black
hole configurations are  embedded  in a natural way in supergravity
theories. Indeed supergravity, being invariant under local
super-Poincar\'e transformations, includes general relativity, i.e. it describes gravitation coupled to other fields in a
supersymmetric framework. Therefore it admits black holes among its classical
solutions. Moreover, as black holes describe a physical
regime where the gravitational field is very strong, a complete
understanding of their physics seems to require a theory of quantum
gravity, like superstring theory is. In this respect, as anticipated
above, extremal black holes have become objects of the utmost relevance
in the context of superstrings after 1995
\cite{blackmicro,review,black,ortino}. This interest, which is just
part of a more general interest in the $p$-brane classical solutions
of supergravity theories in all dimensions $4 \le D \le 11$
\cite{mbrastelle,duffrep}, stems from the interpretation of the
classical solutions of supergravity that preserve a fraction of the
original supersymmetries as non-perturbative states, necessary to
complete the perturbative string spectrum and make it invariant
under the many conjectured duality symmetries
\cite{schw,sesch,huto,gmv,vasch}.
Extremal black holes and their parent
$p$-branes in higher dimensions are then viewed as additional {\it
particle-like} states that compose the spectrum of a fundamental
quantum theory.
As the monopoles in gauge theories, these non-perturbative quantum
states originate from  regular solutions of the classical field
equations, the same Einstein equations one deals with in classical
general relativity and astrophysics. The essential new ingredient,
in this respect, is supersymmetry, which requires the presence of {\it
vector fields} and {\it scalar fields} in appropriate proportions.
Hence the black holes we are going to discuss are solutions of
generalized Einstein--Maxwell-dilaton equations.

Within the superstring framework, supergravity provides an effective description that holds
at lowest order in the string loop expansion and in the limit in which the space-time
curvature  is much smaller than the typical string scale (string tension). The supergravity
description of extremal black holes is therefore reliable when the radius of the horizon is
much larger than the string scale, and this corresponds to the limit of large charges.
Superstring corrections induce higher derivative terms in the low energy action and therefore the B-H entropy formula is expected to be corrected as well by terms which are subleading in
the small curvature limit.   In this paper we will not consider these higher derivative
effects.

\par
Thinking of a black-hole configuration as a particular bosonic
background of an $N$-extended locally supersymmetric theory gives a
simple and natural understanding at the cosmic censorship
conjecture. Indeed, in  theories with extended supersymmetry ($N\geq
2$) the bound \eqn{censur} is just a consequence of the
supersymmetry algebra, and this ensures that in these theories the
cosmic censorship conjecture is always verified, that is there are
no naked singularities. When the black hole is embedded in extended
supergravity, the model depends in general also on scalar fields. In
this case, as we will see,
 the electric charge $Q$ has to be replaced by the maximum eigenvalue
of the central charge appearing in the supersymmetry algebra
(depending on the expectation value of scalar fields and on the
electric and magnetic charges). The R-N metric takes in general a
more complicated form.

However, extremal black holes have a peculiar feature: even when the
dynamics depends on scalar fields, the event horizon loses all
information about the scalars; this is true
 independently of the fact that the solution preserves any supersymmetries
or not. Then, as will be discussed extensively in section
\ref{sec:extremum}, also if the extremal black hole is coupled to
scalar fields, the near-horizon geometry is still described by a
conformally flat, B-R-type geometry, with a mass parameter $M_{\mbox{\small B-R}}$
depending on the given configuration of electric and magnetic
charges, but not on the scalars. The horizon is in fact an attractor
point \cite{fekast,feka,strom3}: scalar fields, independently
of their boundary conditions at spatial infinity, when approaching the
horizon flow to a fixed point given by a certain ratio of electric
and magnetic charges. This may be understood in the context of
Hawking theory. Indeed quantum black holes are not stable: they
radiate a thermic radiation as a black body, and correspondingly
 lose their energy (mass). The only stable black-hole
configurations are
 the extremal ones, because they have the minimal possible
energy compatible with  relation \eqn{censur} and so they cannot
radiate. Indeed, physically they represent the limit case in which
the black-hole temperature, measured by the surface gravity at the
horizon, is sent to zero.
\par
Remembering now that the black-hole entropy is given by the
area/entropy B-H relation \eq{bekhaw}, we see that the entropy of
extremal black holes is a topological quantity, in the sense that it
is fixed in terms of the quantized electric and magnetic charges,
while it does not depend on continuous parameters such as scalars. The
horizon mass parameter $M_{\mbox{\small B-R}}$ turns out to be given in this case
(extremal configurations) by the maximum eigenvalue $Z_{\mbox{max}}$ of the
central charge appearing in the supersymmetry algebra, evaluated at
the fixed point:
\begin{equation}M_{\mbox{\small B-R}}=M_{\mbox{\small B-R}}(p,q)=|Z_{\mbox{max}}(\phi_{\mbox{\small fix}},p,q)|
\end{equation} this gives, for the B-H entropy:
\begin{equation}S_{\mbox{\small B-H}}=\frac{A_{\mbox{\small B-R}}(p,q)}{4}=\pi \vert
Z_{\mbox{max}}(\phi_{\mbox{\small fix}},p,q)\vert ^2 \,. \label{bert} \end{equation}
\vskip 5mm A lot of efffort was
  made in the course of the years to
give an explanation for the topological entropy of extremal black
holes in the context of a quantum theory of gravity, such as string
theory. A particularly interesting problem is finding a microscopic,
statistical mechanics interpretation of this thermodynamic quantity.
Although we will not deal with the microscopic  point of view at all
in this paper, it is important to mention that such an
interpretation became possible after the introduction of D-branes in
the context of string theory \cite{Polchinski:1995mt},
\cite{blackmicro}. Following this approach, extremal black
holes are interpreted as bound states of D-branes in a space-time
compactified to four or five dimensions, and the different
microstates contributing to the B-H entropy are, for instance,
related to the different ways of wrapping branes in the internal
directions. Let us mention that all calculations made in particular
cases using this approach provided values for the B-H entropy
compatible with those obtained with the supergravity, macroscopic
techniques. The entropy formula turns out to be in all cases a
U-duality-invariant expression (homogeneous of degree 2) built out
of electric and magnetic charges and as such it can be in fact also
computed through certain (moduli-independent) topological quantities \cite{FM},
which only depend on the nature of the U-duality groups and the
appropriate representations of electric and magnetic charges \cite{uinvar}.
 We mention for completeness that, as previously pointed out,
superstring corrections that take into account  higher derivative effects  determine  a deviation
from the area law for the entropy \cite{w,cdm}. Recently, a deeper insight into the
microscopic description of black-hole entropy  was gained, in this case, from the fruitful
proposal in \cite{osv}, describing the microscopic degrees of freedom of black holes in terms
of topological strings.

Originally, the attention was mainly devoted to the so-called {\it
BPS-extremal  black holes}, i.e. to solutions which saturate the
bound in \eq{rhor}.  From an abstract viewpoint BPS saturated states
are characterized by the fact that they preserve a fraction, $1/2$
or $1/4$ or $1/8$, of the original supersymmetries. What this
actually means is that there is a suitable projection operator $S^2
=S$ acting on the supersymmetry charge $Q_{\mbox{\small SUSY}}$, such that:
\begin{equation}
 \left(S \cdot Q_{\mbox{\small SUSY}} \right) \, \vert \, \mbox{BPS state} \, \rangle
 \,=  \, 0\,.
 \label{bstato}
\end{equation}
Since the supersymmetry transformation rules of any supersymmetric
field theory are linear in the first derivatives of the fields,
eq. \eqn{bstato} is actually a system of first-order
differential equations, to be combined with the second-order field
equations of the theory.
Translating eq. \eqn{bstato} into an explicit first-order
differential system requires knowledge of the supersymmetry
transformation rules of supergravity. The latter have a rich
geometric structure whose analysis will be the subject of section
\ref{gen4d}. The BPS saturation condition transfers the geometric
structure of supergravity, associated with its scalar sector, into
the physics of extremal black holes. We note that first-order
differential equations $\frac{d\Phi}{dr}= f(\Phi)$ have in general
fixed points, corresponding to the values of $r$ for which
$f(\Phi)=0$. For the BPS black holes, the fixed point is reached
precisely at the black-hole horizon, and this is how the attractor
behavior is realized for this class of extremal black holes.

For BPS configurations, non-renormalization theorems based on
supersymmetry guarantee the validity of the (BPS) bound $M=|Q|$
beyond the perturbative regime: if the bound is saturated in the
classical theory, the same must be true also when quantum
corrections are taken into account and the theory is in a regime
where the supergravity approximation breaks down. That it is
actually an exact state of non-perturbative string theory follows
from supersymmetry representation theory. The classical BPS state is
by definition an element of a short supermultiplet and, if
supersymmetry is unbroken, it cannot be renormalized to a long
supermultiplet. For this class of extremal black holes an accurate
agreement between the macroscopic and microscopic calculations was
found. For example in the $N=8$ theory the entropy was shown to
correspond to the unique quartic $E_{7(7)}$-invariant built in terms
of the $56$ dimensional representation. Actually,
 topological U-invariants constructed in
terms of the (moduli dependent) central charges and matter charges
 can be derived for
all $N \geq 2$ theories; they can be  shown, as expected, to coincide with the squared ADM mass
at fixed scalars.

 Quite recently it has been recognized that
the attractor mechanism, which is responsible for the area/entropy
relation, has  a larger application \cite{ortin,
fegika,gijt,k,tt,g,kss,dst} beyond the BPS cases, being  a
peculiarity of all {\em extremal black-hole configurations}, BPS
or not. The common feature is that extremal black-hole
configurations always belong to some representation of
supersymmetry, as will be surveyed in section \ref{introgen} (this
is not the case for non-extremal configurations, since the action
of supersymmetry generators cannot be defined for non-zero
temperature \cite{gibbons}). Extremal configurations that
completely break supersymmetry will belong to long representations
of supersymmetry.

Even for these more general cases, because of the topological nature of
the extremality condition, the entropy formula turns out to be still
given by a U-duality invariant expression built out of electric and
magnetic charges. We will report  in section \ref{casebycase} on the
classification of all extremal solutions (BPS and non-BPS) of
$N$-extended supergravity in four dimensions.

 For all the $N$-extended
theories in four dimensions, the general feature that allows us to find the B-H entropy as a
topological invariant is the presence of vectors and scalars in the same representation of
supersymmetry. This causes the electric/magnetic duality transformations on the vector field
strengths (which for these theories are embedded into symplectic transformations) to also act
as isometries on the scalar sectors \cite{gaizum} \footnote{We note that symplectic
transformations outside the U-duality group have a non-trivial action on the solutions, allowing one to bring a BPS configuration to a non-BPS one
 \cite{k1}.
}. The symplectic structure of the various $\sigma$-models of
$N$-extended supergravity in four dimensions and the relevant
relations involving the charges obeyed by the scalars will be worked
out in section \ref{gen4d}.

As a final remark, let us observe  that, since the aim of the present review is to calculate
the B-H entropy of extremal black holes, we will only discuss solutions which have
$S_{\mbox{\small B-H}}\neq 0$. For this class of solutions, known as {\it large black holes}, the classical
area/entropy formula is valid, as it gives the dominant contribution to the black-hole entropy.
For these configurations the area of the horizon is in fact proportional to a
duality-invariant expression constructed with the electric and magnetic charges, which for
these states is not vanishing \cite{libro}. This will prove to be a powerful computational tool and will be
the subject of section \ref{u-inv}. As we will see in detail in the following sections,
configurations with non-vanishing horizon area in supersymmetric theories  preserve at most
four supercharges ($N=1$ supersymmetry) in the bulk of space-time. Black-hole solutions
preserving more supercharges do exist, but they do not correspond to classical attractors
since in that case the classical area/entropy formula vanishes. These configurations are named
{\it small black holes} and require,  for finding the entropy, a quantum attractor mechanism
taking into account the presence of higher curvature terms \cite{osv,dw,sen}.

\par The paper is
organized as follows. Section \ref{introgen} treats the supersymmetry structure of extremal
 black-hole solutions of supergravity theories, and the black-hole configurations
 are described as massive representations of
supersymmetry.
  In section \ref{gen4d} we briefly review the
properties of four-dimensional extended supergravity related to its
global symmetries. A particular emphasis is given to the
general symplectic structure characterizing the moduli spaces of
these theories. The presence of this structure allows the global
symmetries of extended supergravities  to be realized as generalized
electric-magnetic symplectic duality transformations acting on the
electric and magnetic charges of dyonic solutions (as black
holes). In section \ref{sec:extremum} we start reviewing
extremal regular black-hole solutions embedded in supergravity
and, for
the BPS case, an explicit solution will be found by solving the
Killing spinor equations.
In section 
\ref{sec:geopot}
we give a general overview of extremal and non-extremal solutions showing how the attractor mechanism comes about in the extremal case only. 
 Then a general tool for calculating the Bekenstein--Hawking entropy for both  BPS and non-BPS extremal black holes
will be given, based on the  observation that
the black-hole potential takes a particularly simple form in the supergravity case,
which is fixed in terms of the geometric properties of the moduli space of the given theory.
Moreover, for theories based on moduli spaces given by symmetric manifolds $G/H$,
which is the case of all supergravity theories with $N\geq 3$ extended supersymmetry,
 but also of several $N=2$ models, the BPS and non-BPS black holes are classified by some
 U-duality invariant expressions, depending on the representation of the isometry group $G$
 under which the electric and magnetic charges are classified.
  Finally in section \ref{casebycase}, by exploiting the supergravity machinery introduced in
  sections \ref{gen4d} and \ref{sec:extremum},  we shall give a detailed analysis of the
   attractor solutions for  the various theories of extended
   supergravity. Section \ref{conclusions} contains some concluding
   remarks.

Our discussion will be confined to four-dimensional black holes.


\section{Extremal Black Holes as massive representations of supersymmetry}
\label{introgen}

We are going to review in the present section the algebraic
structure of the massive representations of supersymmetry, both for
short and long multiplets, in order to pinpoint, for each
supergravity theory, the extremal black-hole configurations
corresponding to a given number of preserved supercharges. The
condition of extremality is in fact independent on the supersymmetry
preserved by the solution, the only difference between the
supersymmetric and non-supersymmetic case being that the
configurations preserving some supercharges correspond to short
multiplets, while the configurations which completely break
supersymmetry  will instead belong to long representations of
supersymmetry. The highest spin of the configuration
 \footnote{We confine our analysis here to the {\em minimal} highest spin allowed for a given theory.}
 depends on the number of supercharges of the theory under consideration \cite{superhiggs}.

 As a
result of our analysis we find for example, as far as large BPS
black-hole configurations are considered, that for $N=2$ theories
the highest spin of the configuration (which in this case is
1/2-BPS) is $J_{MAX}=1/2$,  for $N=4$ theories (1/4-BPS) is
$J_{MAX}=3/2$, while for the $N=8$ case (1/8-BPS) is  $J_{MAX}=7/2$.
On the other hand, $1/2$-BPS multiplets have maximum spin $J_{MAX}=N/4$ ($N=2,4,8$) as for massless representations. They are given in Tables 2, 3, 4. The corresponding black holes (for $N>2$) have vanishing classical entropy (small black holes) \cite{FM}.

The long multiplets corresponding to non-BPS
extremal black-hole configurations have  $J_{MAX}=1$ in the $N=2$
theory, $J_{MAX}=2$ in the $N=4$ theory and $J_{MAX}=4$ in the $N=8$
theory. However, as we will see in detail in the following, for the
non-BPS cases we may have solutions with vanishing or non-vanishing
central charge. Since the central charge $Z_{AB}$ is a complex
matrix, it  is not invariant under CPT symmetry, but transforms as
$Z_{AB}\to \bar Z_{AB}$ \footnote{We use here a different definition
of central charge with respect to \cite{superhiggs}: $Z_{AB} \to \ii
Z_{AB}$. }. The representation then depends on the charge of the
configuration: if the solution has vanishing central charge the
long-multiplet will be neutral (real), while if the solution has non-zero
central charge the long multiplet will be charged (complex),
with a doubled dimension as required for CPT invariance
\cite{superhiggs}.

 We have listed in Tables \ref{spin3/2}, \ref{spin1} and \ref{spin1/2}  all possible massive  representations
 with highest spin $J_{MAX} \leq 3/2$  for $N\leq 8$. The occurence of long spin 3/2 multiplets is only possible for $N=3,2$ and of long spin
   1 multiplets for $N=2$. In $N=1$ there is only one type of massive multiplet (long) since there are
    no central charges. Its structure is
$$\bigl[(J_0+\frac{1}{2}),2(J_0),(J_0-\frac{1}{2})\bigr],$$ except
for $J_0=0$ where we have
 $\bigl[(\frac{1}{2}),2(0)\bigr]$.

In the tables we will denote the spin states by $(J)$ and the
number in front of them is their multiplicity. In the fundamental
multiplet, with spin $J_0=0$ vacuum, the multiplicity of the spin
$(N-q-k)/2$  is the dimension of the  $k$-fold antisymmetric
$\Omega$-traceless  representation of $USp(2(N-q))$. For
multiplets with $J_0\neq 0$ one has to make the tensor product of
the fundamental multiplet with the representation of spin $J_0$.
We also indicate if the multiplet is long or short.

\begin{table}[h]
\begin{center}
\begin{tabular} {|c|l|c|c|}
\cline{1-4}  $N$& massive spin 3/2 multiplet& long  &short
\\ \cline{1-4}&&&\\
8&none&&\\&&&\\
6&$2\times\bigl[(\frac{3}{2}),6(1),14(\frac{1}{2}),
14'(0)\bigr]$&no&$q=3,\,(\frac{1}{2}\mathrm{BPS})$\\&&& \\
5&$2\times\bigl[(\frac{3}{2}),6(1),14(\frac{1}{2}),
14'(0)\bigr]$&no&$q=2,\,(\frac{2}{5}\mathrm{BPS})$\\&&&\\
4&$2\times\bigl[(\frac{3}{2}),6(1),14(\frac{1}{2}),
14'(0)\bigr]$&no&$q=1,\,(\frac{1}{4}\mathrm{BPS})$\\&&&\\
&$2\times\bigl[(\frac{3}{2}),4(1),6(\frac{1}{2}),
4(0)\bigr]$&no&$q=2,\,(\frac{1}{2}\mathrm{BPS})$\\&&&\\
3&$\quad\;\;\;\bigl[(\frac{3}{2}),6(1),14(\frac{1}{2}),
14'(0)\bigr]$&yes&no\\&&&\\
&$2\times\bigl[(\frac{3}{2}),4(1),6(\frac{1}{2}),
4(0)\bigr]$&no&$q=1,\,(\frac{1}{3}\mathrm{BPS})$\\&&&\\
2&$\quad\;\;\;\bigl[(\frac{3}{2}),4(1),6(\frac{1}{2}),
4(0)\bigr]$&yes&no\\&&&\\
&$2\times\bigl[(\frac{3}{2}),2(1),(\frac{1}{2})\bigr]$&no&$q=1,\,(\frac{1}{2}
\mathrm{BPS})$\\&&&\\
1&$\quad\;\;\;\bigl[(\frac{3}{2}),2(1),(\frac{1}{2})\bigr]$&yes&no\\&&&\\
\cline{1-4}
\end{tabular}
\caption{Massive spin 3/2 multiplets.}\label{spin3/2}
\end{center}
\end{table}

\begin{table}[h]
\begin{center}
\begin{tabular} {|c|l|c|c|}
\cline{1-4}  $N$& massive spin 1 multiplet& long  &short
\\ \cline{1-4}&&&\\
8,6,5&none&&\\&&&\\ 4&$2\times\bigl[(1),4(\frac{1}{2}),
5(0)\bigr]$&no&$q=2,\,(\frac{1}{2}\mathrm{BPS})$\\&&& \\
3&$2\times\bigl[(1),4(\frac{1}{2}),
5(0)\bigr]$&no&$q=1,\,(\frac{1}{3}\mathrm{BPS})$\\&&&\\
2&$\quad\;\;\;\bigl[(1),4(\frac{1}{2}), 5(0)\bigr]$&yes&no\\&&&\\
&$2\times\bigl[(1),2(\frac{1}{2}),
(0)\bigr]$&no&$q=1,\,(\frac{1}{2}\mathrm{BPS})$\\&&&\\
1&$\quad\;\;\;\bigl[(1),2(\frac{1}{2}),
(0)\bigr]$&yes&no\\&&&\\\cline{1-4}
\end{tabular}
\caption{Massive spin 1 multiplets.}\label{spin1}

\end{center}
\end{table}

\begin{table}[h]
\begin{center}
\begin{tabular} {|c|l|c|c|}
\cline{1-4}  $N$& massive spin 1/2 multiplet& long  &short
\\ \cline{1-4}&&&\\
8,6,5,4,3&none&&\\&&&\\ 2&$2\times\bigl[(\frac{1}{2}),
2(0)\bigr]$&no&$q=1,\,(\frac{1}{2}\mathrm{BPS})$\\&&& \\
1&$\quad\;\;\;\bigl[(\frac{1}{2}),
2(0)\bigr]$&yes&no\\&&&\\\cline{1-4}
\end{tabular}
\caption{Massive spin 1/2 multiplets.}\label{spin1/2}
\end{center}
\end{table}


\subsection{Massive representations of the supersymmetry algebra} The $D=4$ supersymmetry
algebra
  is given by
\begin{eqnarray}
&\left\{ {\bar Q}_{A  \alpha }\, , \,{\bar Q}_{B  \beta}
\right\}\, = \, - \left( {C} \, \gamma^\mu \right)_{\alpha \beta}
\, P_\mu \, \delta_{AB} \, \, + {\rm i}\, ({ C}\, {\mathbb
Z}_{AB})_{\alpha
\beta} & \nonumber\\
&\left( A,B = 1,\dots, 2p \right)& \label{susyeven2}\,,
\end{eqnarray}
where the SUSY charges ${\bar Q}_{A}\equiv Q_{A}^\dagger \gamma_0=
Q^T_{A} \, { C}$ are Majorana spinors, $ C$ is the charge
conjugation matrix, $P_\mu$ is the 4-momentum operator and
 the antisymmetric tensor $\mathbb{Z}_{AB}$ is defined  as:
\begin{eqnarray}
\mathbb{Z}_{AB}&=&\Re (Z_{AB})+{\rm i}\,\gamma^5\,\Im(Z_{AB})\,,
\end{eqnarray}
the complex matrix ${ Z}_{AB}=-{ Z}_{BA}$ being the central charge
operator. For the sake of simplicity we shall suppress the spinorial
indices in the formulae. Using the symmetries of the theory, it can
always be reduced to normal form \cite{zum}. For $N$ even it reads:
\begin{equation}
{Z}_{AB} ~~~~= \pmatrix{\epsilon Z_1&0& \dots& 0\cr 0&\epsilon Z_2&
\dots &0\cr \dots &\dots & \dots & \dots \cr 0&0& \dots &\epsilon
Z_p\cr} \label{skewZ}\,,
\end{equation}
where $\epsilon$ is the $2\times 2$ antisymmetric matrix, (every
zero is a $2\times2$ zero matrix) and the $p$ skew eigenvalues $Z_m$
of ${ Z}_{AB}$ are the central charges. For $N$ odd the central
charge matrix has the same form as in \eq{skewZ} with $p= (N-1)/2$,
except for one extra zero row and one extra zero column. Note that
it is not always possible to reduce $Z_{AB}$ to its normal form with
real $Z_m$ by means of symmetries of the theory \cite{zum}. This is
the case in particular of $N=8$ supergravity where the $SU(8)$
R-symmetry does not affect the global phase of the skew-eigenvalues
$Z_m $. Therefore we shall consider the general situation in which
$Z_m $ are complex and define for each of them the spinorial
matrices which will enter the supersymmetry algebra:
\begin{eqnarray}
\mathbb{Z}_m&=&\Re (Z_{m})+{\rm
i}\,\gamma^5\,\Im(Z_{m})\,,\nonumber\\
\bar{\mathbb{Z}}_m&=&\Re (Z_{m})-{\rm
i}\,\gamma^5\,\Im(Z_{m})\,,\,\,\,\,\,n=1,\dots, p\,.
\end{eqnarray}
If we identify each index $A,B,\dots$ with the pair of indices
\begin{equation}
 A=(a,m) \quad ; \quad a,b,\dots =1,2 \quad ; \quad m,n,\dots =1,
 \dots, p\,,
 \label{indrang}
\end{equation}
the matrix $\mathbb{Z}_{AB}$ in the normal frame will have the
form:
\begin{eqnarray}
\mathbb{Z}_{AB}&=&\mathbb{Z}_{am,\,bn}=\mathbb{Z}_m\,\delta_{mn}\,\epsilon_{ab}\,,
\end{eqnarray}
and the superalgebra \eqn{susyeven2} can be rewritten as:
\begin{eqnarray}
&\left\{ {\bar Q}_{a m  }\, , \,{\bar Q}_{b n} \right\}\, = \,-
 \left( C \, \gamma^\mu \right) \, P_\mu \, \delta_{ab} \,
\delta_{mn} \, + {\rm i}\, C \, \epsilon_{ab} \,
\mathbb{Z}_m\,\delta_{mn}& \label{susyeven}
\end{eqnarray}
where $\epsilon_{ab}$ is the two-dimensional Levi Civita symbol. Let
us consider a generic unit time-like Killing vector $\zeta^\mu$
($\zeta^\mu\zeta_\mu=1$), in terms of which we define the following
projectors acting on both the internal ($a,m$) and Lorentz indices
($\alpha,\beta$) of the spinors:
\begin{eqnarray}
S^{(\pm)}_{am,\,bn}&=&\frac{1}{2}\,\left(\delta_{ab}\,\delta_{mn}\pm{\rm
i}\,\zeta_\mu\,\gamma^\mu\,\frac{\bar{\mathbb{Z}}_m}{\vert
Z_m\vert}\,\delta_{mn}\,\epsilon_{ab} \right)\,,\nonumber\\
\tilde{S}^{(\pm)}_{am,\,bn}&=&\frac{1}{2}\,\left(\delta_{ab}\,\delta_{mn}\pm{\rm
i}\,\zeta_\mu\,\gamma^\mu\,\frac{\mathbb{Z}_m}{\vert
Z_m\vert}\,\delta_{mn}\,\epsilon_{ab} \right)\,,
\end{eqnarray}
and define the projected supersymmetry generators:
\begin{eqnarray}
\bar{Q}^{(\pm)}&=&\bar{Q}\,S^{(\pm)}\,.
\end{eqnarray}
The anticommutation relation (\ref{susyeven}) can be rewritten in
the following form:
\begin{eqnarray}
\left\{ { Q}_{a m  }^{(\pm)}\, , \,{\bar Q}_{b n}^{(\pm)}
\right\}& =
&\tilde{S}^{(\pm)}_{am,\,bn}\,\zeta_\mu\,\gamma^\mu\,(\zeta_\nu\,P^\nu\mp
\vert Z_m\vert )\,.\label{susypm}
\end{eqnarray}
 In the case in which $\zeta^\mu=(1,0,0,0)$ and we are in the rest
 frame ($P^0=M$) the above relation reads:
\begin{eqnarray}
\left\{ { Q}_{a m  }^{(\pm)}\, , \,{Q}_{b n}^{(\pm)\,\dagger}
\right\}& = &\tilde{S}^{(\pm)}_{am,\,bn}\,(M\mp \vert Z_m\vert
)\,.\label{susypmrf}
\end{eqnarray}
Since the left-hand side of \eq{susypmrf} is non-negative definite,
we deduce   the BPS bound required by unitarity of the
representations:
\begin{equation}
M \, \ge \, \vert \, Z_m \vert \qquad \forall Z_m \, , \,
m=1,\dots,p \,. \label{bogobound}
\end{equation}
It is an elementary consequence of the supersymmetry algebra and of
the identification between central charges and topological charges
\cite{wittenolive}.

\paragraph{\bf Massive BPS multiplets\\}
Suppose that on a given state
$\vert BPS\rangle$ the BPS bound (\ref{bogobound}) is saturated by
 $q$ of the $p$ eigenvalues $Z_m $:
\begin{eqnarray}
M&=&\vert Z_1\vert=\vert Z_2\vert=\dots=\vert Z_q\vert\,\,\,\,q\le
p\,,
\end{eqnarray}
then, from (\ref{susypmrf}) we deduce that:
\begin{eqnarray}
{Q}^{(+)}_{am}\,\vert BPS\rangle&=&0\,\,\,,\,\,\,m=1,\dots,
q\,,\label{susinvbps}
\end{eqnarray}
namely  $q$ of the pairs of creation-annihilation operators, which
have abelian anticommutation relations, annihilate the state. The
 multiplet obtained by acting on $\vert BPS\rangle$ with the remaining supersymmetry generators is said to be $q/N$ BPS. Note that $q_{MAX}=N/2$ for
$N$ even and $q_{MAX}=(N-1)/2$ for $N$ odd. The $USp(2N)$ symmetry
is now reduced to $USp(2(N-q))$. The short multiplet has the same
number of states as a long multiplet of the $N-q$ supersymmetry
algebra. The fundamental multiplet, with $J=0$ vacuum, contains
$2\cdot2^{2(N-q)}$
 states with $J_{MAX}=(N-q)/2$. Note the doubling due to CPT invariance.
 Generic massive short multiplets can be obtained by making the tensor product
 with a spin $J_0$ representation of SU(2).

If we write the infinitesimal generator of a supersymmetry in the
form:
\begin{eqnarray}
\bar{Q}_A\,\epsilon_A&=&\bar{Q}_A^{(+)}\,\epsilon_A^{(+)}+\bar{Q}_A^{(-)}\,\epsilon_A^{(-)}\,,
\end{eqnarray}
the supersymmetries preserved by $\vert BPS\rangle$ are
parametrized by $\epsilon_{am}^{(+)}$ with $m\le q$ and thus
defined by the condition:
\begin{eqnarray}
\epsilon_{am}^{(-)}&=&S^{(-)}_{am,\,bn}\,\epsilon_{bn}=0\,\,\,;\,\,\,\,\,\,m,\,n\le
q\,,\\
\epsilon_{am}&=&0\,\,\,;\,\,\,\,m>q\,,
\end{eqnarray}
which can be written in terms of \emph{Weyl} spinors
$\epsilon_A,\,\epsilon^A$ in the following form:
\begin{eqnarray}
\epsilon_{am}&=&{\rm i}\,\frac{Z_m}{\vert
Z_m\vert}\,\zeta_\mu\,\gamma^\mu\,\epsilon_{ab}\,\epsilon^{b
m}={\rm i}\,\frac{Z_m}{\vert
Z_m\vert}\,\epsilon_{ab}\,\gamma^0\,\epsilon^{b
m}\,\,\,\,;\,\,\,\,\,\,m\le
q\,,\\
\epsilon_{am}&=&0\,\,\,;\,\,\,\,m>q\,.\label{epsiloncond}
\end{eqnarray}
If, in a given supergravity theory, the state $\vert BPS\rangle$
corresponds to a background described by a certain configuration
of fields, eq. \eqn{susinvbps} is translated into the request that
the supersymmetry variations of all the fields are zero in the
background.
 We consider extremal black-hole solutions for which the supersymmetry variations
 of the bosonic fields are identically zero.
 Then the condition \eqn{susinvbps}
yields a set of first order differential equations for the bosonic
fields, called ``Killing spinor'' equations,  to be satisfied on
the given configuration
\begin{equation}
0=\delta \mbox{fermions} = \mbox{SUSY rule} \left(
\mbox{bosons},\epsilon_{am} \right)\,, \label{fermboserule}
\end{equation}
where the supersymmetry transformations are made with respect to
the residual supersymmetry parameter $\epsilon^{(+)}_{am}$ defined
by the conditions (\ref{epsiloncond}). These conditions are
important in order to be able to recast equations
(\ref{fermboserule}) into differential equations involving only
the bosonic fields of the solution.

\paragraph{\bf Massive non-BPS multiplets\\}
Massive multiplets with $Z_m=0$ or $Z_m\neq 0$ but  $M>|Z_m|$ are
called long multiplets or non BPS states. They  are qualitatively
the same, the only difference being that  in the first case the
supermultiplets are real, while in the second one the
representations must be doubled in order to have CPT invariance,
since $Z_m \rightarrow  \bar Z_m$ under CPT.

 In both cases the supersymmetry algebra can be put in a form with $2N$ creation and $2N$ annihilation operators. It shows explicit invariance
  under $SU(2)\times USp(2N)$. The vacuum state is now labeled by the spin representation of SU(2),
   $|\Omega\rangle_J$. If $J=0$ we have the fundamental massive multiplet with $2^{2N}$ states. These
   are organized in  representations of SU(2) with $J_{MAX}=N/2$. With respect to $USp(2N)$ the states
   with fixed $0<J<N/2$ are arranged in the $(N-2J)$-fold $\Omega$-traceless antisymmetric representation,
    $[N-2J]$.

The general multiplet with a spin $J$  vacuum can  be obtained by tensoring the fundamental multiplet
 with spin $J$ representation of SU(2). The total number of states is then $(2J+1)\cdot2^{2N}$.


\section{ The general form of the supergravity action in four-dimensions and its BPS configurations}
 \label{gen4d} In this section we begin
the study of extremal black-hole solutions of extended supergravity
in four space-time dimensions. To this aim we first have to
introduce the main features of four dimensional $N$-extended
supergravities. These theories contain in the bosonic sector,
besides the metric, a number $n_V$ of vectors and $m$ of (real)
scalar fields. The relevant bosonic action is known to have the
following general form:
\begin{eqnarray}
 {\cal S}&=&\int\sqrt{-g}\,
d^4x\left(-\frac{1}{2}\,R+\Im{\cal N}_{\Lambda
\Gamma}F_{\mu\nu}^{\Lambda } F^{\Gamma |\mu\nu}+
\frac{1}{2\,\sqrt{-g}}\,\Re{\cal N}_{\Lambda \Gamma  }
\epsilon^{\mu\nu\rho\sigma}\, F_{\mu\nu}^{\Lambda } F^{\Gamma
}_{\rho\sigma}+\right.\nonumber\\
&+&\left.\frac 12 \,g_{rs}(\Phi) \partial_{\mu}
\Phi^{r}\partial^{\mu}\Phi^{s}\right)\,,\label{bosonicL}
\end{eqnarray}
where $g_{rs}(\Phi)$ ($r,s,\cdots =1,\cdots ,m$) is the scalar
metric on the $\sigma$-model described by the scalar manifold ${\cal
M}_{scalar}$ of real dimension $m$ and the vectors kinetic matrix
${\cal N}_{\Lambda\Sigma}(\Phi) $ is a complex, symmetric, ${n_V} \,
\times \, {n_V}$ matrix depending on the scalar fields. The
 number  of vectors and scalars, namely ${
n_V}$ and $m$, and the geometric properties of the scalar manifold
${\cal M}_{scalar}$ depend on the number $N$ of supersymmetries and
are resumed in Table \ref{topotable}. The imaginary part $\Im{\cal
N}$ of the vector kinetic matrix is negative definite and
generalizes the inverse of the squared coupling constant appearing
in ordinary gauge theories while its real part $\Re{\cal N}$ is
instead a generalization of the \emph{theta}-angle of quantum
chromodynamics.
 In supergravity theories, the
kinetic matrix ${\cal N}$ is in general not a constant, its
components being functions of the scalar fields.
 However, in extended supergravity ($N\geq 2$) the relation between the scalar geometry and the kinetic
matrix ${\cal N}$ has a very general and universal form. Indeed it
is related to the solution of a general problem, namely  how to lift
the action of the scalar manifold isometries from the scalar to the
vector fields. Such a lift is necessary because of supersymmetry
since scalars and vectors generically belong to the same
supermultiplet and must rotate coherently under symmetry operations.
This problem has been solved in a general (non supersymmetric)
framework in reference \cite{gaizum} by considering   the possible
extension of the Dirac electric-magnetic duality to more general
theories involving scalars. In the next subsection we review this
approach and in particular we show how enforcing covariance with
respect to such duality rotations leads to a determination of the
kinetic matrix ${\cal N}$. The structure of ${\cal N}$ enters the
black-hole equations in a crucial way so that the topological
invariant associated with the hole, that is its entropy, is an
invariant of the group of electro-magnetic duality rotations, the
U-duality group.
{\footnotesize
\begin{table}[h]
\begin{center}
\caption{\sl Scalar Manifolds of $N>2$ Extended Supergravities}
\label{topotable}
\begin{tabular}{|c||c|c|c|c|c| }
\hline N &  Duality group $G$ & isotropy $H$ & ${\cal M}_{scalar}$ & $n_V $&$ m$   \\
\hline \hline
$3$  &  $SU(3,n)$ & $SU(3) \times U(n)$ & $\frac{SU(3,n)}
{S(U(3)\times U(n))}$ & $3+n$& $6n$ \\
\hline $4$  &   $SU(1,1)\otimes SO(6,n)$ & $U(4) \times SO(n)$ &
$\frac{SU(1,1)}{U(1)} \otimes \frac{SO(6,n)}{SO(6)\times SO(n)}$ & $6+ n$& $6n +2$ \\
\hline $5$  &  $SU(1,5)$ &$U(5)$  & $\frac{SU(1,5)}
{S(U(1)\times U(5))}$ & 10& 10 \\
\hline $6$  &  $SO^\star(12)$ &$U(6)$ &
$\frac{SO^\star(12)}{U(1)\times SU(6)}$ & 16& 30 \\
\hline $7,8$&  $E_{7(7)}$ & $SU(8)$ &
$\frac{ E_{7(7)} }{SU(8)}$ & 28& 70 \\
\hline
\end{tabular}
\end{center}
{ In the table, $n_V$ stands for the number of vectors and $m$ for
the number of real scalar fields. In all the cases the duality
group $G$ is embedded in ${Sp}(2\,n_V,\mathbb{R})$.}
\end{table}
}


\subsection{Duality Rotations and Symplectic Covariance}
\label{LL1}  Let us review the general structure of an abelian
theory of vectors and scalars displaying covariance under a group
of duality rotations. The basic reference is the 1981 paper by
Gaillard and Zumino \cite{gaizum}. A general presentation in
$D=2p$ dimensions can be found in \cite{castdauriafre}. Here we
fix
 $D=4$.
\par
We consider a theory of  $n_V$ abelian gauge fields
$A^\Lambda_{\mu}$, in a $D=4$  space-time with Lorentz signature
(which we take to be mostly minus). They correspond to a set of
$n_V$ differential $1$-forms
\begin{equation}
A^\Lambda ~ \equiv ~ A^\Lambda_{\mu} \, dx^{\mu} \quad \quad
 \left ( \Lambda = 1,
\dots , {n_V} \right )\,.
\end{equation}
The corresponding field strengths and their Hodge duals are
defined by \footnote{We use, for the $\epsilon$ tensor, the convention: $\epsilon_{0123}=-1$.}:
\begin{eqnarray}
{ F}^\Lambda & \equiv & d \, A^\Lambda \,  \equiv   \,
    F^\Lambda_{\mu  \nu} \,
dx^{\mu } \, \wedge \, dx^{\nu} \nonumber\\
 F^\Lambda_{\mu \nu} & \equiv & {\frac{1}{2 }} \,\left (
\partial_{\mu } A^\Lambda_{\nu} \, - \, \partial_{\nu }
A^\Lambda_{\mu} \right ) \nonumber\\ ({}^{\star}
{F}^{\Lambda})_{\mu\nu} & \equiv &\, {\frac{\sqrt{-g}}{2}}
\varepsilon_{\mu \nu \rho\sigma}\,  F^{\Lambda \vert \rho \sigma}\,.
\label{campfort}
\end{eqnarray}
The dynamics of a system of abelian gauge fields coupled to scalars in a
 gravity theory is encoded in the bosonic action
(\ref{bosonicL}).

 Introducing self-dual and
antiself-dual combinations
\begin{eqnarray}
   F^{\pm} &=& {\frac 12}\left( F\, \pm {\rm i} \,
 {}^ \star
 F\right)\,, \nonumber \\
{}^\star \,  F^{\pm}& =& \mp \mbox{i}  F^{\pm}\,, \label{selfduals}
\end{eqnarray}
the vector part of the Lagrangian defined by (\ref{bosonicL}) can
be rewritten in the form:
\begin{equation}
{\cal L}_{vec} \,  = \, {\mbox{i}} \, \left [ F^{-T} {\bar {\cal
N}}  F^{-}-  F^{+T} {\cal N}  F^{+} \right]\,. \label{lagrapm}
\end{equation}
Introducing further the new tensors
\begin{equation}
  {{}^\star G}_{\Lambda|\mu\nu} \, \equiv \,  {\frac 12 } { \frac{\partial
{\cal L}}{\partial  F^\Lambda_{\mu\nu}}}= \Im{\cal
N}_{\Lambda\Sigma}\,F^\Sigma_{\mu\nu}  +\Re{\cal
N}_{\Lambda\Sigma}\, {}^\star F^\Sigma_{\mu\nu} \leftrightarrow
G^{\mp }_{\Lambda |\mu\nu} \, \equiv \, \mp{\frac {\rm i}{2} } {
\frac{\partial {\cal L}}{\partial  F^{\mp \Lambda}_{\mu\nu}}}\,,
\label{gtensor}
\end{equation}
the Bianchi identities and field equations associated with the
Lagrangian (\ref{bosonicL}) can be written as \bea
\nabla^{\mu }{{}^\star  F}^{\Lambda}_{\mu\nu} &=& 0 \nonumber\\
\nabla^{\mu }{{}^\star G}_{\Lambda |\mu\nu} &=& 0 \label{biafieq}
\eea or equivalently \bea
 \nabla^{\mu }
{\rm Im} F^{\pm \Lambda}_{\mu\nu} &=& 0 \\
\nabla^{\mu } {\rm Im} G^{\pm }_{\Lambda |\mu\nu} &=& 0 \ .
\label{biafieqpm} \eea This suggests that we introduce the $2{n_V}$
column vector
\begin{equation}
{\bf V} \, \equiv \, \pmatrix{ {}^{\star}  F\cr {}^{\star} G\cr }
\label{sympvec}
\end{equation}
and that we consider general linear transformations on such a
vector
\begin{equation}
\left ( \matrix{ {}^{\star}  F\cr {}^{\star}  G\cr }\right
)^\prime \, =\, \left (\matrix{ A & B \cr C & D \cr } \right )
\left ( \matrix{ {}^{\star}  F\cr {}^{\star}  G\cr }\right ).
\label{dualrot}
\end{equation}
For any constant matrix ${\cal S} =\left (\matrix{ A & B \cr C & D
\cr } \right ) \, \in \, GL(2{ n_V},\mathbb{R} )$ the new vector of
magnetic and electric field-strengths ${\bf V}^\prime ={\cal S}
\cdot {\bf V}$ satisfies the same equations ~\eqn{biafieq} as the
old one. In a condensed notation we can write
\begin{equation}
\partial \, {\bf V}\, = \, 0 \quad \Longleftrightarrow \quad
\partial \, {\bf V}^\prime \, = \, 0.
\label{dualdue}
\end{equation}
Separating the self-dual and antiself-dual parts
\begin{equation}
 F=\left ( F^+ + F^- \right ) \qquad ; \qquad
 G=\left ( G^+ + G^- \right ) \label{divorzio}
\end{equation}
and taking into account that   we have
\begin{equation}
 G^+ \, = \, {\cal N} F^+  \qquad ; \qquad  G^- \, = \, {\bar
{\cal N}} F^- \label{gigiuno}
\end{equation}
the duality rotation of eq.~\eqn{dualrot} can be rewritten as
\begin{equation}
\left ( \matrix{    F^+ \cr
  G^+\cr }\right )^\prime  \, = \,
\left (\matrix{ A & B \cr C & D \cr } \right ) \left ( \matrix{
F^+\cr {\cal N}  F^+\cr }\right ) \qquad ; \qquad \left ( \matrix{
  F^- \cr
  G^-\cr }\right )^\prime \, = \,
\left (\matrix{ A & B \cr C & D \cr } \right ) \left ( \matrix{
F^-\cr {\bar {\cal N}}  F^-\cr }\right ) . \label{trasform}
\end{equation}
Now, let us note that, since in the system we are considering (eq.
\eq{bosonicL}) the gauge fields are coupled to the scalar sector via
the scalar-dependent kinetic matrix $\mathcal{N}$,  when a duality
rotation is performed on the vector field strengths and their duals,
we have to assume that the scalars get transformed correspondingly,
through the action of some diffeomorphism on the scalar manifold
$\cM_{scalar}$. In particular, the kinetic matrix $\cN(\Phi)$
transforms under a duality rotation. Then, a duality transformation
$\xi$ acts in the following way on the supersymmetric system:
\begin{equation}
\xi : \left\{\matrix{V & \rightarrow & V^{\prime\mp}=S_\xi
V^\mp \cr
\Phi &\rightarrow &\Phi^\prime
=\xi(\Phi) \cr \cN(\Phi) & \rightarrow & \cN^\prime
\left(\xi(\Phi)\right)  }\right. \label{xi}
\end{equation}
Thus, the transformation laws of the equations of motion and of
$\cN$, and so also the matrix ${\cal S}_\xi$, will be induced by a
diffeomorphism of the scalar fields.

Focusing in particular on the first relation in \eq{xi}, that
explicitly reads:
\begin{equation}\left(\matrix{ F^{\pm\prime} \cr
G^{\pm\prime}\cr }\right)= \left(\matrix{A_\xi F^\pm +B_\xi G^\pm
\cr C_\xi F^\pm +D_\xi G^\pm\cr }\right),
\end{equation}we note that it contains the magnetic field
strength $G^\mp_\Lambda$ introduced in \eq{gtensor}, which is defined as a variation  of the
kinetic lagrangian. Under the transformations (\ref{xi}) the
lagrangian transforms in the following way:
\begin{eqnarray}
{\cal L}^\prime &=&{\rm i} \Bigl[
 \left(A_\xi + B_\xi\cN\right)^{\ \Lambda}_{\Gamma}
\left(A_\xi + B_\xi \cN\right)^\Sigma_{\ \Delta}
 \cN^\prime_{\Lambda\Sigma}(\Phi ) F^{+\Gamma}  F^{+\Delta} \nn\\
&-& \left(A_\xi + B_\xi\bar\cN\right)^{\ \Lambda}_{\Gamma}
\left(A_\xi + B_\xi\bar\cN\right)^\Sigma_{\ \Delta}
\bar\cN^\prime_{\Lambda\Sigma}(\Phi ) F^{-\Gamma}  F^{-\Delta}
\Bigr] ; \label{lageven}
 \end{eqnarray}
Equations \eq{xi} must be consistent with the definition of
$G^\mp$ as a variation of the lagrangian  \eq{lageven}:
\begin{equation}
G^{\prime +}_{\Lambda} = \left(C_\xi +
D_\xi\cN\right)_{\Lambda\Sigma} F^{+\Sigma} \equiv - \frac{\rm i
}{ 2} \frac{\partial {\cal L}^\prime}{  \partial  F^{\prime +
\Lambda}}= \left(A_\xi + B_\xi\cN\right)^\Delta_{\ \Sigma}
 \cN^\prime_{\Lambda\Delta} F^{+\Sigma}
\end{equation}
that implies:
\begin{equation}
\cN^\prime_{\Lambda\Sigma}(\Phi^\prime) = \left[\left(C_\xi +
D_\xi\cN\right) \cdot \left(A_\xi +
B_\xi\cN\right)^{-1}\right]_{\Lambda\Sigma} ; \label{ntrasfeven}
\end{equation}
The condition that the matrix $\cN$ is symmetric, and that this
property must be true also in the duality transformed system,
  gives the constraint:
  \begin{equation}  {\cal S}\in Sp(2n_V, \mathbb{R})\subset GL(2n_V,
  \mathbb{R})\,,
  \end{equation}
  that is:
\begin{eqnarray}
\cS^T\,\mathbb{C}\,\cS&=&\mathbb{C}\,,\label{syms}
\end{eqnarray}
where $\mathbb{C}$ is the symplectic invariant $2n_V\times 2 n_V$
matrix:
\begin{eqnarray}
\mathbb{C}&=&\left(\matrix{0 & -\bfone\cr \bfone &
0}\right)\,.\label{C}
\end{eqnarray}
It is useful to rewrite the symplectic condition (\ref{syms}) in
terms of the $n_V\times n_V$ blocks defining $\cS$:
\begin{eqnarray}
A^T\,C- C^T\,A&=& B^T\,D-D^T\,B=0\,\,\,;\,\,\,\,
A^T\,D-C^T\,B=\bfone\,.\label{abcd}
\end{eqnarray}
 The above observation has important implications on the scalar
manifold ${\cal M}_{scalar}$. Indeed, it
implies that on the scalar manifold the following homomorphism is
defined:
\begin{equation}Diff(\cM_{scalar}) \to
Sp(2n,\mathbb{R})\label{homo}\,.
\end{equation}
In particular, the presence on the manifold of a function of
scalars transforming with a fractional linear transformation under
a duality rotation  on the scalars, induces
the existence on $\cM_{scalar}$ of a linear structure (inherited
from the vectors).
As we are going to discuss in detail in section \ref{sugras},
this may be rephrased by saying that the scalar manifold is endowed with a symplectic bundle. As the transition functions of this bundle are given in terms of the {\it constant} matrix $\cS$, the symplectic bundle is flat.
In particular, as we will see in section
\ref{sugras}, for the $N=2$ four dimensional theory this implies
that the scalar manifold be a {\it special manifold}, that is a
K\"ahler--Hodge manifold endowed with a flat symplectic bundle.

If we are interested in the global symmetries of the theory (i.e.
global symmetries of the field equations and Bianchi identities) we
will need to restrict the duality transformations, namely the
homomorphism in (\ref{homo}), to the isometries of the scalar
manifold, which leave the scalar sector of the action invariant. The
transformations \eq{xi}, which are  duality symmetries of the system
field-equations/Bianchi-identities, cannot be extended in general to
be symmetries of the lagrangian. The scalar part of the lagrangian
\eq{bosonicL} is invariant under the action of the isometry group of
the metric $g_{rs}$, but the vector part is in general not
invariant. The transformed  lagrangian under the action of ${\cal
S}\in Sp(2n_V, \mathbb{R})$  can be rewritten:
\begin{eqnarray}
{\mbox Im}\left( F^{-\Lambda}G^-_\Lambda\right)& \rightarrow &
{\mbox Im}\left( F^{\prime-\Lambda}G^{\prime-}_\Lambda\right)\nonumber\\
&=& {\mbox Im}\bigl[  F^{-\Lambda}G^-_\Lambda +2(C^TB)_\Lambda^{\
\Sigma}
 F^{-\Lambda}G^-_\Sigma + \nonumber\\
&+&(C^TA)_{\Lambda\Sigma} F^{-\Lambda} F^{-\Sigma}+
(D^TB)^{\Lambda\Sigma} G^-_\Lambda G^-_\Sigma
\bigr]\label{trasfvett}\,.
\end{eqnarray}
It is evident from \eq{trasfvett} that only the transformations
with $B=C=0$
are symmetries.\\
If $C\not= 0$, $B=0$ the lagrangian varies for a topological term:
\begin{equation}
(C^TA)_{\Lambda\Sigma} F^\Lambda_{\mu\nu}{}^\star
 F^{\Sigma\vert{\mu\nu} }
\end{equation}
 corresponding to a redefinition of the function
$\Re\cN_{\Lambda\Sigma}$; such a transformation  being a total
derivative it leaves classical physics invariant,  but it is
relevant in the quantum theory. It is a symmetry of the partition
function only if
 $\Delta\Re\cN_{\Lambda\Sigma}=\frac{1 }{ 2}(C^TA)$
is an integer multiple of $2\pi$, and this implies that  $S\in
Sp(2n_V,\ZZ) \subset Sp(2n_V,\mathbb{R})$.\\
For $B\not= 0$ neither the action  nor the perturbative partition
function are invariant. Let us observe that in this case the
transformation law \eq{ntrasfeven} of the kinetic matrix $\cN$
contains the transformation $\cN\rightarrow -\frac{1}{ \cN}$ that is
it exchanges the weak and strong coupling regimes of the theory. One
may then think of such a quantum field theory as being described by
a collection of local lagrangians, each defined in a local patch.
They are all equivalent once one defines for each of them what is
{\it electric} and what is {\it magnetic}. Duality transformations
map this set of lagrangians one into the other.
\par
At this point we observe that the supergravity bosonic lagrangian
(\ref{bosonicL}) is exactly of the form considered in this section
as far as the matter content is concerned, so that we may apply the
above considerations about duality rotations to the supergravity
case. In particular, the U-duality acts in all theories with $N\geq
2$ supersymmetries, where the vector supermultiplets contain both
vectors and scalars. For $N=1$ supergravity, instead, vectors and
scalars are still present but they are not related by supersymmetry,
and as a consequence they are not related by U-duality rotations, so
that the previous formalism does not necessarily apply
\footnote{There are however  $N=1$ models where the scalar moduli
space is given by a special-K\"ahler manifold. This is the case for
example for the compactification  of the heterotic theory on
Calabi--Yau manifolds.}. In the next subsection we will discuss in a
geometric framework the structure of the supergravity theories for
$N\geq 2$. In particular, for theories whose $\sigma$-model is a
coset space (which includes all theories with $N>2$) we will give
the expression for the kinetic vector matrix $\cN_{\Lambda\Sigma}$
in terms of the $Sp(2n_V)$ coset representatives embedding the
U-duality group. Furthermore we will show that in the $N=2$ case,
although the $\sigma$-model of the scalars is not in general a coset
space, yet it may be treated in a completely analogous way.


\subsection{Duality symmetries and central charges} \label{sugras}  Let us restrict our attention to
$N$-extended supersymmetric theories coupled to the gravitational
field, that is to supergravity theories, whose bosonic action has
been given in \eqn{bosonicL}. For each theory we are going to
analyze the group theoretical structure and to find the expression
of the central charges, together with the properties they obey.
 As already mentioned,
 with the exception of the $N=1$ and $N=2$ cases, all  supergravity
theories in four dimensions contain scalar fields whose kinetic
Lagrangians are described by $\sigma$-models of the form ${ G}/H$
(we have summarized these cases in Table \ref{topotable}).  We will
first examine the theories with $N
> 2$,  extending then the results to the $N=2$ case.
Here and in the following, ${ G}$ denotes a non compact group
acting as  isometry group on the scalar manifold while $H$, the
isotropy subgroup, is of the form:
\begin{equation}
H=H_{Aut} \otimes H_{matter}
\end{equation}
 $H_{Aut}$ being the automorphism group of the supersymmetry algebra
 while $H_{matter}$ is related to the matter multiplets.
 (Of course $H_{matter}=\bfone$ in all cases where supersymmetric matter
 does not exist, namely $N>4$).

We will see that in all the  theories the fields are in some
representation of the isometry group ${ G}$  of the scalar fields or
of its maximal compact subgroup $H$. This is just a consequence of
the Gaillard--Zumino duality acting on the 2-form field strengths
and their duals, discussed in the preceding section.

The scalar manifolds and the automorphism groups of supergravity
theories for any $D$ and $N$  can be found in the literature
 (see for instance \cite{cj},
\cite{sase}, \cite{castdauriafre}, \cite{adf96}).
 As it was discussed in the previous section,
 the group ${  G}$ acts linearly in a symplectic representation on
 the electric and magnetic
 field strengths   appearing
 in the gravitational and matter
 multiplets. Here and in the following the index $\Lambda$ runs over
 the dimensions of some representation of the duality group ${\rm G}$.
Since consistency of the quantum theory requires the electric and
magnetic charges to satisfy a quantization condition, the true
duality symmetry at the quantum level (U-duality), acting on
 quantized charges,
 is a suitable discrete version of  the continuous group ${G}$
 \cite{huto}. The moduli space of these theories is $ { G}(\ZZ)
 \backslash {G}/H$.
 \par
 All the properties of the given supergravity theories for $N\geq 3$
 are completely fixed in terms of the geometry of ${G}/H$,
 namely in terms of the coset representatives $L$ satisfying the
 relation:
 \begin{equation}
 L(\Phi^\prime) =g L(\Phi) h (g,\Phi)
\end{equation}
where $g\in {G}$, $h\in H$  and $\Phi ^\prime =   \Phi ^\prime
 (\Phi)$,
 $\Phi$ being the coordinates of ${G}/H$.
Note that the  scalar fields in ${G}/H$ can be assigned, in the
linearized theory,  to linear representations
 $R_H$ of the local isotropy group  $H$ so that dim $R_H$ = dim ${G}$
 $-$ dim $H$ (in the full theory, $R_H $ is the representation
which the vielbein of ${G}/H$ belongs to).
\par With any field-strength $ F^\Lambda$ we may associate a magnetic
charge $p^\Lambda$ and  an electric charge $q_\Lambda$ given
respectively by:
\begin{equation}
p^\Lambda = \frac{1}{4\pi}\,\int _{S^2}  F^\Lambda \qquad \qquad
q_\Lambda = \frac{1}{4\pi}\,\int _ {S^2}  G _\Lambda\,,\label{pq}
\end{equation}
where $S^2$ is a spatial two-sphere in the space-time geometry of
the dyonic solution (for instance, in Minkowski space-time the
two-sphere at radial infinity $S^2_\infty$). Clearly the presence of
dyonic solutions requires the Maxwell equations  (\ref{biafieq})  to
be completed by adding corresponding electric and magnetic currents
on the right hand side. These charges however are not the physical
charges of the {\it interacting theory}; these latter can be
computed by looking at the transformation laws of the fermion
fields, where the physical field strengths appear dressed with the
scalar fields \cite{adf96},\cite{noi1}. It is in terms of these
interacting dressed field strengths that the field theory
realization of the central charges occurring in the supersymmetry
algebra \eq{susyeven2}
  is
given. Indeed, let us first introduce the central charges: they
are associated with the dressed 2-form $T_{AB}$ appearing in the
supersymmetry transformation law of the gravitino 1-form. The
physical graviphoton may be identified from the supersymmetry
transformation law of the gravitino field in the interacting
theory, namely:
\begin{equation}
\delta \psi_A = \nabla\epsilon_A + \alpha T_ {AB\vert
\mu\nu}\gamma^a\gamma^{ \mu\nu} \epsilon^B V_a+ \cdots
\label{tragra}
\end{equation}
Here $\nabla$ is the covariant derivative in terms of the
space-time spin connection and the composite connection of the
automorphism group $H_{Aut}$, $\alpha$ is a coefficient fixed by
supersymmetry, $V^a$ is the space-time vielbein, $A=1,\cdots ,N$
is the index acted on by the automorphism group. Here and in the
following
 the dots denote trilinear fermion terms which are characteristic
 of any supersymmetric theory but do not play any role in the following discussion.
The 2-form field strength $T_{AB}$ will be constructed by dressing
the bare field strengths  $ F^\Lambda$ with the coset representative
$L(\Phi)$ of ${G}/H$, $\Phi$ denoting a set of coordinates of ${\rm
G}/H$.
\par
 Note that the same field strength $T_{AB}$ which appears in the gravitino
 transformation law is also present in the dilatino transformation law
  in the following way:
  \begin{equation}
\delta \chi_{ABC} = P_{ABCD,\ell}\partial_\mu \phi^\ell \gamma^\mu
\epsilon^D +\beta T_{[AB\vert \mu\nu}\gamma^{\mu\nu} \epsilon_{C]}
+\cdots   \label{tradil}
\end{equation}
 Analogously, when vector multiplets are present,
 the matter vector field
 strengths $T_I$ appearing in the transformation laws of the gaugino
 fields, which are named matter vector field strengths,
 are linear combinations of the field strengths dressed with a different combination of the scalars:
  \begin{equation}
\delta \lambda_{IA}= {\rm i} P_{IAB,i}\partial_\mu \Phi^i \gamma^\mu
\epsilon^B +\gamma T_{I\vert \mu\nu}\gamma^{\mu\nu} \epsilon_A
+\cdots \,.\label{tragau}
\end{equation}
  Here $P_{ABCD}= P_{ABCD,\ell}d\phi^\ell$ and  $ P^I_{AB }=P^I_{AB,i }d\Phi^i$
  are the vielbein of the  scalar manifolds
  spanned by the scalar fields of the gravitational and vector multiplets respectively
  (more precise definitions are given below),
  and $\beta$ and $\gamma$ are constants fixed by supersymmetry.
  \par
In order to give the explicit dependence on scalars of $T_{AB}$,
$T^I$, it is necessary to recall from the previous subsection that,
according to the Gaillard--Zumino construction, the isometry group
${G}$ of the scalar manifold acts on the
  vector $( F^{- \Lambda},G^{-}_\Lambda)$
  (or its complex conjugate) as a subgroup of
  $Sp(2 n_V,\mathbb{R})$ ($n_V$ is the number of vector fields)
with duality transformations interchanging electric and magnetic
 field strengths:
 \begin{equation}
{\cal S} \left(\matrix{ F^{-\Lambda} \cr G^-_\Lambda\cr }\right)=
\left(\matrix{ F^{-\Lambda} \cr G^-_\Lambda\cr }\right)^\prime .
\end{equation}
\\
Let now $L(\Phi)$ be the coset representative of ${G}$ in the
symplectic representation, namely as a $2\,n_V\times 2\,n_V$ matrix
belonging to  $Sp(2n_V,\mathbb{R})$ and therefore, in each theory,
it can be described in terms of $n_V\times n_V$ blocks
$A_L,B_L,C_L,D_L$ satisfying the same relations (\ref{abcd}) as the
corresponding blocks of the generic symplectic transformation ${\cal
S}$.

Since the fermions of supergravity theories transform in a complex
representation of the R-symmetry group $H_{Aut}\subset G$, it is
useful to introduce a complex basis in the vector space of $Sp(2
n_V,\mathbb{R})$, defined by the action of following unitary
matrix:\footnote{We adopt here and in the following a condensed
notation where $\bfone$ denotes the $n_V$ dimensional identity
matrix $\bfone^M_N = \delta^M_N$. In supergravity calculations,
the index $M$ is often decomposed as $M=(AB, I)$, $AB = -BA$
labelling the two-times antisymmetric representation of the
R-symmetry group $H_{Aut}$ and $I$ running over the $H_{matter}$
representation of the matter fields. We use the convention that
the sum over the antisymmetric couple $AB$  be free and therefore
supplemented by a factor $1/2$ in order to avoid repetitions. In
particular with these conventions, when restricted to the $AB$
indices, the identity reads: $\bfone^{AB}_{CD}\equiv
2\,\delta^{AB}_{CD}=\delta^A_C\delta^B_D-\delta^A_D\delta^B_C$.}
\begin{eqnarray}
\cA&=& \frac{1}{\sqrt{2}}\,\left(\matrix{\bfone & i\,\bfone \cr
\bfone & -i\,\bfone\cr }\right)\,,\nonumber
\end{eqnarray}
and to introduce a new matrix ${\bf V}(\Phi)$ obtained by
complexifying the right index of the coset representative
$L(\Phi)$, so as to make its transformation properties under right
action of $H$ manifest:
\begin{equation}
{\bf V}(\Phi) = \left(\matrix{{\bf f} & \bar {\bf f} \cr {\bf h}
&\bar {\bf h} \cr
 }\right) =
   L(\Phi) \cA^\dagger\,,
  \label{defu}
\end{equation}
where:
\begin{eqnarray}
  {\bf f}&=&\frac{1}{\sqrt{2}} (A_L-{\rm i} B_L)\,\,;\,\,\,
{\bf h} =\frac{1}{\sqrt{2}} (C_L-{\rm i} D_L)\,, \nonumber
\end{eqnarray}
From the properties of $L(\Phi)$ as a symplectic matrix, it is
easy to derive the following properties for ${\bf V}$:
\begin{eqnarray}
{\bf V}\,\eta\,{\bf V}^\dagger &=&
-\ii\mathbb{C}\,\,\,;\,\,\,\,\,{\bf V}^\dagger\,\mathbb{C}\,{\bf
V} = \ii\eta\,,\label{propu}
\end{eqnarray}
where the symplectic invariant matrix $\mathbb{C}$ and $\eta$ are
defined as follows:
\begin{eqnarray}
\mathbb{C}&=&\left(\matrix{0& -\bfone \cr \bfone & 0\cr
}\right)\,\,;\,\,\,\eta = \left(\matrix{\bfone & 0 \cr 0 & -
\bfone}\right)\,,
\end{eqnarray}
and, as usual, each block is an $n_V\times n_V$ matrix. The above
relations imply on the matrices ${\bf f}$ and ${\bf h}$ the
following properties:
 \begin{equation}
\left\lbrace\matrix{{\rm i}({\bf f}^\dagger {\bf h} - {\bf
h}^\dagger {\bf f}) &=& \bfone \cr ({\bf f}^t  {\bf h} - {\bf h}^t
{\bf f}) &=& 0\cr } \right. \label{specdef}
\end{equation}
The $n_V\times n_V$  blocks $ {\bf f},\, {\bf h}$ of ${\bf V}$ can
be decomposed with respect to the isotropy group $H_{Aut} \times
H_{matter}$ as:
\begin{eqnarray}
  {\bf f}&=& (f^\Lambda_{AB} , \bar{f}^\Lambda _{\bar{I}})\equiv ({\bf f}^\Lambda{}_M) \,, \nonumber\\
{\bf h}&=& (h_{\Lambda AB} , \bar{h}_{\Lambda \bar{I}})\equiv
({\bf h}_{\Lambda\,M})\,, \label{deffh}
\end{eqnarray}
where $AB$ are indices in the antisymmetric representation of
$H_{Aut}= SU(N) \times U(1)$, $I$ is an index of the fundamental
representation of $H_{matter}$ and $M=(AB,\,\bar{I})$. Upper
$SU(N)$ indices label objects in the complex conjugate
representation of $SU(N)$: $(f^\Lambda_{AB})^* = \bar{f}^{\Lambda
AB}$, $(f^{\Lambda}_ {I})^* = \bar{
f}^\Lambda_{\bar{I}}=\bar{f}^{\Lambda\, I}$ etc...

Let us remark that,  in order to make contact with the notation used for the $N=2$ case,
 in the definition \eq{deffh} some of the entries ($\bar{f}^\Lambda _{\bar{I}}$ and $\bar{h}_{\Lambda \bar{I}}$) have been written as complex conjugates of other quantities ($f^\Lambda_I$ and $h_{\Lambda I}$ respectively). In this way, $f^\Lambda_{AB}$ and $f^\Lambda_I$ are characterized by having  K\"ahler weight of the same sign. Indeed, for  all the matter coupled theories ($N=2,3,4$) we have, as a general feature, that the entries of the blocks ${\bf f}$ and ${\bf h}$ carrying $H_{matter}$ indices have a K\"ahler  weight with an opposite sign with respect to the corresponding entries with $H_{Aut}$ indices.
This may be seen from the supersymmetry transformation rules of the supergravity fields, in virtue the fact that gravitinos and gauginos with the same chirality have opposite K\"ahler weight.
We note however that this notation differs  from the one in previous papers, where the upper and lower parts of the symplectic section were defined
instead as $(f^\Lambda_{AB} , {f}^\Lambda _{I})\,,\; (h_{\Lambda \,AB}, h_{\Lambda \, I})$.
\par
It is useful to introduce the following quantities
\begin{eqnarray}
{\bf V}_M&=& (V_{AB},\,\overline{V}_{\bar{I}})\,,\quad \mbox{where:}\nonumber\\
V_{AB}&\equiv &(f^\Lambda_{AB}, h_{\Lambda
AB})\,\,\,;\,\,\,\,V_{I}\equiv(f^\Lambda_{I},\, {h}_{\Lambda
{I}})\,.\label{defum}
\end{eqnarray}
The vectors ${\bf V}_M$ are  (complex) symplectic sections of a
$Sp(2n_V, \mathbb{R})$ bundle over ${G}/H$. As anticipated in the
previous subsection, this bundle is actually flat.
 The real
embedding given by $L(\Phi )$ is appropriate for duality
transformations of $ F^\pm$
 and their duals $G^\pm$, according to equations (\ref{trasform}),
  while
the complex embedding in the matrix ${\bf V}$ is appropriate in
writing down the fermion transformation laws and supercovariant
field strengths. The kinetic matrix $\cN$, according to
Gaillard--Zumino \cite{gaizum}, can be written in terms of the
subblocks ${\bf f}$, ${\bf h}$, and turns out to be:
\begin{equation}
\cN= {\bf h}\,{\bf f}^{-1}, \quad\quad \cN = \cN^t\,,
\label{nfh-1}
\end{equation}
transforming projectively under $Sp(2n_V ,\mathbb{R})$ duality
rotations as already shown in the previous section. By using
(\ref{specdef})and (\ref{nfh-1}) we find that
   \begin{equation}
   ({\bf f}^t)^{-1} = {\rm i} (\cN - \bar \cN)\bar {\bf f}\,,
\end{equation}
 that is
 \begin{eqnarray}
({\bf f}^{-1})^{AB}{}_{\Lambda} &=&
{\rm i} (\cN - \bar \cN)_{\Lambda\Sigma} \bar{f}^{\Sigma\,AB}\,,\\
({\bf f}^{-1})^{\bar{I}}{}_{\Lambda} &=& {\rm i} (\cN - \bar
\cN)_{\Lambda\Sigma} f^{\Sigma\,\bar{I}}\,.
\end{eqnarray}
 It can be shown \cite{adf96} that the dressed graviphotons and matter self-dual
field strengths appearing in the transformation
 law of gravitino (\ref{tragra}), dilatino (\ref{tradil})
 and gaugino (\ref{tragau})
  can be constructed as a symplectic invariant  using the ${\bf f}$ and ${\bf h}$ matrices, as follows:
\begin{eqnarray}
T^-_{AB}&=& -{\rm i} (\bar {\bf f}^{-1})_{AB \Lambda} F^{-
\Lambda} = f^\Lambda_{AB}\,(\cN - \bar \cN)_{\Lambda\Sigma}\,
F^{-\Sigma}=
 h_{\Lambda AB}\,  F^{-\Lambda} -f^\Lambda_{AB} \,G^-_\Lambda \,, \nonumber\\
  \bar{T}^-_{\bar{I}}&=&-{\rm i} (\bar {\bf f}^{-1})_{\bar{I}\Lambda} \,F^{- \Lambda} =
  \bar{f}^\Lambda_{\bar{I}}\,(\cN - \bar \cN)_{\Lambda\Sigma}\, F^{-\Sigma}=
 \bar{h}_{\Lambda \bar{I}} \, F^{-\Lambda} - \bar{f}^\Lambda_{\bar{I}}\, G^-_\Lambda \,, \nonumber\\
\bar T^{+ AB} &=& (T^-_{AB})^*\,,  \nonumber\\
 T^{+ }_I &=& (\bar{T}^-_{\bar{I}})^* \,, \label{gravi}
\end{eqnarray}
(for $N>4$, supersymmetry does not allow matter multiplets and $f
^{\Lambda}_I=0= T_I$).
 To construct the dressed charges one integrates
 $T_{AB} = T^+_{AB} + T^ -_{AB}  $ and  (for $N=3, 4$)
 $\bar{T}_{\bar{I}} =\bar{T}^+_{\bar{I}} + \bar{T}^ -_{\bar{I}}$ on a large 2-sphere.
 For this purpose we note that
 \begin{eqnarray}
 T^+_{AB} & = &  h_{\Lambda
 AB}\, F^{+\Lambda} - f^\Lambda_{AB} \,G_\Lambda^+  =0  \,,\label{tiden0}\\
   \bar{T}^+_{\bar{I}} & = &  \bar{h}_{\Lambda
 {\bar{I}}}\, F^{+\Lambda}-\bar{f}^\Lambda_{{\bar{I}}}\, G_\Lambda^+   =0 \,,\label{tiden}
\end{eqnarray}
as a consequence of eqs. (\ref{nfh-1}), (\ref{gigiuno}). Therefore
we can introduce the central and matter charges as the dressed
charges obtained by integrating the 2-forms $T_{AB}$ and
$T_{\bar{I}}$:
\begin{eqnarray}
Z_{AB}& = & -\frac{1}{4\pi}\,\int_{S^2} T_{AB} =
-\frac{1}{4\pi}\,\int_{S^2} (T^+_{AB} + T^ -_{AB}) =
-\frac{1}{4\pi}\,\int_{S^2} T^ -_{AB} =\nonumber\\ &=&
 f^\Lambda_{AB} \,q_\Lambda -h_{\Lambda AB} \,p^\Lambda\,,
\label{zab}\\
\overline{Z}_{\bar{I}} & = & -\frac{1}{4\pi}\,\int_{S^2}
\bar{T}_{\bar{I}} = -\frac{1}{4\pi}\,\int_{S^2}
(\bar{T}^+_{\bar{I}} + \bar{T}^ -_{\bar{I}}) =-\frac{1}{4\pi}\,
\int_{S^2} \bar{T}^ -_{\bar{I}} =\nonumber\\ &=&
  \bar{f}^\Lambda_{\bar{I}} \,q_\Lambda-\bar{h}_{\Lambda {\bar{I}}} \,p^\Lambda    \quad (N\leq
 4)\,,\nonumber\\&&
\label{zi}
\end{eqnarray}
where $p^\Lambda$ and $q_\Lambda$ were defined in (\ref{pq})
 and
the sections $(f^\Lambda, h_\Lambda)$ on the right hand side now
depend on the v.e.v.'s $\Phi_\infty\equiv \Phi(r=\infty )$ of the
scalar fields $\Phi^r$. We see that because of the
electric-magnetic duality,  the central and matter charges are
given in this case by symplectic invariant expressions.
\par
 The scalar field dependent combinations of
fields strengths appearing in the fermion supersymmetry
transformation rules have a profound meaning and, as we are going
to see in the following, they play a key role in the physics of
extremal black holes. The integral of the graviphoton
$T_{AB\mu\nu}$  gives the value of the central charge $Z_{AB}$ of
the   supersymmetry algebra, while by integration of the matter
field strengths $T_{I| \mu\nu}$ one obtains the so called matter
charges $Z_I$.
\par
We are now able to derive some  differential relations among the
central and matter charges using the Maurer--Cartan equations
obeyed by the scalars through the embedded coset representative
${\bf V}$. Indeed, let $\Gamma = {\bf V}^{-1} d{\bf V}$ be the
$Sp(2n_V,\mathbb{R})$ Lie algebra left invariant one form
satisfying:
\begin{equation}
  d\Gamma +\Gamma \wedge \Gamma = 0\,.
\label{int}
\end{equation}
In terms of $({\bf f},{\bf h})$, $\Gamma$ has the following form:
\begin{equation}
  \label{defgamma}
  \Gamma \equiv {\bf V}^{-1} d{\bf V} =
\left(\matrix{{\rm i} ({\bf f}^\dagger d{\bf h} - {\bf h}^\dagger
d{\bf f}) & {\rm i} ({\bf f}^\dagger d\bar {\bf h} - {\bf
h}^\dagger d\bar {\bf f}) \cr -{\rm i} ({\bf f}^t d{\bf h} - {\bf
h}^t d{\bf f}) & -{\rm i}({\bf f}^t d\bar {\bf h} - {\bf h}^t
d\bar {\bf f}) \cr }\right) \equiv \left(\matrix{\Omega^{(H)}&
\bar \cP \cr \cP & \bar \Omega^{(H)} \cr  }\right)\,,
\end{equation}
where the $n_V \times n_V$ sub-blocks $ \Omega^{(H)}$ and  $\cP$
embed the $H$ connection and the vielbein of ${G}/H$ respectively.
This identification follows from the Cartan decomposition of the
$Sp(2n_V,\mathbb{R})$ Lie algebra.

From  (\ref{defu}) and (\ref{defgamma}),  we obtain the $(n_V
\times n_V)$ matrix equation:
\begin{eqnarray}
 D(\Omega) {\bf f} &=& \bar {\bf f} \, \cP \,,\nonumber \\
D(\Omega) {\bf h} &=&   \bar {\bf h} \,\cP \,, \label{nablafh}
 \end{eqnarray}
together with their complex conjugates. Explicitly, if we define
the $H_{Aut} \times H_{matter}$-covariant derivative of the ${\bf
V}_M$ vectors, introduced in (\ref{defum}), as:
\begin{equation}
D {\bf V}_M = d{\bf V}_M - {\bf V}_N\,\omega^N{}_M , \quad \omega
= \left(\matrix{\omega^{AB}_{\ \ CD} & 0 \cr 0 & \omega^I_{\ J}
\cr }\right)\,, \label{nablav}
\end{equation}
we have:
\begin{equation}
\Omega^{(H)} = {\rm i} [ {\bf f}^\dagger (D {\bf h}+ {\bf h}
\omega) - {\bf h}^\dagger (D {\bf f} + {\bf f} \omega)] = \omega
\bfone\,, \label{defomega}
\end{equation}
where we have used:
\begin{equation}
D {\bf h} = \bar \cN D {\bf f} ; \quad {\bf h} = \cN {\bf f}\,,
\end{equation}
which follow from \eq{nablafh} and  the fundamental identity
(\ref{specdef}). Furthermore, using the same relations, the
embedded vielbein  $\cP$ can be written as follows:
\begin{equation}
\cP = - {\rm i} ( {\bf f}^t D {\bf h} - {\bf h}^t D {\bf f}) =
{\rm i} {\bf f}^t (\cN - \bar \cN) D {\bf f} \,.
\end{equation}
 Using further the
definition (\ref{deffh}) we have:
\begin{eqnarray}
  D(\omega) f^\Lambda_{AB} &=&   f^\Lambda_{I}  P^I_{AB} + \frac{1}{2}\bar f^{\Lambda CD} P_{ABCD} \,,\nonumber \\
D(\omega) \bar{f}^\Lambda_{\bar{I}} &=&  \frac{1}{2} \bar
f^{\Lambda AB} P_{AB \bar{I}} +f^{\Lambda \bar{J}}
P_{\bar{J}\bar{I}}\,,
 \label{df}
\end{eqnarray}
where we have decomposed the embedded vielbein $\cP$ as follows:
\begin{equation}
  \label{defp}
  \cP = \left(\matrix{P_{ABCD} & P_{AB \bar{J}} \cr P_{\bar{I} CD} & P_{\bar{I}\bar{J}}\cr
  }\right)\,,
\end{equation}
 the subblocks being related to the vielbein of ${G}/H$,
 written in terms of the indices of $H_{Aut} \times H_{matter}$.
 In particular, the component $P_{ABCD}$ is completely
 antisymmetric in its indices.
Note that, since ${\bf f}$ belongs to the unitary matrix ${\bf
V}$, we have: $\overline{{\bf V}}^M=(  f^\Lambda_{AB},
\bar{f}^\Lambda_{\bar{I}})^\star = (\bar f^{\Lambda AB},
f^{\Lambda \bar{I}})$. Obviously, the same differential relations
that we wrote for ${\bf f}$ hold true for the dual matrix ${\bf
h}$ as well.
\par
Using the definition of the charges (\ref{zab}), (\ref{zi}) we
then get the following differential relations among charges:
\begin{eqnarray}
  D(\omega) Z_{AB} &=&  Z_{I}  P^I_{AB} +  \frac{1}{2} \bar Z^{CD} P_{ABCD}\,, \nonumber \\
D(\omega) \bar{Z}_{\bar{I}} &=& \frac{1}{2}  \bar Z^{AB}  P_{AB
\bar{I}} + Z^{\bar{J}}  P_{\bar{I}\bar{J}}\,.
 \label{dz1}
\end{eqnarray}
Depending on the coset manifold, some of the subblocks of
(\ref{defp}) can be actually zero. For example in $N=3$ the
vielbein of ${G}/H = \frac{SU(3,n)}{SU(3)
 \times SU(n)\times U(1)}$ \cite{ccdffm} is $P_{\bar{I} AB}$ ($AB $ antisymmetric),
 $I=1,\cdots,n;
A,B=1,2,3$ and it turns out that $P_{ABCD}= P_{\bar{I}\bar{J}} =
0$.
\par
In $N=4$, ${G}/H= \frac{SU(1,1)}{U(1)} \times \frac{ O(6,n)}{ O(6)
\times O(n)}$ \cite{bekose}, and we have $P_{ABCD}=
\epsilon_{ABCD} P$, $P_{\bar{I}\bar{J}} =  P \delta_{IJ}$, where
$P$ is the K\"ahlerian vielbein of $\frac{SU(1,1)}{U(1)}$,
($A,\cdots , D $ $SU(4)$ indices and $I,J$ $O(n)$ indices) and
$P_{I AB}$ is the vielbein of $\frac{ O(6,n)}{O(6) \times O(n)}$.
\par
For $N>4$ (no matter indices) we have that $\cP $ coincides with
the vielbein $P_{ABCD}$ of the relevant ${G}/H$.
\par
For the purpose of comparison of the previous formalism with the
$N=2$ supergravity case, where the $\sigma$-model is in general
not a coset, it is interesting to note that, if the connection
$\Omega^{(H)}$ and the vielbein $\cP$ are regarded as data of
${G}/H$, then the Maurer--Cartan equations (\ref{df}) can be
interpreted as an integrable system of differential equations for
the section $(V_{AB} ,\, \bar{V}_{\bar{I}} ,\, \bar V^{AB}  ,\,
V^{\bar{I}} )$ of the symplectic fiber bundle constructed over
${G}/H$. Namely the integrable system (\ref{nablav}) that we
explicitly write in the following equivalent matrix form
\begin{equation}
 D \left(\matrix{V_{AB} \cr \bar{V}_{\bar{I}} \cr \bar V^{AB} \cr  V^{\bar{I}}  }\right) =
  \left(\matrix{0 & 0 & \frac{1}{2} P_{ABCD} & P_{AB \bar{J}} \cr
0 & 0 & \frac{1}{2} P_{\bar{I} CD} & P_{\bar{I}\bar{J}} \cr
\frac{1}{2}\bar{P}^{ABCD} & \bar{P}^{ AB \bar{J}} & 0 & 0 \cr
\frac{1}{2}\,\bar{P}^{\bar{I} CD} & \bar{P}^{\bar{I}\bar{J}} & 0 &
0 \cr }\right)
 \left(\matrix{V_{CD} \cr \bar{V}_{\bar{J}} \cr \bar V^{CD} \cr V^{\bar{J}}
 }\right)\,,
 \label{intsys}
\end{equation}
 has $2n_V$ solutions given by ${\bf V}_M$.
The integrability condition (\ref{int}) means that $\Gamma$ is a
flat connection of the symplectic bundle. In terms of the geometry
of ${G}/H$ this in turn implies that the $\IH$-curvature
associated to the connection $\Omega^{(H)}$ (and hence, since the
manifold is a symmetric space, also the Riemannian curvature) is
constant, being proportional to the wedge product of two vielbein.
\par
Furthermore, besides the differential relations (\ref{dz1}) the
charges also satisfy sum rules.
\par
The sum rule has the following form:
\begin{equation}
 \frac{1}{2} Z_{AB} \bar Z^{AB} + Z_I \bar Z^I=-\frac{1}{2} Q^t \cM (\cN)
 Q\,, \label{sumrule}
\end{equation}
where $\mathbb{C}$ is the symplectic metric while $\cM(\cN)$ and
$Q$ are:
\begin{eqnarray}
\cM (\cN) &=&  \pmatrix{ \bfone & - \Re \cN \cr 0 &\bfone\cr }
\cdot  \pmatrix{ \Im \cN & 0 \cr 0 &\Im \cN^{-1}\cr } \cdot
\pmatrix{ \bfone
& 0 \cr - \Re \cN & \bfone \cr } \nonumber\\
&=& \pmatrix{  \Im \cN  +  \Re \cN \Im \cN^{-1}\Re \cN & - \Re \cN
\Im \cN^{-1}\cr- \Im \cN^{-1}\Re \cN & \Im \cN^{-1}\cr
}=\mathbb{C}\,{\bf V}\,{\bf V}^\dagger\,\mathbb{C}\,,\nonumber\\&&
\label{m+}
\end{eqnarray}
and
\begin{equation}
Q=\left(\matrix{p^\Lambda \cr q_\Lambda \cr } \right)
\label{eg}\,.
\end{equation}
This result is obtained from the
 fundamental identities (\ref{specdef}) and  from the definition of ${\bf V}$ and of the
 kinetic matrix given in (\ref{defu}) and (\ref{nfh-1}). Indeed one can verify that  \cite{adf96,fk2006}:
 \begin{eqnarray}
{\bf f}\,{\bf f}^\dagger &= &-{\rm i} \left( \cN - \bar \cN \right)^{-1}\,,\nonumber \\
{\bf h}\,{\bf h}^\dagger &= &-{\rm i} \left(\bar \cN^{-1} -
\cN^{-1} \right)^{-1}\equiv
-{\rm i} \cN \left( \cN - \bar \cN \right) ^{-1}\bar \cN \,,\nonumber\\
{\bf h}\,{\bf f}^\dagger &= & \cN {\bf f}\,{\bf f}^\dagger\,,\nonumber  \\
{\bf f}\,{\bf h}^\dagger & = & {\bf f}\,{\bf f}^\dagger \bar
\cN\,,
\end{eqnarray}
so that, using the explicit expression for the charges in eqs.
\eq{zab} and \eq{zi}, eq. \eq{sumrule} is easily retrieved.

In the following, studying the applications of these formulas to
extremal black holes, other relations coming from the same
identities listed above  will also be useful, in particular:
\begin{eqnarray}
\frac 12\, \left(\cM + {\rm i}\,\mathbb{C}\right) &=& \pmatrix{-
{\bf h}\,{\bf h}^\dagger & {\bf h}\,{\bf f}^\dagger \cr {\bf
f}\,{\bf h}^\dagger & -{\bf f}\,{\bf f}^\dagger}=\frac
12\,\mathbb{C}\, {\bf V}\,(\bfone+\eta)\, {\bf V}^\dagger\,
\mathbb{C}=\,\nonumber\\
&=&-(\mathbb{C}\,{\bf V})_M\,(\mathbb{C}\,\bar{{\bf V}})^M\,,\label{mmic}\\
\frac 12 \left(\cM +\ii  \mathbb{C}\right) \,  {\bf V}_M &=& \ii \,\mathbb{C} \,{\bf V}_M\,,\\
\frac 12 \left(\cM -\ii  \mathbb{C}\right) \,  {\bf V}_M &=& 0\,,\\
\cM \,Q &=& \mathbb{C}\,{\bf V}\,{\bf
V}^\dagger\,\mathbb{C}\,Q=-2\,\Re
\left(\mathbb{C}\,{\bf V}_M \,<Q , \bar{{\bf V}}^M>\right)\,,\label{inversecharge1}\\
\mathbb{C}\,Q&=&- \ii\,\mathbb{C}\,{\bf V}\,\eta\,{\bf
V}^\dagger\mathbb{C}\,Q=-2\,\Im\left(\mathbb{C}\,{\bf V}_M\,<Q ,
\bar{{\bf V}}^M>\right)\label{inversecharge2}\,.
\end{eqnarray}
The symplectic scalar product appearing in \eq{inversecharge1},
\eq{inversecharge2} is defined as:
\begin{eqnarray}
<V,\,W >&\equiv &V^t\,\mathbb{C}\,W\,,\label{wedge}
\end{eqnarray}
moreover $\bar{{\bf V} }^M=({\bf V}_M)^*$. Using eqs.
(\ref{defum}), (\ref{zab}) and (\ref{zi}) we can use the following
short-hand notation for the central charge vector:
\begin{eqnarray}
Z_M&=&(Z_{AB},\,\bar{Z}_{\bar{I}})= < Q,\,{\bf V}_M>\,.
\end{eqnarray}
From the above expression and from eq. (\ref{mmic}) equation
(\ref{sumrule}) follows.

\subsection{The $N=2$ theory}\label{sec:n=2} The formalism we have developed so
far for the $D=4$, $N>2$ theories is completely determined by the
embedding of the coset representative of ${G}/H$ in
$Sp(2n,\mathbb{R})$ and by the embedded Maurer--Cartan equations
(\ref{df}). We want now to show that this formalism, and in
particular the identities (\ref{specdef}), the differential
relations among charges (\ref{dz1}) and the sum rules
(\ref{sumrule}) of $N=2$ matter-coupled supergravity
\cite{spegeo},\cite{str} can be obtained in a way completely
analogous to the $N>2$ cases discussed in the previous subsection,
where the $\sigma$-model was a coset space. This follows essentially
from the fact that, though the scalar manifold $\cM_{scalar}$ of the
$N=2$ theory is not in general a coset manifold, nevertheless it has
a symplectic structure identical to the $N>2$ theories, as a
consequence of the Gaillard--Zumino duality.
\par
In the case of $N=2$ supergravity the requirements imposed by
supersymmetry on the scalar manifold ${\cal M}_{scalar}$ of the
theory dictate that it should be the following direct product: $
{\cal M}_{scalar}={\cal M}^{SK}\, \otimes \, {\cal M}^Q$ where
${\cal M}^{SK}$ is a special K\"ahler manifold of complex
dimension $n$ and ${\cal M}^Q$ a qua\-ter\-nio\-nic manifold
of real dimension $4n_H$. Note that $n$ and $n_H$ are respectively
the number of vector multiplets and hypermultiplets
contained in the theory. The direct product structure imposed by
supersymmetry precisely reflects the fact that the quaternionic and
special K\"ahler scalars belong to different supermultiplets. In the
construction of extremal black holes it turns out that the
hyperscalars are spectators playing no dynamical role. Hence we do
not discuss here the hypermultiplets any further and we confine our
attention to an $N=2$ supergravity where the graviton multiplet,
containing besides the graviton $g_{\mu \nu}$ also a graviphoton
$A^0_{\mu}$, is coupled to $n$  vector multiplets. Such a
theory has an action of type \eqn{bosonicL} where the number of
gauge fields is ${n_V}=1+n$ and the number of (real) scalar fields
is $m=2 \, {n}$. We shall use capital Greek indices to label the
vector fields: $\Lambda,\,\Sigma\dots=0,\dots, n$. To make the
action \eqn{bosonicL} fully explicit, we need to discuss the
geometry of the manifold $\cM^{SK}$ spanned by the vector-multiplet
scalars, namely special K\"ahler geometry. Since $\cM^{SK}$ is in
particular a complex manifold, we shall describe the corresponding
scalars as complex fields: $z^i,\,\bar{z}^{\bar \imath}$, $i,\,{\bar
\imath}=1,\dots, n$. We refer to \cite{n=2} for a detailed analysis.
A special K\"ahler manifold $\cM^{SK}$ is a K\"ahler--Hodge manifold
endowed with an extra symplectic structure. A K\"ahler manifold
${\cal M}$ is a Hodge manifold if and only if there exists a $U(1)$
bundle ${\cal L} \, \longrightarrow \, {\cal M}$ such that its first
Chern class equals the cohomology class of the K\"ahler 2-form K:
\begin{equation}
c_1({\cal L} )~=~\left [ \, K \, \right ]\,. \label{chernclass25}
\end{equation}
In local terms we can write
\begin{equation}
K\, =\, {\rm i} \, g_{i\bar\jmath} \, dz^{i} \, \wedge \, d{\bar
z}^{\bar \jmath}\,, \label{chernclass26}
\end{equation}
where $z^i$ are $n$ holomorphic coordinates on ${\cal M}^{SK}$ and
$g_{i\bar\jmath}$ its metric.\\ In this case the $U(1)$ K\"ahler
connection is given by:
\begin{equation}
\cQ = -\frac{\rm i}2 \left(\partial_i \cK dz^i -
\partial_{\bar\imath}\cK d\bar z^{\bar\imath}\right)\,,
\end{equation}
where  $\cK$ is the K\"ahler potential, so that $K = d\cQ$.
\par
 Let now
 $\Phi (z, \bar z)$ be a section of the $U(1)$ bundle of  weight $p$.  By definition its
covariant derivative is
\begin{equation}
D \Phi = (d + i p {\cal Q}) \Phi\,,
\end{equation}
or, in components,
\begin{equation}
\begin{array}{ccccccc}
D_i \Phi &=&
 (\partial_i + \frac{1}{2} p \partial_i {\cal K}) \Phi &; &
D_{\bar\imath}\Phi &=&(\partial_{\bar\imath}-\frac{1}{2} p
\partial_{\bar\imath} {\cal K}) \Phi \cr
\end{array}\,.
\label{scrivo2}
\end{equation}
A covariantly holomorphic section is defined by the equation: $
D_{\bar\imath} \Phi = 0  $. Setting:
\begin{equation}
\tilde{\Phi} = e^{-p {\cal K}/2} \Phi  \,   , \label{mappuccia}
\end{equation}
we get:
\begin{equation}
\begin{array}{ccccccc}
D_i    \tilde{\Phi}&    =& (\partial_i   +   p   \partial_i
{\cal K}) \tilde{\Phi}& ; & D_{\bar \imath}\tilde{\Phi}&=&
\partial_{\bar \imath} \tilde{\Phi}\cr
\end{array}\,,
\end{equation}
so that under this map covariantly holomorphic sections $\Phi$
become truly holomorphic sections.
\par
There are several equivalent ways of defining what a special
K\"ahler manifold is. An intrinsic definition is the following. A
special K\"ahler manifold can be given by constructing a
$2n+2$-dimensional flat symplectic bundle over the K\"ahler--Hodge
manifold whose generic sections (with weight $p=1$)
\begin{equation}V = (f^\Lambda,h_\Lambda)
\,, \label{sezio} \end{equation}are covariantly holomorphic
\begin{equation}D_{\bar\imath} V =
(\del_{\bar\imath}-\half\del_{\bar\imath} {\cal K}) V=0\,,
\label{ddue}
\end{equation}and satisfy the further condition
\begin{equation}
\ii <V,\bar V
>=\ii (\bar f^\Lambda h_\Lambda-\bar h_\Lambda f^\Lambda)=1\ ,
\label{dtre} \end{equation}
where the $<\,,\,>$ product was
defined in (\ref{wedge}).
 Defining
 \begin{equation}
 V_i =D_i
V=(f_i^\Lambda,h_{\Lambda i}),\label{ui}
\end{equation}
 and introducing a symmetric
three-tensor $ C_{ijk}$ by
\begin{equation}D_i V_j=i
C_{ijk} g^{k\bar k}\bar V_{\bar k}\ , \label{dqua}
\end{equation}
the set of differential equations \ba
D_i V &=& V_i\,,\nonumber\\
D_i V_j &=& i C_{ijk} g^{k\bar k}\bar V_{\bar k}\,,\nonumber\\
D_i V_{\bar\jmath} &=& g_{i \bar\jmath } \bar V\,,\nonumber\\
D_{i} \bar V &=& 0\,, \label{geospec} \ea defines a symplectic
connection. Requiring that the differential system \eqn{geospec}
is integrable is equivalent to requiring that the symplectic
connection
 is flat.
Since the integrability condition of \eqn{geospec} gives
constraints on the base K\"ahler--Hodge manifold, we define
special-K\"ahler a manifold whose associated symplectic
connection is flat. At the end of this section we will give the
restrictions on the manifold imposed by the flatness of the
connection.
\par
It must be noted that, for special K\"ahler manifolds, the
K\"ahler potential can be computed as a symplectic invariant from
eq. (\ref{dtre}). Indeed, introducing also the holomorphic
sections \ba
\Omega &=& e^{-{\cal K}/2} V=e^{-{\cal K}/2} (f^\Lambda, h_\Lambda)=(X^\Lambda, F_\Lambda)\,,\nonumber\\
\del_{\bar\imath}\Omega &=& 0\,, \ea eq. (\ref{dtre})  gives
\begin{equation}
{\cal K}=-\ln {\rm i}<\Omega,\bar \Omega >=-\ln \ii (\bar X^\Lambda
F_\Lambda- X^\Lambda \bar F_\Lambda)\ . \label{dtrd}
\end{equation}
If we introduce the complex symmetric $(n+1)\times (n+1)$ matrix
$\cN_{\Lambda\Sigma}$  defined through the relations
\begin{equation}h_\Lambda=\cN_{\Lambda\Sigma} f^\Sigma\ \ , \ \
h_{\Lambda\,\bar \imath}=\cN_{\Lambda\Sigma}
\bar f^\Sigma_{\bar\imath}\ , \label{defi}
\end{equation} then we have:
\begin{equation}
 <V,\bar V > =
\left(\cN- \bar \cN \right)_{\Lambda\Sigma} f^\Lambda \bar
f^\Sigma = -{\rm i}\,,\label{VV}
\end{equation}
so that \begin{equation} {\cal K}=-\ln[{\rm i}(\bar X^\Lambda
\left(\cN- \bar \cN \right)_{\Lambda\Sigma} X^\Sigma)]
\,,\end{equation} and \ba g_{i \bar\jmath} &=& -\ii  <V_i, V_{
\bar\jmath} >=-2 f^\Lambda_i \mbox{Im} \cN_{\Lambda\Sigma}
\bar{f}^\Sigma_{\bar\jmath}\ ,\label{metri}\\
C_{ijk} &=& <D_i V_j, V_k> = 2{\rm i} {\rm Im}\cN_{\Lambda\Sigma}
f^\Lambda_i D_j f^\Sigma_k \ . \label{dott} \ea We shall also use
the following identity which follows from the previous ones
\begin{eqnarray}
f^\Lambda_i\,g^{i\bar{\jmath}}\,\bar{f}_{\bar{\jmath}}^\Sigma&=&-\frac{1}{2}\,({\rm
Im}\mathcal{N})^{-1\,\Lambda\Sigma}-\bar{L}^\Lambda\,L^\Sigma\,.\label{Uu}
\end{eqnarray}
The matrix $\cN_{\Lambda\Sigma}$ turns out to be the matrix
appearing in the kinetic lagrangian of the vectors in $N=2$
supergravity. Under coordinate transformations, the sections
$\Omega$ transform as
\begin{equation}\tilde\Omega =e^{-f_{\cal S} (z)}{\cal S}\Omega\ ,
\label{ddoc} \end{equation}where ${\cal S} =\left(\matrix{A&B\cr C
& D\cr }\right)$ is an element of $Sp(2n_V,\mathbb{R})$ and the
factor $e^{-f_{\cal S} (z)}$ is a $U(1)$ K\"ahler transformation.
We also note that, from the definition of $\cN$, eq. (\ref{defi}):
\begin{equation}\tilde\cN (\tilde X,\tilde F)=[C+D\cN
(X,F)][A+B\cN (X,F)]^{-1}\ . \label{dvdu} \end{equation}
\par
 We can now
define a matrix ${\bf V}$ as in (\ref{defu}) satisfying the relations
(\ref{propu}), in terms of the quantities $(f^\Lambda,\,\bar
f^\Lambda{}_{\bar\imath},\,h_\Lambda,\,\bar
h_{\Lambda\,{\bar\imath}})$ introduced in \eq{sezio} and \eq{ui}. In
order to identify the blocks ${\bf f}$ and ${\bf h}$ of ${\bf V}$  in (\ref{defu}), we
note that in $N=2$ theories $H_{Aut}=SU(2)\times U(1)$, so that the
$f^\Lambda_{AB}$ and $h_{\Lambda \,AB}$ entries in (\ref{deffh}) are
actually $SU(2)$-singlets. We can therefore consistently write ${\bf f}$
and ${\bf h}$ as the following $n_V \times n_V$ matrices:
\begin{equation}
{\bf f} \equiv \left( f^\Lambda_{AB},  \bar
f^{\Lambda}_{\bar{I}}\right); \quad {\bf h} \equiv \left(
h_{\Lambda\,AB},
 \bar h_{\Lambda\,\bar{I}} \right)\,,
\label{matrici}
\end{equation}
where $f^\Lambda_{AB},\,h_{\Lambda\,AB}$ and $f^{\Lambda}_{I},\,
h_{\Lambda\,I}$ are defined as follows:
\begin{eqnarray}
f^\Lambda_{AB}&=&f^\Lambda\,\epsilon_{AB}\,\,\,;\,\,\,h_{\Lambda\,AB}=h_\Lambda\,\epsilon_{AB}\,,\nonumber\\
f^{\Lambda}_{I}&=&f_i^\Lambda\,P^i_I\,\,\,;\,\,\,h_{\Lambda\,I}=h_{\Lambda\,i}\,P^i_I\,,
\end{eqnarray}
$P^i_I,\,P^{\bar{\imath}}_{\bar{I}}$ being the inverse of the
K\"ahlerian vielbein $P_i^I,\, \bar{P}^{\bar{I}}_{\bar{\imath}}$
defined by the relation:
\begin{eqnarray}
g_{i\bar{\jmath}}&=&P_i^I\,
\bar{P}^{\bar{J}}_{\bar{\jmath}}\,\eta_{I\bar{J}}\,,
\end{eqnarray}
and $\eta_{I\bar{J}}$ is the flat metric. From the definition
(\ref{matrici}) and the properties (\ref{VV}), (\ref{metri}) it is
straightforward to verify that the ${\bf f}$ and ${\bf h}$ blocks satisfy the
relations (\ref{specdef}), or equivalently that the matrix ${\bf V}$
satisfies the conditions (\ref{propu}). The relations
(\ref{specdef}) therefore encode the set of algebraic relations of
special geometry.
\par
Let us now consider the analogous of the embedded Maurer--Cartan
equations of ${G}/H$. We introduce, as before, the matrix one-form
$\Gamma= {\bf V}^{-1} d{\bf V}$ satisfying the relation $d\Gamma+
\Gamma\wedge\Gamma =0$. We further introduce the covariant
derivative of the symplectic section $(f^\Lambda,\bar
f^\Lambda_{\bar I}, \bar f^\Lambda , f^\Lambda_I)$ with respect to
the $U(1)$-K\"ahler connection $\cQ$ and the spin connection
$\omega^{IJ}$ of $\cM^{SK}$:
 \begin{eqnarray}
& D (f^\Lambda,\bar  f^\Lambda_{\bar I}, \bar f^\Lambda,  f^\Lambda_I) = & \nonumber\\
& d (f^\Lambda,\bar f^\Lambda_{\bar I},  \bar f^\Lambda,
f^\Lambda_{ I}) -(f^\Lambda,\bar f^\Lambda_{\bar J},  \bar
f^\Lambda, f^\Lambda_{J})\left(\matrix{-{\rm i} \cQ   & 0 & 0 & 0
\cr 0 & {\rm i} \cQ \delta^{\bar J} _{\bar I} + \omega^{\bar J}_{\
\bar I} & 0 & 0 \cr  0 & 0 &  {\rm i} \cQ & 0 \cr 0 & 0 & 0 &-{\rm
i} \cQ \delta^{ J} _{ I} + \omega^{J}_{\ I} \cr }\right)
&\nonumber\\&&
\end{eqnarray}
 the K\"ahler weight of $(f^\Lambda,\, f^\Lambda_I)$ and
$(\bar f^\Lambda,\,\bar f^\Lambda_{\bar I})$ being $p=1$ and
$p=-1$ respectively. Using the same decomposition as in equation
(\ref{defgamma}) and eq.s (\ref{nablav}), (\ref{defomega}) we have
in the $N=2$ case:
\begin{eqnarray}
  \label{gammaspec}
  \Gamma &=& \left(\matrix{\Omega & \bar \cP \cr \cP & \bar \Omega \cr }\right)
  ,\nonumber\\
  \Omega &=& \omega = \left(\matrix{-{\rm i}  \cQ  & 0 \cr 0 &
  {\rm i}   \cQ \delta^I_J+ \bar \omega^I_{\ J} \cr }\right)\,.
   \end{eqnarray}
For the subblock $\cP$ we obtain:
\begin{equation}
\cP = - {\rm i} (f^t D h - h^t D f)  = {\rm i} f^t (\cN
- \bar \cN) D f = \left(\matrix{0 &  P_{\bar I} \cr P^{ J} &
P^{ J}_{\ \bar I} \cr }\right)\,,  \label{viel2}
\end{equation}
where $ P^J \equiv \eta^{J\bar I} P_{\bar I} $ is the
$(1,0)$-form K\"ahlerian vielbein while
\begin{equation}
 P^{ J}_{\ \bar I}
\equiv {\rm i} \left( f^t (\cN - \bar \cN) D f \right)^{
J}_{\ \bar I}
\end{equation}
 is a one-form which in general, in the cases where the manifold is
not a coset, represents a new geometric quantity on $\cM^{SK}$. Note
that we get zero in the first entry of equation (\ref{viel2}) by
virtue of the fact that the identity (\ref{specdef}) implies
$f^\Lambda(\cN - \bar \cN)_{\Lambda\Sigma}f^\Sigma_I =0 $ and that
$f^\Lambda$ is covariantly holomorphic. If $\Omega $ and $\cP $ are
considered as data on $\cM^{SK}$ then we may interpret $\Gamma
=V^{-1} dV$ as an integrable system of differential equations,
namely:
\begin{equation}
  \label{picfuc}
  D (V ,\bar V_{\bar I} , \bar V ,   V_I ) =  (V ,\bar V_{\bar J} , \bar V ,  V_J )
  \left(\matrix{0& 0 & 0 & \bar P_{I} \cr
0 & 0 & \bar P^{\bar J} & \bar P^{\bar J}_{\ I} \cr 0 &  P_{\bar
I} & 0 & 0 \cr  P^J & P^J_{\ \bar I} & 0 & 0 \cr }\right)\,,
\end{equation}
where the flat K\"ahler indices $I,\bar I, \cdots $ are raised and
lowered with the flat K\"ahler metric $\eta_{I\bar J}$. Note that
the equation \eqn{picfuc} coincides with the set of relations
\eqn{geospec} if we trade world indices $i, \bar\imath$ with flat
indices $I,\bar I$, provided we also identify:
\begin{equation}
\bar P^{\bar J}_{\ I}=\bar P^{\bar J}_{\ I k}dz^k= P^{\bar J, i}
P_I^{\ j} C_{ijk} dz^k .
\end{equation}
Then, the integrability condition $d\Gamma + \Gamma\wedge
\Gamma=0$ is equivalent to the flatness of the special K\"ahler
symplectic connection and it gives the following three constraints
on the K\"ahler base manifold:
\begin{eqnarray}
  d( {\rm i} \cQ) + \bar P_I \wedge  P^{I} &=& 0
  \to \partial_{\bar\jmath}\partial_i \cK = P^I_{\ ,i} \bar P_{  I,\bar
  \jmath}=g_{i\bar\jmath}\,,
  \label{spec1} \\
  (d\omega +\omega\wedge\omega)_{\ \bar I}^{\bar J} &=& P_{\bar I}
  \wedge\bar P^{\bar J} -{\rm i} d\cQ
  \delta_{\bar I}^{\bar J}
-\bar P^{\bar J}_{\ L} \wedge  P^L_{\ \bar I}\,,
\label{spec2}\\
D P_{\ \bar I}^{J} &=& 0\,,
\label{spec3}\\
\bar P_J \wedge P^J_{\ \bar I}&=&0\,. \label{spec4}
\end{eqnarray}
Equation (\ref{spec1}) implies that $\cM^{SK}$ is a
K\"ahler--Hodge manifold. Equation (\ref{spec2}), written with
holomorphic and antiholomorphic curved indices, gives:
\begin{equation}
  \label{curvspec}
  R_{\bar\imath j \bar k l} = g_{\bar\imath l} g_{j \bar k} + g_{\bar k l}
  g_{\bar\imath j} - \bar C_{\bar\imath\bar k \bar m}
C_{jln} g^{\bar m n}\,,
\end{equation}
which is the usual constraint on the Riemann tensor of the special
geometry. The further special geometry constraints on the three
tensor $C_{ijk}$  are then consequences of
 equations (\ref{spec3}), (\ref{spec4}), which imply:
\begin{eqnarray}
  D_{[l} C_{i]jk} &=& 0\,, \nonumber\\
D_{\bar l} C_{ijk}&=&0\,. \label{cijk}
\end{eqnarray}
In particular, the first of eq. (\ref{cijk}) also implies that $
C_{ijk} $ is a completely symmetric tensor.
\par
In summary, we have seen that the $N=2$ theory and the higher $N$
theories have essentially the same symplectic structure, the only
difference being that since the scalar manifold of $N=2$ is not in
general a coset manifold the symplectic structure allows the
presence of a new geometric quantity which physically corresponds to
the anomalous magnetic moments of the $N=2$ theory. It goes without
saying that, when $\cM^{SK}$ is itself a coset manifold \cite{cp},
then the anomalous magnetic moments $C_{ijk}$ must be expressible in
terms of the vielbein of ${G}/H$. \par To complete the analogy
between the $N=2$ theory and the higher $N$ theories in $D=4$,
 we also give for completeness the supersymmetry transformation laws, the central and matter
 charges, the differential
 relations among them and the sum rules.
\par
The transformation laws for the chiral gravitino $\psi_A$ and
gaugino $\lambda^{iA}$ fields are:
\begin{equation}
 \delta \psi_{A \mu}={ \nabla}_{\mu}\,\epsilon _A\,+ \epsilon _{AB}
T_{\mu\nu} \gamma^\nu \epsilon^B + \cdots\,, \label{trasfgrav}
 \end{equation}
 \begin{equation}
 \delta\lambda^{iA} = {\rm i} \partial_\mu z^i \gamma^\mu\epsilon^A +
 {\frac{\rm i} 2}
\bar T_{\bar \jmath\mu\nu} \gamma^{\mu \nu}
g^{i\bar\jmath}\epsilon^{AB}\epsilon_B + \cdots\,,
\label{gaugintrasfm}
\end{equation}
where:
\begin{equation}
T \equiv  h_\Lambda F ^{\Lambda}  - f^\Lambda G_\Lambda\,,
\label{T-def}
\end{equation}
\begin{equation}
\bar T_{\bar \imath} \equiv  \bar T_{\bar I} \bar P^{\bar I}_{\bar\imath}\,,\mbox{ with: }  \bar T_{\bar I} \equiv
 \bar h_{\Lambda {\bar I}}  F^{\Lambda}  - \bar f^\Lambda_{\bar I}
  G_\Lambda\,,
  \label{G-def}
\end{equation}
are respectively the graviphoton and the matter-vectors, and the
position of
 the $SU(2)$ automorphism index A (A,B=1,2) is related to chirality
 (namely $(\psi_A, \lambda^{iA})$ are chiral, $(\psi^A,
 \lambda^{\bar\imath}_A)$ antichiral).
 In principle only the (anti) self-dual part of $ F$ and $G$ should
 appear in the transformation laws of the (anti)chiral fermi fields; however,
 exactly as in eqs. (\ref{tiden0}),(\ref{tiden}) for $N>2$ theories,
 from equations (\ref{geospec}) it follows that :
 \begin{eqnarray}
T^+ &=& h_\Lambda   F^{+\Lambda} - f^\Lambda G_\Lambda ^+
=0\,,\nonumber\\
T^-_I &=& h_{\Lambda\, I}\,   F^{-\Lambda} - f^\Lambda_I\,
G_\Lambda ^- =0\,,
\end{eqnarray}
so that $T=T^-$ and $T_I=T^+_I$ (i.e. $\bar T = \bar T^+$, $\bar
T_{\bar I} = \bar T^-_{\bar I}$).
 Note that both the graviphoton and the matter vectors are symplectic
  invariant according to the fact that the fermions do not
 transform under the duality group (except for a possible R-symmetry
 phase).
To define the physical charges let us recall the definition of the
moduli-independent charges in (\ref{pq}). The central charges and
the matter charges are now defined as the integrals over $S^2$ of
the physical graviphoton and matter vectors:
\begin{eqnarray}
Z&=&-\frac{1}{4\pi}\, \int_{S^2} T= -\frac{1}{4\pi}\,\int_{S^2} (
h_\Lambda  F ^{\Lambda} - f^\Lambda G_\Lambda ) = f^\Lambda(z,\bar
z) q_\Lambda -h_\Lambda (z,\bar z) p^{\Lambda}
\,,\nonumber\\
 Z_I&=&-\frac{1}{4\pi}\,\int_{S^2}
T_I= -\frac{1}{4\pi}\,\int_{S^2} ( h_{\Lambda\, I}\,  F ^{\Lambda}
- f^\Lambda_I G_\Lambda ) = f^\Lambda_I (z,\bar z) q_\Lambda -
h_{\Lambda\, I} (z,\bar z) p^{\Lambda} \,.\nonumber\\&& \label{cc}
\end{eqnarray}
where $z^i, \bar z^{\bar \imath}$ denote the v.e.v. of the moduli
fields in a given background. In virtue of eq. (\ref{geospec}) we
get immediately:
\begin{equation}
Z_I=P^i_I\, Z_i\,\,\,;\,\,\,\, Z_i\equiv D_i Z\label{fundeq}\,.
\end{equation}
 As a consequence of the symplectic structure, one can derive
  two sum
 rules for  $Z$ and $Z_I$:
  \begin{equation}
  \vert Z \vert ^2 +  \vert Z_I \vert ^2 \equiv
  \vert Z \vert ^2 +   Z_i g^{i\bar \jmath} \bar Z_{\bar \jmath} =
  -\frac{1}{2} Q^t \cM Q
  \label{sumrules}
\end{equation}
  where the symmetric matrix $\cM$ was defined in (\ref{m+}) and
  $Q$ is the symplectic vector of electric and magnetic charges defined in (\ref{eg}).

Equation (\ref{sumrules}) is obtained by using exactly the same
procedure as in (\ref{sumrule}).


\section{Supersymmetric black holes: General discussion}
 \label{sec:extremum}
We are going to study in this section the peculiarities of extremal
black holes that are solutions of extended supergravity theories.
\par
As anticipated in the introduction, for black-hole configurations
that are particular bosonic backgrounds of  $N$-extended locally
supersymmetric theories,  the cosmic censorship conjecture
(expressing the request that the space-time singularities are always
hidden by event horizons) finds a simple and natural understanding.
For the Reissner-Nordstrom black holes this is codified in the bound
\eq{censur} on the mass $M$ and charge $Q$ of the solution, that we
recall here
\begin{equation}
M \, \ge \, |Q| \label{censur2}.
\end{equation}
In extended supersymmetric theories this bound is just a consequence
of the supersymmetry algebra \eq{susypm},  as a consequence of the
fact that
\begin{equation}
\left\{ { Q}_{a m  }^{(\pm)}\, , \,{  Q}_{am}^{\dagger(\pm)}
\right\}\geq 0
\end{equation}
so that the cosmic censorship conjecture is always verified.

Another general property of extremal black holes, that will be
surveyed in section \ref{sec:geopot},
 is encoded in the so-called no-hair
theorem. It states that the end point of the gravitational
collapse of a black hole is independent of the initial conditions. Then,
if one tries to perturb an extremal black hole with
whatever additional hair (some slight mass anisotropy, or a
long-range field, like a scalar) all these features disappear near
the horizon, except for those associated with the conserved
quantities of general relativity, namely, for a non-rotating black
hole, its mass and charge. When the black hole is embedded in an
$N$-extended supergravity theory, the solution depends in general
also on scalar  fields. In this case,
 the electric charge $Q$ has to be replaced by the  central charge appearing in the supersymmetry algebra
(which is dressed with the expectation value of scalar fields). The
black-hole  metric takes a generalized form with respect to the
Reissner--Nordstr\"om one.
However, for the extremal case the event horizon looses all
information about the scalar 'hair'. As for the
Reissner--Nordstr\"om case, the near-horizon geometry is still
described by a conformally flat, Bertotti--Robinson-type geometry,
with a mass parameter $M_{\mbox{\small B-R}}$ which only depends on the
distribution of charges and not on the scalar fields. As  will be
discussed extensively in section \ref{sec:geopot}, this follows from
the fact
 that the differential equations on the metric and scalars fields
 of the extremal black hole \eq{geoeq'}, \eq{geoeq}
are solved under the condition that the horizon be an attractor
point \cite{moore} (see equation \eq{min}).
 Scalar fields, independently of their boundary conditions at
spatial infinity, approaching the horizon flow to a fixed point
given by a certain ratio of electric and magnetic charges. Since the
dominant contribution to the black-hole entropy is given (at least
for large black holes) by the area/entropy Bekenstein--Hawking
relation \eq{bekhaw}, it follows that the entropy of extremal black
holes is a topological quantity fixed in terms of the quantized
electric and magnetic charges while it does not depend on continuous
parameters like scalars.

It  will be shown   that the request that the scalars $\Phi^r$ be
regular at the fixed point (reached at the horizon $\tau \to
\infty$) implies two important conditions which have both to be
satisfied:
\begin{eqnarray}
\left(\frac{d\Phi^r}{d \tau}\right)_{hor} &=&0\\
  \left(\frac{\partial
V_{\mbox{\small B-H}}(\Phi)}{\partial \Phi_i}\right)_{hor}&=&0 .\label{vbhhor}
\end{eqnarray}
where the function $V_{\mbox{\small B-H}}(\Phi ,p,q)$, called the black-hole
potential, will be introduced in \eq{geopot}.

Exploiting \eq{vbhhor}, a decade ago  a general rule was given
\cite{feka} for finding the values of fixed scalars,
and then the Bekenstein--Hawking entropy, in $N=2$ theories, through
an {\it extremum principle} in moduli space. This follows from the
observation that, when the scalar fields are evaluated at spatial
infinity ($\tau =0$), $V_{\mbox{\small B-H}}$ coincides with the squared ADM mass
of the black hole.
 Then, since equation \eq{vbhhor} does not
 depend explicitly on the radial variable $\tau$ (as the extremization is
done with respect to the scalar fields at any given point)  the
expectation values $\Phi_\infty$ may be chosen as independent
variables. Equation \eq{vbhhor} is then reformulated as the
statement that the fixed scalars $\Phi_{\mbox{\small fix}}$ are the ones, among
all the possible expectation values taken by scalar fields, that
extremize the ADM mass of the black hole in moduli space:
\begin{equation}
 \Phi_{\mbox{\small fix}} \, : \quad {\frac{\partial
M_{ADM}(\Phi_\infty)}{\partial
\Phi^r_\infty}}\left\vert_{\Phi_{\mbox{\small fix}}}=0 \right. \label{fixed}
\end{equation}

 Correspondingly, the
Bekenstein--Hawking entropy is given in terms of that extremum among
the possible ADM masses (given by all possible boundary conditions
that one can impose on scalars at spatial infinity), this last being
identified with the Bertotti--Robinson mass $M_{\mbox{\small B-R}}$:
 \begin{equation}
M_{\mbox{\small B-R}}\equiv M_{ADM}(\Phi_{\mbox{\small fix}}) .\end{equation} The solutions with
the scalar fields  constant and everywhere equal to the fixed value
 $\Phi_{\mbox{\small fix}}$ are called {\it double extremal black holes}.

  The approach outlined above  will prove to be a very useful
  computational tool
to calculate the B-H entropy since, as will be explained in section \ref{sec:geopot}, in
extended supergravity the explicit dependence of $V_{\mbox{\small B-H}}$
  on the moduli is given.


\subsection{BPS extremal black holes}\label{sec:bpsextr}
 For the case of BPS extremal black holes, the extremum
principle \eq{fixed} may be explained by means of the Killing
spinor equations near the horizon and these are encoded  in  some
relations on the scalars moduli spaces, discussed in detail in
section \ref{sugras} and \ref{sec:n=2}, which express the embedding
of the scalar geometry in a  symplectic  representation of the
U-duality group \cite{frac}. For definiteness,  to present the argument we will
refer, for the sequel of this subsection, to the case $N=2$,  which
is the model originally considered in \cite{fekast,feka}.

The Killing-spinor equations expressing the existence of unbroken
supersymmetries are obtained, for the gauginos in the $N=2$ case
\cite{n=2}, by setting   to zero the r.h.s. of equation
\eqn{gaugintrasfm} that is, using flat indices:
\begin{equation}
\delta \lambda^I_{A}=P^I_{,i}\partial_\mu z^i \gamma^\mu
\varepsilon_{AB} \epsilon^B +\bar T^I_{ \mu\nu}\gamma^{\mu\nu}
\epsilon_A +\cdots =0. \label{kill}
\end{equation}
As we will see in detail in the next subsection, approaching the
black-hole horizon the scalars $z^i$ reach their fixed values
$z_{\mbox{\small fix}}$ \footnote{A point $x_{\mbox{\small fix}}$ where the phase velocity is
vanishing is named \emph{fixed point} and represents the system in
equilibrium $v(x_{\mbox{\small fix}})=0$ \cite{feka,strom3}. The fixed point is said to
be an attractor if $\lim_{t\rightarrow \infty}\, x(t)=x_{\mbox{\small fix}}$.\label{fixedpoint}} so
that
\begin{equation}\partial_\mu z^i =0
\end{equation}and equation \eq{kill} is satisfied for
\begin{equation}T_I =0 \end{equation}which implies, using integrated quantities:
 \begin{equation}Z_I = Z_i P_I^i =-\frac{1}{4\pi}\,\int_{S^2} T_I = \left(f^\Lambda_I q_\Lambda -h_{\Lambda
I}p^\Lambda \right)\,|_{\mbox{\small fix}} = 0.\label{mat}
\end{equation} What we have found is that the Killing spinor equation imposes the
vanishing of the matter charges near the horizon.
Then,  remembering eq. \eq{fundeq}, near the horizon we have:
\begin{equation}
Z_I= D_I Z =0 \label{charges1}
\end{equation}
where $Z$ is the central charge appearing in the $N=2$
supersymmetry algebra, so that:
\begin{equation}\partial_i \vert Z \vert =0.\label{extrbps} \end{equation}

For an extremal BPS black hole ($\vert Z \vert=M_{ADM}$),
\eq{extrbps} coincides with  eq. \eq{fixed} giving the fixed scalars
$\Phi_{\mbox{\small fix}}\equiv z_{\mbox{\small fix}}$ at the horizon. We then see that the
entropy of the black hole is related to the central charge, namely
to the integral of the graviphoton field strength evaluated for very
special values of the scalar fields $z^i$. These special values, the
{\it fixed scalars} $z^i_{\mbox{\small fix}}$, are functions solely of the
electric and magnetic charges $\{ q_ \Sigma,p^\Lambda\}$ of the
black hole  and are attained by the scalars $z^i(r)$ at the black
hole horizon $r=0$.
\par
Let us discuss in detail the explicit solution of the Killing spinor
equation and the general properties of $N=2$ BPS saturated black holes \cite{fekast,bls,ortin2,fre97}.
 As our analysis will reveal,
these properties are completely encoded in the special K\"ahler
geometric structure of the mother theory.

Let us consider a black-hole ansatz for the metric\footnote{This
ansatz is dictated by the general p-brane solution of supergravity
bosonic equations in any dimensions \cite{mbrastelle}.}, restricting
the attention to static, spherically symmetric solutions:
\begin{equation}
ds^2 = e^{2U(r)} \, dt^2 - e^{-2U(r)} G_{ij}(r)\, dx^i dx^j  ;
\quad \left( r^2 = G_{ij}x^i x^j \right) , \quad i,j=1,2,3
\label{ds2U}
\end{equation}
and  for the vector field strengths:
\begin{equation}
 F^{\Lambda}\,=\,\frac{p^{\Lambda}}{2r^3}\epsilon_{abc} x^a
dx^b\wedge dx^c-\frac{\ell^{\Lambda}(r)}{r^3}e^{2 U}dt \wedge
\vec{x}\cdot d\vec{x}. \label{flambda}
\end{equation}
 Note that here $r$ parametrizes the distance from the horizon.
 
 It is convenient to rephrase the same ansatz in the complex
formalism well-adapted to the $N=2$ theory. To this effect we begin
by constructing a 2-form which is anti-self-dual in the
background of the metric \eqn{ds2U} and whose integral on the
$2$-sphere at infinity $ S^2_\infty$ is normalized to $ 4 \pi $. A
short calculation yields:
\begin{eqnarray}
E^-   &=&   \mbox{i} \frac{e^{2U(r)}}{r^3} \, dt \wedge {\vec
x}\cdot d{\vec x}  +  \frac{1}{2}   \frac{x^a}{r^3} \, dx^b
\wedge
dx^c   \epsilon_{abc} \nonumber\,,\\
 \int_{S^2_\infty} \, E^- &=& 4 \, \pi    \label{eaself}
\end{eqnarray}
from which one obtains:
\begin{equation}
E^-_{\mu\nu} \, \gamma^{\mu\nu}  = 2 \,\mbox{i}
\frac{e^{2U(r)}}{r^3}  \, \gamma_a x^a \, \gamma_0 \,
\frac{1}{2}\left[ {\bf 1}+\gamma_5 \right] \label{econtr}
\end{equation}
which will simplify the unfolding of the supersymmetry
transformation rules. Next, introducing the following complex
combination:
\begin{equation}
\label{tlamb} t^\Lambda(r)\,=\,\frac 12 \,(p^\Lambda+{\rm i}\ell
^\Lambda (r))
\end{equation}
 of the magnetic charges $ p^\Lambda =\frac{1}{4 \pi}
\int_{S^2} F^\Lambda$ and of the functions
$\ell^\Lambda(r)=-\frac{1}{4 \pi} \int_{S^2}{}^\star F^\Lambda$
introduced in  eq. \eqn{flambda}, we can rewrite the ansatz
\eqn{flambda} as:
\begin{eqnarray}
\label{strenghtsans}
 F^{-\vert \Lambda}\,&=&\, t^\Lambda \, E^-
\end{eqnarray}
and we retrieve the original formulae from:
\begin{equation}
\begin{array}{rcccl}
 F^{\Lambda}\,&=&\,2{\rm Re} F^{-\vert \Lambda}&= &
\frac{p^{\Lambda}}{2r^3}\epsilon_{abc} x^a dx^b\wedge
dx^c-\frac{\ell^{\Lambda}(r)}{r^3}e^{2 U}dt
\wedge \vec{x}\cdot d\vec{x} \\
^\star{F}^{\Lambda}&=&-2{\rm Im} F^{-\vert \Lambda}& = &
-\frac{\ell^{\Lambda}(r)}{2r^3}\epsilon_{abc} x^a dx^b\wedge dx^c
-\frac{p^{\Lambda}}{r^3}e^{2 U}dt\wedge \vec{x}\cdot d\vec{x}.
\end{array}
\label{fedfd}
\end{equation}
Before proceeding further it is convenient to define the electric
and magnetic charges of the black hole as it is appropriate in any
abelian gauge theory. Recalling the general form of the field
equations and of the Bianchi identities as given in \eqn{biafieq},
we see that on-shell the field strengths ${F}_{\mu \nu}$ and ${
G}_{\mu \nu}$ are both closed 2-forms, since their duals are
divergenceless. Hence, for Gauss theorem, their integral on a
closed space-like $2$-sphere does not depend on the radius of the
sphere. These integrals are the (constant) electric and magnetic
charges of the black hole defined in (\ref{pq}) that, in a quantum
theory, we expect to be quantized. Using the ansatze (\ref{fedfd})
and the definition  \eq{gtensor}, we find
\begin{eqnarray}
q_\Lambda = \frac{1}{4 \pi} \int_{S^2} G_\Lambda =\Im
\cN_{\Lambda\Sigma} \ell^\Sigma + \Re \cN_{\Lambda\Sigma} p^\Sigma
= 2 \Re\left(\cN_{\Lambda\Sigma} \bar t^\Sigma\right).\label{cu}
\end{eqnarray}
From the above equation we can obtain the field dependence of the
functions $\ell^\Lambda(r)$
\begin{eqnarray}
\ell^\Lambda(r)&=&({\rm
Im}\mathcal{N})^{-1\,\Lambda\Sigma}\,\left(q_\Sigma-{\rm
Re}\mathcal{N}_{\Sigma\Gamma}\,p^\Gamma\right)\,.
\end{eqnarray}
 Consider now the Killing spinor
equations obtained from the supersymmetry transformations rules
\eqn{trasfgrav}, \eqn{gaugintrasfm}:
\begin{eqnarray}
0 &=& {\nabla}_{\mu}\,\xi _A\, + \epsilon_{AB} \, T^-_{\mu \nu}\,
\gamma^{\nu}\xi^B\,,
 \label{kigrav} \\
0 &=& {\rm i}\,    \nabla _ {\mu}\, z^i  \, \gamma^{\mu} \xi^A
+\frac{\ii}{2}\,g^{i\bar{\jmath}}\,\bar{T}_{\,\bar{\jmath}|\mu
\nu}^{-}\, \gamma^{\mu \nu} \epsilon^{AB}\,\xi_B \,,\label{kigaug}
\end{eqnarray}
where the Killing spinor $\xi _A(r)$  is of the form of a single
radial function times a constant spinor satisfying:
\begin{eqnarray}
\xi_A (r)\,&=&\,e^{f(r)} \chi _A~~~~~~~~~\chi_A=\mbox{constant}\nonumber\\
\gamma_0 \chi_A\,&=&\, {\rm i}\,\frac{Z}{\vert Z\vert}\,
\epsilon_{AB}\chi^B \label{quispi}
\end{eqnarray}
We observe that the condition \eq{quispi} halves the number of
supercharges preserved by the solution. Inserting eq.s
\eqn{T-def},\eqn{G-def},\eqn{quispi} into eq.s\eqn{kigrav},
\eqn{kigaug} and using  the result \eqn{econtr}, with a little
work we obtain the first order differential equations:
\begin{eqnarray}
\frac{dz^i}{dr}\, &=&\,  -\left(\frac{e^{U(r)}}{
r^2}\right)\frac{Z}{\vert Z\vert}\, g^{i\bar\jmath}\bar{f}_{\bar
\jmath}^\Lambda
({\cal N}-\bar{{\cal N}})_{\Lambda\Sigma}t^\Sigma\,=\,\nonumber\\
&=& \left(\frac{e^{U(r)}}{r^2}\right)\, \frac{Z}{\vert Z\vert}\,g^{i\bar \jmath} D_{\bar
\jmath}\bar{Z}(z,\bar{z},{p},{q})= 2\,\left(\frac{e^{U(r)}}{r^2}\right)\,g^{i\bar
\jmath}\,\partial_{\bar \jmath}\vert
Z(z,\bar{z},{p},{q})\vert\,, \nonumber\\&&\label{zequa}\\
\frac{dU}{dr}\, &=&\, \left(\frac{e^{U(r)}}{r^2}\right) \vert
h_\Sigma {p}^\Sigma- f^\Lambda {q}_\Lambda\vert\,=\,
\left(\frac{e^{U(r)}}{r^2}\right)\vert Z(z,\bar{z},{p},{q})\vert\,,
\label{Uequa}
\end{eqnarray}
where ${\cal N}_{\Lambda\Sigma}(z,\bar{z})$ is the kinetic matrix
of special geometry defined by eq.\eqn{defi},
 the vector $V= \left(f^\Lambda(z,\bar{z}),
h_\Sigma(z,\bar{z})\right)$, according to eq. \eq{sezio}, is the covariantly holomorphic section
of the symplectic bundle entering the definition of a Special
K\"ahler manifold.
Moreover, according to eq. \eq{cc},
\begin{equation}
 Z(z,\bar{z},{p},{q}) \equiv
f^\Lambda {q}_\Lambda- h_\Sigma {p}^\Sigma \,,\label{zentrum}
\end{equation}
is the local realization on the scalar manifold ${\cal SM}$ of the
central charge of the $N=2 $ superalgebra,
\begin{equation}
\bar Z^i(z,\bar{z},{p},{q}) \equiv g^{i\bar \jmath}D_
{\bar \jmath}\bar{Z}(z,\bar{z},{p},{q})\,, \label{zmatta}
\end{equation}
are the  charges associated with the matter vectors, the
so-called matter central charges, written with world indices of the special-K\"ahler manifold. In terms of the complex charge
vector $t^\Lambda$ introduced in (\ref{tlamb}), the central and
matter  charges have the following useful expressions
\begin{eqnarray}
Z&=&-2\ii\,f^\Lambda\,{\rm
Im}\mathcal{N}_{\Lambda\Sigma}\,t^\Sigma\,,\label{zz}\\
\overline{Z}_{\bar\imath}&=&-2\ii\,\bar{f}^\Lambda_{\bar\imath}\,{\rm
Im}\mathcal{N}_{\Lambda\Sigma}\,t^\Sigma\,,\label{zzi}
\end{eqnarray}
\par
In summary, we have reduced the  condition that the black hole
should be a BPS saturated state to the pair of first order
differential equations \eq{zequa}, \eq{Uequa} for the metric scale
factor $U(r)$ and for the scalar fields $z^i(r)$. To obtain explicit
solutions one should specify the special K\"ahler manifold one is
working with, namely the specific Lagrangian model. There are,
however, some very general and interesting conclusions that can be
drawn in a model-independent way. They are just consequences of the
fact that these BPS conditions are first order differential
equations. Because of that there are fixed points (see footnote \ref{fixedpoint}), 
namely values either of the metric or of
the scalar fields which, once attained in the evolution parameter
$r$ (= the radial distance), will persist indefinitely. The fixed
point values are just the zeros of the right hand side in either of
the coupled eq.s \eqn{Uequa} and \eqn{zequa}. The fixed point for
the metric equation \eq{Uequa} is $r=\infty $, which corresponds to its
asymptotic flatness. The fixed point for the moduli equation \eq{zequa} is $r=0$.  So,
independently from the initial data at  $r=\infty$ that determine
the details of the evolution,  the scalar fields flow into their
fixed point values at $r=0$, which, as we will show, turns out to be
a horizon. Indeed in the vicinity of $r=0$ also the metric takes the
universal form of the Bertotti--Robinson  $AdS_2 \, \times \, S^2$
metric.
\par
Let us see this more closely.  To begin with we consider the
equations determining the fixed point values for the moduli and the
universal form attained by the metric at the moduli fixed point. Using eq. \eq{zzi}, we find:
\begin{eqnarray}
0 &=&\left.g^{i\bar \jmath}
\,\bar{Z}_{\bar\jmath}\right\vert_{\mbox{\small fix}}=
-2\ii\,\left.g^{i\bar \jmath} \, {\bar f}^\Gamma_{\bar \jmath}
\left( \mbox{Im}{\cal N}\right)_{\Gamma\Lambda} \, t^{\Lambda
}\right\vert_{\mbox{\small fix}}\,,
\label{zequato} \\
\left.\left(\frac{dU}{dr}\right)\right\vert_{\mbox{\small fix}}\,
& = & \left.\left(\frac{e^{U(r)}}{r^2}\right)\,\vert Z\left
(z,{\bar{z}} ,{p},{q} \right)\vert\right\vert_{\mbox{\small
fix}}\,. \label{Uequato}
\end{eqnarray}
Multiplying eq.\eqn{zequato} by $f^\Sigma_i$, using the identity
\eqn{Uu} and the definition \eqn{zz} of the central charge
 we conclude that
at the fixed point the following condition is true:
\begin{equation}
0=\,\left.\left(t^\Lambda+\ii\,\bar{f}^\Lambda\,Z\right)\right\vert_{\mbox{\small
fix}} \,.\label{passetto}
\end{equation}
In terms of the previously defined electric and magnetic charges
(see eq.s \eqn{pq}, \eqn{cu})  eq. \eqn{passetto} can be rewritten
as:
\begin{eqnarray}
p^\Lambda & = & -\mbox{i}\left.\left( Z\,{\bar f}^\Lambda
- {\bar Z}\,f^\Lambda \right)\right\vert_{\mbox{\small fix}}\,,\label{minima1}\\
q_\Sigma & = & -\mbox{i}\left.\left( Z\,{\bar h}_\Lambda - {\bar
Z}\,h_\Lambda \right)\right\vert_{\mbox{\small
fix}}\,.\label{minima2}
\end{eqnarray}
Eq.s \eq{zequato}, or equivalently eq.s \eq{minima1}, \eq{minima2}, can be regarded as algebraic equations determining the value
of the scalar fields at the fixed point as functions of the
electric and magnetic charges $p^\Lambda, q_\Sigma$. Note
therefore that, at the horizon, also the central charge depends
only on the quantized charges: $Z(z,\bar{z},p,q)|_{\mbox{\small
fix}}\equiv Z(p,q)$.
\par In the vicinity of the fixed point the differential equation
for the metric becomes:
\begin{equation}
 \, \frac{dU}{dr}=\frac{\vert Z(p,q)\vert}{  \, r^2} \, e^{U(r)}
\end{equation}
which has the approximate solution:
\begin{equation}
\exp[-U(r)]\, {\stackrel{r \to 0}{\longrightarrow}}\,
\frac{\vert Z(p,q)\vert}{  \, r} \label{approxima}
\end{equation}
Hence, near $r=0$   the metric \eqn{ds2U} becomes of the Bertotti
Robinson type (see eq.\eqn{br} ) with Bertotti Robinson mass given
by:
\begin{equation}
M_{\mbox{\small B-R}}^2 = \vert  { Z(p,q)}  \vert^2 \label{brmass}
\end{equation}
In the metric \eqn{br} the surface $r=0$ is light-like and
corresponds to a horizon since it is the locus where the Killing
vector generating time translations $\frac{\partial}{\partial t} $,
which is time-like at spatial infinity $r=\infty$, becomes
light-like. The horizon $r=0$ has a finite area given by:
\begin{equation}
\mbox{Area}_H = \int_{r=0} \, \sqrt{g_{\theta\theta}\,g_{\phi\phi}}
\,d\theta \,d\phi \, = \, 4\pi \, M_{\mbox{\small B-R}}^2 \label{horiz}
\end{equation}
Hence, independently from the details of the considered model, the
BPS saturated black holes in an N=2 theory have a
Bekenstein--Hawking entropy given by the following horizon area:
\begin{equation}
 \frac{\mbox{Area}_H}{4\pi} = \,     \vert Z(p,q) \vert^2,
 \label{ariafresca}
\end{equation}
where \eq{brmass} was used, the value of the central charge being
determined by eq.s \eqn{minima1}, \eq{minima2}. Such equations, as we shall see in the next secton, can also be seen as
the variational equations for the minimization of the horizon area
as given by \eqn{ariafresca}, if the central charge is regarded as a
function of both the scalar fields and the charges:
\begin{eqnarray}
 \mbox{Area}_H (z,{\bar z})&=& \,  4\pi \, \vert Z(z,{\bar z},p,q) \vert^2
 \nonumber\\
 \frac{\delta \mbox{Area}_H }{\delta z}&=&0 \, \longrightarrow \,  z =
 z_{\mbox{\small fix}}\,.
\end{eqnarray}


\section{BPS and non-BPS attractor mechanism: The geodesic potential} \label{sec:geopot}
Quite recently it was noticed that the attractor behavior of extremal black holes in
supersymmetric theories is not peculiar of BPS solutions preserving some supersymmetries
\cite{fegika}, and examples of non-supersymmetric extremal black holes  exhibiting the
attractor phenomenon were found \cite{tt,kss,ae,misra,bfmy,aste}.\par It is then appropriate
to introduce  an alternative approach to extremality which does not rely on the existence of
supersymmetry \cite{gkk,fegika,kss}. Let us start by writing the space-time metric of a black
hole in terms of a new radial parameter $\tau$:
\begin{eqnarray}
ds^2&=&e^{2U}\,dt^2-e^{-2U}\,\left(\frac{c^4}{\sinh^4(c\tau)}\,d\tau^2+\frac{c^2}{\sinh^2(c\tau)}\,d\Omega^2\right)\,.\label{dstau}
\end{eqnarray}
The coordinate $\tau$ is related to the radial coordinate $r$ by
the following relation:
\begin{eqnarray}
\frac{c^2}{\sinh^2(c\tau)}&=&(r-r_0)^2-c^2=(r-r^-)\,(r-r^+)\,.
\end{eqnarray}
Here $c^2\equiv 2ST$ is the extremality parameter of the solution,
with $S$ the entropy and $T$ the temperature of the black hole.
When $c$ is non vanishing the black hole has two horizons located
at  $r^{\pm}=r_0\pm c$. The outer horizon is located at $r_H=r^+$
corresponding to  $\tau\rightarrow -\infty$. The extremality limit
 at which the two horizons coincide, $r_H=r^+=r^-=r_0$, is $c\rightarrow 0$. In this case
 the metric (\ref{dstau}) takes the simple form in the $r$ coordinate
 \begin{eqnarray}
ds^2&=&e^{2U}\,dt^2-e^{-2U}\,\left(dr^2+(r-r_H)^2\,d\Omega^2\right)\,.\label{dsrc0}
 \end{eqnarray}
In the general case, if we require the horizon to have a finite area $A$, the scale
function $U$ in the near-horizon limit should behave as follows
\begin{eqnarray}
e^{-2U}\,&\stackrel{\tau\rightarrow -\infty}{\longrightarrow}
&\,\frac{A}{4\pi}\,\frac{\sinh^2(c\tau)}{c^2}=\frac{A}{4\pi}\,\frac{1}{(r-r^-)(r-r^+)}\,,
\label{nonextremal}
\end{eqnarray}
so that the near--horizon
metric reads
\begin{eqnarray}
ds^2&=&\frac{4\pi}{A}\,(r-r^-)(r-r^+)\,dt^2-\frac{A}{4\pi}\,\frac{dr^2}{(r-r^-)(r-r^+)}-\frac{A}{4\pi}\,d\Omega^2\,.
\end{eqnarray}
The above metric coincides with the near--horizon metric of a
Reissner--Nordstr\"om solution with horizons located at $r^\pm$.
It is useful to introduce the radial coordinate $\rho$ defined as
$\rho=2\,e^{c\tau}$, in terms of which, in the near-horizon
limit, we can write $e^{-2U}\sim \left(\frac{r_H}{\rho
c}\right)^2$,  where $r_H=\sqrt{A/4\pi}$
is the radius of the (outer) horizon,  and the metric becomes
\begin{eqnarray}
ds^2=\left(\frac{\rho
c}{r_H}\right)^2\,dt^2-(r_H)^2\,(d\rho^2+d\Omega^2)\,.
\end{eqnarray}
The coordinate $\rho$ measures the \emph{physical distance} from the
horizon, which is located at $\rho=0$, in units of $r_H$. It is
important to note that the distance of a point at some finite
$\rho_0$ from the horizon is finite:
\begin{eqnarray}
d=\int_0^{\rho_0}\,r_H\,d\rho=r_H\,\rho_0<\infty\,.
\end{eqnarray}
 Using this feature, in \cite{kss} an
intuitive argument was given in order to justify the absence of a
universal behavior for the scalar fields near the horizon of a
non-extremal black hole: the distance from the horizon is not ``long
enough'' in order for the scalar fields to ``loose memory'' of their
initial values at infinity.
\par
Let us now consider the extremal case $c=0$.
The relation
between $\tau$ and $r$ becomes $\tau=-1/(r-r_H)$.
In order to have a
finite horizon area, $U$ should behave near the horizon as:
\begin{eqnarray}
e^{-2U}&\sim& \left(\frac{r_H}{r-r_H}\right)^2\,,
\end{eqnarray}
The physical distance from the horizon is now measured in units
$r_H$ by the coordinate $\omega=\ln (r-r_H)$ in terms of which the
near-horizon metric reads:
\begin{eqnarray}
ds^2=\frac{1}{(r_H)^2}\,e^{2\omega}\,dt^2-(r_H)^2\,(d\omega^2+d\Omega^2)\,.
\end{eqnarray}
Since now the horizon is located at $\omega\rightarrow -\infty$, the
distance of a point at some finite $\omega_0$ from the horizon is
always infinite, as opposite to the non-extremal case:
\begin{eqnarray}
d&=&\int_{-\infty}^{\omega_0}\,r_H\,d\omega=\infty\,.
\end{eqnarray}
 Therefore, as observed in \cite{kss},  the infinite distance from the
 horizon in the extremal case justifies the fact that the scalar
 fields at the horizon ``loose memory'' of their initial values at
 infinity and therefore exhibit a universality behavior. In order to simplify the notation,  in the following
 we shall use the coordinate $r$ to denote the distance from the horizon,
 consistently with our previous treatment of the BPS black hole solutions.
 \par
  Let
us consider the field equations for the metric components
 (see eq. \eq{dstau}) and for the scalar fields $\Phi^r$ coming from
 the lagrangian \eq{bosonicL}. By eliminating the vector fields through their
 equations of motion, the resulting equations for the metric  and the scalar fields,
written in terms of the evolution parameter $\tau$, take the following simple form \cite{gkk}:
\begin{eqnarray}\frac{d^2 U}{d\tau^2}  &=&
V_{\mbox{\small B-H}}(\Phi,p,q)e^{2U}\,,\label{geoeq'}\\ \frac{D^2 \Phi^r}{D\tau^2} &=&
g^{rs}(\Phi)\,\frac{\partial V_{\mbox{\small B-H}}(\Phi,p,q)}{\partial \Phi^s} e^{2U} \,, \label{geoeq}
\end{eqnarray}
with the constraint
\begin{equation}
\left(\frac{d U}{d\tau}\right)^2 + \frac{1}{2}\,g_{rs}(\Phi)\,\frac{d \Phi^r}{d\tau}\frac{d
\Phi^s}{d\tau} - V_{\mbox{\small B-H}}(\Phi,p,q)e^{2U}=c^2 \,,\label{bhconstr}
\end{equation}
where  $V_{\mbox{\small B-H}}(\Phi,p,q)$ is a function of the scalars and of the electric and
magnetic charges of the theory defined by:
\begin{equation}
V_{\mbox{\small B-H}}=-\frac{1}{2}\,Q^t\cM(\cN)Q\,, \label{geopot}
\end{equation}
where as usual $Q$ is the symplectic vector of quantized electric
and magnetic charges and $\cM(\cN)$ is the  symplectic matrix
defined in (\ref{m+}) in terms of the matrix
$\cN_{\Lambda\Sigma}(\Phi)$. Let us note that the field equations
\eq{geoeq} can be extracted from the effective one-dimensional
lagrangian:
\begin{equation}\cL_{eff}= \left(\frac{d U}{d\tau}\right)^2 +\frac{1}{2}\,
g_{rs}\frac{d \Phi^r}{d\tau}\frac{d \Phi^s}{d\tau} + V_{\mbox{\small B-H}}(\Phi,p,q)e^{2U}, \label{effect}
\end{equation}
 constrained with  equation \eq{bhconstr}.
 The extremality condition is $c^2 \to 0$.

 From equation
\eq{effect} we see that the properties of extremal black holes are
completely encoded in the metric of the scalar manifold $g_{rs}$
and on the scalar effective potential $V_{\mbox{\small B-H}}$,
known as black-hole potential or geodesic potential
\cite{gkk,fegika}. In particular, as it was shown in
\cite{gkk,fegika,kss} and as we shall review below, the area of
the event horizon is proportional to the value of $V_{\mbox{\small
B-H}}$ at the horizon:
\begin{equation}\frac{A}{ 4\pi}=V_{\mbox{\small B-H}}(\Phi_h,p,q) \label{mingeo} \end{equation}
where $\Phi_h$ denotes the value taken by the scalar fields at the
horizon \footnote{For the sake of clarity in the comparison with
equivalent formulas in \cite{kss}, let us note that in \cite{kss}
the definition $\Sigma^r = \frac{d \Phi^r}{d\tau}$  has been used.}.
This follows from the property that there is an attractor mechanism
at work in the extremal case.
 To see this  let us consider the set of equations
\eq{geoeq} at $c=0$. Regularity of the scalar fields at the horizon, which is located, with
respect to the physical distance parameter $\omega$, at $\omega\rightarrow -\infty$, implies
that at the horizon the first derivative of $\Phi^r$ with respect to $\omega$ vanishes:
$\partial_\omega\Phi^r_{|h}=0$. Near the horizon a solution to eqs. (\ref{geoeq}), under the
hypothesis that $\left(\partial V_{\mbox{\small B-H}}/\partial\Phi^r\right)_h$ be finite, behaves as
follows:
\begin{equation}
\Phi^r \sim \left.\frac{1}{ 2\,(r_H)^2}\,g^{rs}(\Phi_h)\,\frac{\partial V_{\mbox{\small
B-H}}}{
\partial \Phi^s}\right\vert_{ \Phi_h}\,\omega^2 + \Phi^r_h .
        \label{phih}
\end{equation}
Regularity of $\Phi^r$ at $\omega \rightarrow -\infty$ then further requires that $(\partial
V_{\mbox{\small B-H}}/\partial\Phi^r)_{|h}=0$, implying that the horizon be an attractor point for the
scalar fields.
%
We conclude that in the extremal case the scalar fields tend in the near-horizon limit  to
some fixed values $\Phi^r_h$ which extremize the potential $V_{\mbox{\small B-H}}$:
\begin{equation}
\omega \rightarrow - \infty : \quad \Phi^r (\omega)\mbox{ regular } \Rightarrow \quad
\left.\frac{\partial V_{\mbox{\small B-H}}}{\partial \Phi^r}\right\vert_{\Phi_h} \to 0 \quad ;
\quad \frac{d\Phi^r}{ d \omega}\to 0 . \label{min}
\end{equation} These values are functions of the quantized  electric and magnetic charges only: $\Phi^r_h=\Phi^r_h(p,q)$.
Furthermore, let us consider eq. \eq{bhconstr}. In the extremal
limit $c =0$, near the horizon it becomes:
\begin{equation}\left(\frac{dU}{d\tau}\right)^2 \sim
V_{\mbox{\small B-H}}(\Phi_h(p,q),p,q)e^{2U} \label{hor0}  \end{equation}from which it follows, for the
metric components near the horizon:
\begin{equation}e^{2U} \sim \frac{r^2}{V_{\mbox{\small B-H}}(\Phi_h)} =
\left(\frac{r}{r_H}\right)^2\, ,
\end{equation}that is: \begin{equation}ds^2_{hor} =
\frac{r^2}{V_{\mbox{\small B-H}}(\Phi_h)} dt^2 - \frac{V_{\mbox{\small B-H}}(\Phi_h)}{r^2}\left( dr^2+r^2 d\Omega \right).
\label{hor} \end{equation}

From eqs. \eq{hor0}  and \eq{hor} we immediately see that the value
of the potential at the horizon measures its area, as anticipated in
eq. \eq{mingeo}. The metric  \eq{hor} describes a Bertotti--Robinson
geometry $AdS_2 \times S^2$, with mass parameter $M_{\mbox{\small B-R}}^2
=V_{\mbox{\small B-H}}(\Phi_h)$.

To summarize, we have just shown that the area of the event horizon
of an extremal black hole (and hence its B-H entropy) is given by
the black-hole potential evaluated at the horizon, where it gets an
extremum. This justifies our assertion at the end of the previous section.\par Let us briefly comment on the non-extremal case $c\neq 0
$. For these solutions the physical distance is measured by the
coordinate $\rho$ introduced in eq. \eq{nonextremal} and the horizon
is located at $\rho=0$. The requirement of regularity of the scalar
fields at the horizon is less stringent. It just means that the
scalars should admit a Taylor expansion in $\rho$ around $\rho =0$
and thus it poses no constraints, aside from finiteness, on their
derivatives at the horizon:
\begin{equation}
\Phi^r \sim \Phi^r_h+\left.\frac{\partial\Phi^r}{\partial
\rho}\right\vert_0\,\rho+\left.\frac{1}{ 2\,(r_H)^2}\,g^{rs}(\Phi_h)\,\frac{\partial
V_{\mbox{\small B-H}}}{
\partial \Phi^s}\right\vert_{ \Phi_h}\,\rho^2+ O(\rho^3)\,.
        \label{phinonextr}
\end{equation}
The horizon is therefore not necessarily an attractor point, since at $\rho =0$
$\left(\partial V_{\mbox{\small B-H}}/\partial\Phi^r\right)_{\Phi_h}$ can now be a non
vanishing constant.

\subsection{Extremal black holes in supergravity}
For supergravity theories, supersymmetry fixes the black-hole
potential $V_{\mbox{\small B-H}}$ defined in eqs. \eq{geopot} to take a
particular form that allows to find its extremum in an easy way.
Indeed, an expression exactly coinciding with \eq{geopot} has been
found in section 3 in an apparently different context, as the result
of a sum rule among central and matter charges in supergravity
theories \eq{m+}. So, in every supergravity theory, the black-hole
potential has the general form:
 \begin{equation}V_{\mbox{\small B-H}}\equiv -\frac 12 Q^t\cM(\cN)Q =\frac{1}{2} Z_{AB}\bar Z^{AB}+Z_I \bar
 Z^I\label{bhpotential}.
\end{equation}
By making use of the geometric relations of section 3, the value of
the charge vector $Q=\pmatrix{p^\Lambda \cr q_\Lambda }$ in terms of
the moduli $\Phi$  is given by equations \eq{inversecharge1},
\eq{inversecharge2}.
Then, to find the extremum of $V_{\mbox{\small B-H}}$ we can apply the differential relations (\ref{dz1})
among central and matter charges found in Section 3.

Let us now analyze more in detail, for the case of supergravity
theories, the extremality condition $c=0$ as it comes from the
constraint \eq{bhconstr} which has to be imposed on the solution all
over space-time.
According to the discussion given in the previous section, the  existence of solutions to equation \eq{bhconstr}  does not not rely on
supersymmetry, therefore also non supersymmetric extremal black holes   still exhibit an attractor behavior  \eq{min} (found at $c=0$).

At spatial infinity $\tau \to 0$, where the macroscopic features of the black hole are well
defined, we have $U \to M_{ADM} \tau$, as it follows from the general definition of ADM mass
in General Relativity (see for example \cite{wald}). The metric \eq{ds2U} reduces to the
Minkowski one and the constraint 
\eq{bhconstr} becomes:
\begin{equation}
M_{ADM}^2 = | Z(\Phi_\infty,p,q)|^2+ |Z_I (\Phi_\infty,p,q)|^2 -\frac{1}{2}\,g_{rs}\frac{d
\Phi_\infty^r}{d\tau}\frac{d \Phi_\infty^s}{d\tau} .\label{constrinfinity}\end{equation} These solutions
do not necessarily saturate the BPS bound, since  in general, from \eq{constrinfinity},
$M_{ADM}^2 \neq |Z(\Phi_\infty)|^2$. They then completely break supersymmetry. The behavior
at the horizon may nevertheless be easily found thanks to the expression \eq{bhpotential} that
the black-hole potential takes in supergravity theories, by exploiting the condition \eq{min}
and in particular $\frac{\partial V_{\mbox{\small B-H}}}{\partial \Phi^r}|_{\Phi_h} \to 0$.

For the cases where the black-hole solution preserves some
supersymmetries, we are going to find that
 the constraint
\eq{bhconstr} yields the BPS bound on the mass of the solution. Indeed in that
case one may apply the results of section \ref{sec:bpsextr}. Let us restrict to the case of $N=2$ supergravity, where
$V_{\mbox{\small B-H}}=|Z|^2+|Z_I|^2$. The Killing-spinor equation $\delta_\epsilon \lambda =0$ gives
equation \eq{zequa} that implies
\begin{equation}
\left| \frac{d z^i}{d\tau}\right|^2
=e^{2U}\,|g^{i\bar{\jmath}}\,D_{\bar\jmath}Z|^2.
\label{zibps}
\end{equation}
 By making use of \eq{zibps}, the constraint  \eq{bhconstr} reduces in the extremal limit $c=0$ to
 the following equation, valid all over space-time
\begin{equation}
\left( \frac{d U}{d\tau}\right)^2 =e^{2U}\,|Z|^2 .\label{zbps}
\end{equation}
At spatial infinity $\tau \to 0$, equations \eq{zibps} and \eq{zbps} become
\begin{equation}
M_{ADM}^2 = | Z(\Phi_\infty,p,q)|^2\,; \qquad |Z_I (\Phi_\infty,p,q)|^2 = g_{rs}\frac{d
\Phi_\infty^r}{d\tau}\frac{d \Phi_\infty^s}{d\tau}\,.\label{fix}
\end{equation}
The first equation in \eq{fix} may be recognised as the saturation
of the BPS bound on the mass of the solution.
 On the other hand, near
the horizon the attractor condition holds
\begin{equation}\left.\frac{d
\Phi^r}{d\tau} \right\vert_{h} =0\,,\label{attr}
\end{equation}
 and from \eq{zibps} it gives $Z_I|_{h}=0$, which may be solved to find $\Phi_{\mbox{\small fix}}(p,q)$ leaving,
  for the
 mass parameter at the horizon
\begin{equation}
\left( \frac{d U}{d\tau}\right)_{h}^2=M_{\mbox{\small B-R}}^2(p,q) = | Z(\Phi_{\mbox{\small
fix}},p,q)|^2.
\end{equation}

Actually, the extrema of the  black-hole potential may be systematically studied, both for the
BPS and non-BPS case, by use of the geometric relations \eq{dz1}.
 One finds that
the extrema are given by:
\begin{eqnarray}
dV_{\mbox{\small B-H}} &=& \frac{1}{2}D Z_{AB}\bar Z^{AB}+D Z_I \bar Z^I  +c.c. =\nn\\
   &=& \frac{1}{2}\left(\frac{1}{2} \bar Z^{AB} \bar Z^{CD} P_{ABCD}+
     \bar Z^{AB}    \bar Z^{I}  P_{AB I} + c.c. \right) \nn\\
   &+& \left(\frac{1}{2}  \bar Z^{AB} \bar Z^I  P_{AB I} + \bar Z^I \bar Z^{J}  P_{IJ}   + c.c. \right)
      =0.\label{operativeway}
        \end{eqnarray}

Let us remark that  the one introduced in \eq{operativeway} is a covariant procedure, not
referring explicitly to the horizon properties for finding the entropy, so it is not necessary
to specify explicitly horizon parameters (like the metric and the fixed values of scalars at
that point), $V_{\mbox{\small B-H}}$ being a well defined quantity over all the space-time.

The conditions \eq{operativeway}, defining the extremum of the black-hole potential and thus
the fixed scalars, when restricted to the BPS case have the same content as, and are therefore
completely equivalent to, the relations \eq{zequa} and \eq{Uequa} found in the previous
subsection from the Killing-spinor conditions. In particular, extremal   black holes
preserving   one supersymmetry correspond  to $N$-extended multiplets with
\begin{equation}
M_{ADM} = \vert Z_1 \vert >  \vert Z_2 \vert  \cdots > \vert
Z_{[N/2]} \vert
         \end{equation}
         where $Z_m$, $m =1,\cdots, [N/2]$, are the skew-eigenvalues of
 the
         central charge antisymmetric matrix introduced in \eq{skewZ}
 \cite{fesaz},\cite{CDFp},\cite{adf96},\cite{noi1}: $Z_1=Z_{12},\,Z_2=Z_{34},\dots$.  At the attractor point, where
  $M_{ADM}$ is extremized, supersymmetry requires the vanishing of each term on the right hand side of
  eq. (\ref{operativeway}). In particular we find $Z_I=0$ (recall that $Z_I$ do not exist for $N>4$) and
\begin{eqnarray}
\bar Z^{AB} \bar Z^{CD} P_{ABCD}&=&\,\,\,\Rightarrow\,\,\,\,\bar Z^{[AB} \bar Z^{CD]}=0\,.
\end{eqnarray}
The above condition is satisfied taking $Z_1=Z_{12}\neq 0$ and $Z_m=0,\,m>1$. A general
property of regular BPS black hole solutions is that supersymmetry doubles at the horizon.
This  is consistent with the fact that the near horizon geometry is a Bertotti--Robinson
space--time of the form $AdS_2\times S^2$ which is known to have an unbroken $N=2$
supersymmetry \cite{black}. Let us now give an argument for the vanishing of the supersymmetry
variation along $\epsilon_1,\,\epsilon_2$ of the fermion fields
 at the horizon. As far as the dilatino
fields are concerned, it is sufficient to remember that, since $(d\Phi^r/d\tau)_h=0$, at the
horizon the supersymmetry variation is proportional to $Z_{[AB} \epsilon_{C]}$. However this
expression is also zero since the only non--vanishing central charge is $Z_1\equiv Z_{12}$ and
furthermore $Z_{[12}\epsilon_{1]}=Z_{[12}\epsilon_{2]}=0$. As for the gaugini their
supersymmetry variation at the horizon is automatically zero being $Z_I=0$. Finally let us
remark that  the gravitino variation is not actually zero, however the variation of its field
strength along $\epsilon_1,\,\epsilon_2$ vanishes because of the property of the
Bertotti--Robinson solution of being conformally flat and the fact that the graviphoton field
strength $T_{AB}$ is Lorentz-covariantly constant at the horizon \cite{feka}.

A case by case analysis of the BPS and non-BPS black holes in the various supergravity models,
by inspection of the extrema of $V_{\mbox{\small B-H}}$, will be given in section
\ref{casebycase}. As an exemplification of the method, let us anticipate here the detailed
study of the BPS solution of $D=4$, $N=4$ pure supergravity. The field content is given by the
gravitational multiplet, that is by the graviton $g_{\mu\nu}$, four gravitini $\psi_{\mu A}$,
$A=1,\cdots ,4 $, six vectors $A_\mu^{[AB]}$, four dilatini $\chi^{[ABC]}$ and a complex
scalar $\phi =a+{\rm i}e^\varphi$ parametrizing the coset manifold $G/H=SU(1,1)/U(1)$. The
symplectic $Sp(12)$-sections $(f^\Lambda_{AB}, h_{\Lambda AB})$ ($\Lambda \equiv[AB]=1,\cdots
,6$) over the scalar manifold are given by: \ba
f^\Lambda_{AB}&=& e^{-\varphi /2} \delta^\Lambda_{AB} \nn\\
h_{\Lambda AB}&=& \phi e^{-\varphi /2} \delta_{\Lambda AB} \ea so
that: \begin{equation}\cN_{\Lambda\Sigma}=({\bf h}\cdot
{\bf f}^{-1})_{\Lambda\Sigma}=\phi \delta_{\Lambda\Sigma}
\end{equation}The central charge matrix is then given by:
\begin{equation}Z_{AB}=f^\Lambda_{AB}q_\Lambda -h_{\Lambda AB}p^\Lambda
=-e^{-\varphi /2} (\phi p_{AB}-q_{AB})\,.
\end{equation}
The black-hole potential is therefore: \bea V(\phi , p,q)&=&\half
e^{-\varphi} (\phi p_{AB}-q_{AB})
                (\bar\phi p^{AB}-q^{AB}) \nn\\
              &=&\half (a^2 e^{-\varphi}+e^{\varphi} ) p_{AB}q^{AB}
                  + e^{-\varphi} q_{AB}q^{AB}-2ae^{-\varphi}q_{AB}p^{AB}\nn\\
              &\equiv &\half  (p,q)\left(\matrix{1&0 \cr -a &
              1\cr }\right)
                     \left(\matrix{e^\varphi & 0 \cr 0 & e^{-\varphi} \cr }\right)
                     \left(\matrix{1& -a \cr 0 & 1\cr }\right)
                     \left(\matrix{p \cr q }\right)
\eea By extremizing the potential in the moduli space we get: \bea
\frac{\del V}{\del a}=0 & \to & a_{h}=\frac{q_{AB}p^{AB}}{p_{AB}p^{AB}}\nn\\
\frac{\del V}{\del \varphi}=0 & \to & e^{\varphi_{h}} = \frac{\sqrt{|q_{AB}q^{AB}p_{CD}p^{CD}-
(q_{AB}p^{AB})^2}|}{p_{AB}p^{AB}} \eea from which it follows that the entropy is:
\begin{equation}S_{\mbox{\small B-H}}=4\pi V(\phi_h, p,q)=4\pi \sqrt{|q_{AB}q^{AB}p_{CD}p^{CD}-
(q_{AB}p^{AB})^2|}
\end{equation}\vskip 5mm As a final observation, let us note,
following \cite{fegika}, that the extremum reached by the black-hole
potential at the horizon is in particular a minimum, unless the
metric of the scalar fields change sign, corresponding to some sort
of phase transition, where the effective lagrangian description
\eq{effect} of the theory breaks down. This can be seen from the
properties of the Hessian of the black-hole potential. It was shown
in \cite{fegika} for the $N=2$, $D=4$ case that at the critical
point $\Phi=\Phi_{\mbox{\small fix}}\equiv \Phi_h$, from the special geometry
properties it follows: \begin{equation}\left(
\partial_{\bar\imath}
\partial_j |Z| \right)_{\mbox{\small fix}}= \frac{1}{2}g_{\bar \imath j}|Z|_{\mbox{\small fix}}
\end{equation}and then, remembering, from the above discussion, that
$V_{\mbox{\small fix}}=|Z_{\mbox{\small fix}}|^2$: \begin{equation}\left(
\partial_{\bar\imath}
\partial_j V \right)_{\mbox{\small fix}}= {2}g_{\bar \imath j}|Z_{\mbox{\small fix}}|^2
\label{wein} \end{equation}From eq. \eq{wein} it follows, for the
$N=2$ theory, that the minimum is unique.
 \\
 In the next section we will show one more technique for finding
 the entropy, exploiting the fact that it is a `topological quantity' not
 depending on scalars. This last procedure is particularly interesting
 because it refers only to group theoretical properties of the coset manifolds
 spanned by scalars, and do not need the knowledge of any details of the
 black-hole horizon.


\subsection{B-H entropy as a U-invariant for symmetric spaces}\label{u-inv}
For theories based on moduli spaces given by symmetric manifolds
$G/H$, which is the case of all supergravity theories with $N\geq 3$
extended supersymmetry, but also of several $N=2$ models, the BPS
and non-BPS black holes are classified by some U-duality invariant
expressions depending on the representation of the group $G$ of
$G/H$ under which the electric and magnetic charges are classified.
In this respect, the classification of the $N=2$ invariants is
entirely similar to the $N>2$ cases, where all scalar manifolds are
symmetric spaces.

For theories that have a quartic invariant $I_4$ \cite{kako} (this includes all $N=2$ symmetric spaces
based on cubic prepotentials \cite{fgk06,gp} and $N= 4, 6,8$ theories), the B-H entropy turns out to be
proportional to its square root \begin{equation} S_{\mbox{\small B-H}} \propto \sqrt{|I_4|}.
\end{equation}
The BPS solutions have $I_4 >0$ while the non-BPS ones (with non
vanishing central charge) have instead $I_4<0$. For all the above
theories with the exception of the $N=8$ case,  there is also a
second non-BPS solution with vanishing central charge and $I_4>0$.

For theories based on symmetric spaces with only a quadratic invariant $I_2$  (this includes
$N=2$ theories with quadratic prepotentials  as well as $N=3$ and $N=5$ theories), the B-H
entropy is
\begin{equation}
S_{\mbox{\small B-H}} \propto |I_2|.
\end{equation}
In these cases, beyond the BPS solution which has $I_2>0$ there is
only  one non-BPS solution, with vanishing central charge and
$I_2<0$.

All the solutions discussed here give $S_{\mbox{\small B-H}}\neq 0$ and then fall
in the class of the so-called large black holes, for which the
classical area/entropy formula is valid as it gives the dominant
contribution to the black-hole entropy. Solutions with $I_4,I_2 =0$
do exist but they do not correspond to classical attractors since in
that case the classical area/entropy formula vanishes. In this case
one deals with small black holes, and a quantum attractor
mechanism, including higher curvature terms, has to be considered
for finding the entropy.

 The main purpose of this subsection
is to provide  particular expressions which
 give the
entropy formula as a moduli-independent quantity in the entire
moduli space and not just at the critical points. Namely, we are
looking for quantities $S\left(Z_{AB}(\phi), \bar Z^{AB}
 (\phi),Z_{I}(\phi), \bar Z^{I} (\phi)\right)$
such that $\frac{\partial}{\partial \phi ^i} S =0$, $\phi ^i$
being the moduli
 coordinates
 \footnote{The Bekenstein-Hawking entropy $S_{\mbox{\small B-H}} =\frac{A}{4}$ is actually
$\pi S$ in our notation.}. To this aim, let us first consider
invariants $I_\alpha$ of the isotropy group $H$ of the scalar
manifold $G/H$, built with the central and matter charges. We will
take all possible $H$-invariants up to quartic ones for four
dimensional theories (except for the $N=3$ case, where the
invariants of order higher than quadratic are not irreducible).
Then, let us consider a linear combination $S^2=\sum_\alpha
C_\alpha I_\alpha$ of the $H$-invariants, with arbitrary
coefficients $C_\alpha$. Now, let us extremize $S$ in the moduli
space $\frac{\del S}{\del \Phi^i}=0$, for some set of
$\{C_\alpha\}$. Since $\Phi^i \in G/H$, the quantity found in this
way (which in all cases turns out to be unique) is a
U-invariant, and is indeed proportional to the
Bekenstein--Hawking entropy.

 These formulae generalize the quartic $E_{7(-7)}$ invariant of $N=8$
supergravity \cite{kako} to all other cases. \footnote{Our analysis
is based on general properties of scalar coset manifolds. As a
consequence, it can be applied straightforwardly also to the $N=2$
cases, whenever one considers special coset manifolds.}
\par
Let us first consider the theories $N=3,4$, where  matter can be
present \cite{ccdffm},\cite{bekose}.
\par
The U-duality groups \footnote{Here we denote by U-duality group the
isometry
 group $U$
acting on the scalars in a symplectic representation, although only
a restriction of it to integers is the
 proper U-duality group \cite{huto}.}
 are, in these cases, $SU(3,n)$ and $SU(1,1)
\times SO(6,n)$ respectively. The central and matter charges
$Z_{AB}, Z_I$ transform in an obvious way under the isotropy
groups
\begin{eqnarray}
H&=& SU(3) \times SU(n) \times U(1) \qquad (N=3) \\
 H&=& SU(4) \times
 SO(n) \times U(1) \qquad (N=4)
\end{eqnarray}
Under the action of the elements of $G/H$ the charges may get mixed with their complex conjugate. The infinitesimal transformation
 can be read from the differential relations satisfied by the charges
 \eq{dz1} \cite{adf96} .
\par
For $N=3$:
\begin{equation}
P^{ABCD}=P_{IJ}=0 , \quad P_{ AB I} \equiv \epsilon_{ABC}P^C_I
\quad Z_{AB}
 \equiv \epsilon_{ABC}Z^C
 \label{viel3}
\end{equation}
Then  the variations are:
\begin{eqnarray}
\delta Z^A  &=& \xi^{A I}  Z_I  \\
\delta Z_{I}  &=&\bar\xi_{A I}  Z^A \label{deltaz3}
\end{eqnarray}
where $\xi^{AI}$ are infinitesimal parameters of $K=G/H$.
\par
The possible quadratic $H$-invariants are: \ba
I_1 &=& Z^A\bar Z_A \nn\\
I_2 &=& Z_I \bar Z^I \ea So, the U-invariant expression is:
\begin{equation}
S =  | Z^A \bar Z_A - Z_I \bar Z^I| \label{invar3}
\end{equation}
In other words, $D_i S = \partial_i S =0 $, where the
covariant derivative is defined in ref. \cite{adf96}.
\par
Note that at the attractor point ($Z_I =0$) it coincides with the
moduli-dependent potential (\ref{bhpotential}) computed at its
extremum.
\par
For $N=4$
\begin{equation}
P_{ABCD} = \epsilon_{ABCD}P ,\quad P_{IJ} = \eta_{IJ}P,\quad P_{AB
I}=\frac{1}{2} \eta_{IJ} \epsilon_{ABCD}\bar P^{CD J}
\label{viel4}
\end{equation}
and the transformations of $K= \frac{SU(1,1)}{U(1)} \times
\frac{O(6,n)}{O(6) \times O(n)}$ are:
 \begin{eqnarray}
\delta Z_{AB}  &=& \frac{1}{2} \epsilon_{ABCD} \, \xi \,\bar Z^{CD}  +
 \xi^I_{AB }  Z_I \\
\delta Z_{I}  &=&\bar \xi \,\eta_{IJ}\, \bar Z^J + \frac{1}{2}\,\bar \xi^{AB}_I\,
 Z_{AB} \label{deltaz4}
\end{eqnarray}
with $\bar \xi^{AB }_I =  \frac{1}{2} \eta_{IJ}   \epsilon^{ABCD}
\xi_{CD }^J$.
\par
The possible $H$-invariants are: \ba
I_1 &=& Z_{AB} \bar Z^{AB}\nn\\
I_2 &=& Z_{AB} \bar Z^{BC} Z_{CD} \bar Z^{DA} \nn\\
I_3 &=& \epsilon^{ABCD}  Z_{AB}   Z_{CD} \nn\\
I_4 &=& Z_I Z^I \ea There are three $O(6,n)$ invariants given by
$S_1$, $S_2$, $\bar S_2$ where:
\begin{eqnarray}
  S_1 &=& \frac{1}{2}  Z_{AB} \bar Z^{AB} - Z_I \bar Z_I
\label{invar41} \\
S_2 &=& \frac{1}{4} \epsilon^{ABCD}  Z_{AB}   Z_{CD} - Z_I
Z_I \label{invar42}
\end{eqnarray}
and the unique   $SU(1,1)  \times O(6,n) $  invariant $S$, $D
S =0$, is given by:
\begin{equation}
S= \sqrt{|(S_1)^2 - \vert S_2 \vert ^2 |} \label{invar4}
\end{equation}
At the attractor point $Z_I =0 $ and $\epsilon^{ABCD} Z_{AB}
Z_{CD}
=0$ so that $S$ reduces to the square of the BPS mass.\\
Note that, in absence of matter multiplets, one recovers the
expression found in the previous subsecion by extremizing the black
hole potential.
\par
For $N=5,6,8$ the U-duality invariant expression $S$ is the square
root of a unique invariant under the corresponding U-duality
groups $SU(5,1)$, $O^*(12)$ and $E_{7(-7)}$. The strategy is to
find a quartic expression $S^2$    in terms of $Z_{AB}$ such that
$D S=0$, i.e. $S$ is moduli-independent.
\par
As before, this quantity is a particular combination of the $H$
quartic invariants.
\par
For $SU(5,1)$ there are only two  $U(5)$ quartic invariants. In
terms of the matrix $A_A^{\ B} = Z_{AC} \bar Z^{CB}$ they are:
$(Tr A)^2$, $Tr(A^2)$, where
\begin{eqnarray}
 Tr A & = & Z_{AB} \bar Z^{BA} \\
 Tr (A^2) & = & Z_{AB} \bar Z^{BC} Z_{CD} \bar Z^{DA}
\end{eqnarray}
As before, the relative coefficient is fixed by the transformation
properties of $Z_{AB}$ under $\frac{SU(5,1)}{U(5) } $ elements of
infinitesimal parameter $\xi^C$:
\begin{eqnarray}
  \delta Z_{AB} = \frac{1}{2} \xi^C \epsilon_{CABPQ} \bar Z^{PQ}
\end{eqnarray}
It then follows that the required invariant is:
\begin{equation}
S= \frac{1}{2} \sqrt{  |4 Tr(A^2) - (Tr A)^2| } \label{invar5}
\end{equation}

The $N=6$ case is the more complicated because under $U(6)$ the
left-handed spinor of $O^*(12)$ splits into:
\begin{equation}
32_L \to 15_1 +  \bar {15}_{-1} + 1_{-3} + 1_{3}
\end{equation}
The transformations of $\frac{O^*(12)}{U(6)}$ are:
\begin{eqnarray}
\delta Z_{AB} &=& \frac{1}{4} \epsilon_{ABCDEF} \xi^{CD} \bar
Z^{EF} +
\xi_{AB} \bar X \\
\delta X &=& \frac{1}{2} \xi _{AB} \bar Z^{AB}
\end{eqnarray}
 where we denote by $X$ the $SU(6)$ singlet.

The quartic $U(6)$ invariants are:
\begin{eqnarray}
I_1&=& (Tr A)^2 \label{invar61}\\
I_2&=& Tr(A^2)\label{invar62} \\
I_3 &=& Re (Pf \, ZX)= \frac{1}{2^3 3!}
Re( \epsilon^{ABCDEF}Z_{AB}Z_{CD}Z_{EF}X)\label{invar63}\\
I_4 &=& (Tr A) X \bar X\label{invar64}\\
I_5&=& X^2 \bar X^2\label{invar65}
\end{eqnarray}
where the matrix $A$ is, as for the $N=5$ case,
$A_A^{\ B} = Z_{AC} \bar Z^{CB}$.

The unique $O^*(12)$ invariant is:
\begin{eqnarray}
S&=&\frac{1}{2} \sqrt{|4 I_2 - I_1 + 32 I_3 +4I_4 + 4 I_5| }
\label{invar6}   \\
D S &=& 0
\end{eqnarray}
Note that at the BPS attractor point $Pf\,Z =0$, $X=0$ and $S$ reduces
to the square of the BPS mass.

For $N=8$ the $SU(8)$ invariants are \footnote{The Pfaffian of an
 $(n\times n)$ ($n$ even) antisymmetric
matrix is defined as $Pf Z=\frac{1}{ 2^n n!} \epsilon^{A_1 \cdots
A_n}
 Z_{A_1A_2}\cdots Z_{A_{N-1}A_N}$, with the property:
$ \vert Pf Z \vert = \vert det Z \vert ^{1/2}$.}:
\begin{eqnarray}
I_1 &=& (Tr A) ^2 \\
I_2 &=& Tr (A^2) \\
I_3 &=& Pf \, Z
 =\frac{1}{2^4 4!} \epsilon^{ABCDEFGH} Z_{AB} Z_{CD} Z_{EF} Z_{GH}
\end{eqnarray}
The $\frac{E_{7(-7)}}{ SU(8)}$ transformations are:
\begin{equation}
\delta Z_{AB} =\frac{1}{2} \xi_{ABCD} \bar Z^{CD}
\end{equation}
where $\xi_{ABCD}$ satisfies the reality constraint:
\begin{equation}
\xi_{ABCD} = \frac{1}{24} \epsilon_{ABCDEFGH} \bar \xi^{EFGH}
\end{equation}
One finds the following $E_{7(-7)}$ invariant \cite{kako}:
\begin{equation}
S= \frac{1}{2} \sqrt{|4 Tr (A^2) - ( Tr A)^2 + 32 Re (Pf \, Z)| }
\end{equation}


\section{Detailed analysis of attractors in extended supergravities: BPS and non-BPS critical points}\label{casebycase}

 The extremum principle was
found originally in the context of  $N=2$ four-dimensional black holes.
However, as we have described in section \ref{sec:extremum}, it has
 a more general validity, being true for all $N$-extended
supergravities in four dimensions (in the cases where the Bekenstein--Hawking
entropy is different from zero)
\cite{adf96}. Indeed, the general discussion of section \ref{sugras}
shows that the coset structure of extended supergravities in four
dimensions (for $N>2$) induces the existence, in every theory, of
differential relations among central and matter charges that
generalize the ones existing for the $N=2$ case. Furthermore, as far
as BPS solutions are considered, Killing-spinor equations for
gauginos and dilatinos analogous to eq. \eq{dz1} are obtained by
setting to zero the supersymmetry transformation laws of the
fermions. Correspondingly, at the fixed point $\partial_\mu
\Phi^i=0$, for any extended supergravity theories one gets some
conditions that allow to find the value of fixed scalars and hence
of the B-H entropy both for BPS and non-BPS black
hole solutions.

We will first discuss in section \ref{n=2attractors} the case of $N=2$ supergravity, then in section \ref{n=34attractors} the case of the other extended theories allowing matter couplings to the supergravity multiplet, that is $N=3,4$ extended supergravities, and finally we will pass to analyze in section \ref{n=568attractors} $N=5,6,8$ theories, which are pure supergravity models.

For every theory, the strategy adopted to find the extrema will be to solve the equation
$dV_{\mbox{\small B-H}}=0$, as given in general in \eq{operativeway}, by setting to zero all the independent
components in the decomposition on a basis of vielbein of the moduli space \cite{adf96}.

We confine our analysis to large black holes, with finite horizon area.


\subsection{$N=2$ attractor equations}\label{n=2attractors}

In the original paper \cite{fegika} the $N=2$ attractor conditions
were introduced via an extremum condition on the black-hole
potential \eq{geopot}
\begin{equation}
V_{\mbox{\small B-H}} = -\frac 12 Q^T \mathcal{M} Q = |Z|^2 + |D_i Z|^2\end{equation} discussed in section
\ref{sec:geopot}. Indeed, by making use of properties of $N=2$ special geometry, the extremum
condition was written in the form
\begin{equation}
\partial_i V_{\mbox{\small B-H}}= 2 \bar Z D_i Z + \ii C_{ijk}g^{j\jbar}g^{k\kbar} D_{\jbar} \bar Z D_{\kbar} \bar Z =0 \,,\label{extremum}
\end{equation}
where use of the special geometry relations \eq{geospec}  was made.

Given \eq{extremum}, it is useful to write the attractor equations in a different form. Indeed, recalling equations \eq{inversecharge1},
\eq{inversecharge2}  \cite{bfm1,bcf,k} (which are true all over the moduli space) we may write:
\begin{equation}
Q -\ii \,\mathbb{C} \,\cM(\cN ) \cdot Q =- 2\ii \,\bar {\bf V}^M
Z_M = -2 \ii\left( Z \bar V + g^{i\jbar} D_{\jbar} \bar Z D_{i}
V\right) \, ,\label{extremumbis}
\end{equation}
where $V$ is the symplectic section introduced in \eq{sezio};   substituting the extremum condition from \eq{extremum}, eq. \eq{extremumbis} gives the value of the charges in terms of the fixed scalars:
\begin{eqnarray}
\left.\left[Q -\ii \,\mathbb{C} \,\cM(\cN ) \cdot
Q\right]\right\vert_{\mbox{\small fix}} &=& - 2 \ii\left.\left( Z
\bar V +
\frac\ii{2Z} \bar C^{ijk} D_{i}V \,D_j Z \,D_k Z\right)\right\vert_{\mbox{\small fix}}  \mbox{ for } Z_{\mbox{\small fix}} \neq 0 \,,\nonumber\\
\left.\left[Q -\ii \,\mathbb{C} \,\cM(\cN ) \cdot
Q\right]\right\vert_{\mbox{\small fix}}& =& -2  \ii\,\left.\left(
g^{i\jbar} D_{\jbar} \bar Z D_{i}
V\right)\right\vert_{\mbox{\small fix}}\,\,\,\, \mbox{ for }
Z_{\mbox{\small fix}} = 0 \,. \label{extremumter}
\end{eqnarray}
The BPS solution corresponds to set $D_i Z=0$, in which case, for
large black holes ($Z_{\mbox{\small fix}} \neq 0$), eq.
\eq{extremumter} reduces to \eq{cu}.

The attractive nature of the extremum was further seen to come
from the fact that the mass matrix at that point is strictly
positive since
\begin{equation}
\partial_i \partial_j V_{\mbox{\small B-H}}|_{(\partial_i V_{\mbox{\tiny B-H}}=0)} =0\,;\quad
\partial_i \partial_\jbar V_{\mbox{\small B-H}}|_{(\partial_i V_{\mbox{\tiny B-H}}=0)} = 2 |Z|^2 g_{i\jbar}\,.
\end{equation}

Non supersymmetric extremal black holes with finite  horizon area correspond to solutions of \eq{extremum} with
\begin{equation}
D_i Z\neq 0\,.
\end{equation}
These solutions may be divided in two classes
\begin{itemize}
\item
$D_i Z\neq 0,  Z\neq 0 $\,,
\item $D_i Z\neq 0, Z = 0$\,.
\end{itemize}
For these more general cases, the horizon mass parameter $M_{\mbox{\small B-R}}$
which extremizes the ADM mass in moduli space is then  given by
\begin{equation}
M^2_{\mbox{\small B-R}}=V_{\mbox{\small B-H}}|_{(\partial_i V_{\mbox{\tiny B-H}}=0)} =\left[|Z|^2 + |D_i
Z|^2\right]_{(\partial_i V_{\mbox{\tiny B-H}}=0)} > |Z|^2_{(\partial_i
V_{\mbox{\tiny B-H}}=0)} \,. \label{bpsbound}
\end{equation}
Equation \eq{bpsbound} is nothing but the BPS bound on the mass.

If  the central charge $Z$ vanishes on the extremum, then $D_i Z$ have to satisfy
\begin{equation}
 C_{ijk}g^{j\jbar}g^{k\kbar} D_{\jbar} \bar Z D_{\kbar} \bar Z =0 \qquad \forall i
\end{equation}
in order to fulfill \eq{extremum}.
Solutions to the above equation, for the case of special geometries based on symmetric spaces, have been given in \cite{bfgm}.

When $Z\neq 0, D_i Z \neq 0$,  one may obtain some further consequences of \eq{extremum}. Let us define
\begin{equation}
Z^\ibar \equiv  g^{i\ibar} D_i Z \,,\quad \bar Z^i \equiv g^{i\ibar} D_\ibar \bar Z
.\end{equation}
From \eq{extremum} we get, by multiplication with $g^{i\ibar}$
\begin{equation}
Z^\ibar = -\frac\ii{2\bar Z} C^\ibar_{\ jk}\bar Z^j  \bar Z^k\label{ridefz}
\end{equation}
and, by multiplication with $\bar Z^i$
\begin{equation}
|D_i Z|^2 = -\frac\ii{2\bar Z} N_3(\bar Z^k)=\frac \ii{2 Z} N_3 (Z^\kbar) \label{ridef2}
\end{equation}
where we have introduced the definition $N_3(\bar Z^k)\equiv C_{ijk}\bar Z^i \bar Z^j  \bar Z^k$.  Note that, if at the attractor point $N_3=0$, then $Z=0$ (or $Z\neq 0$ but then $Z^\ibar=0$).

The complex conjugate of \eq{extremum} may be rewritten, using \eq{ridefz} as
\begin{equation}
2Z D_\ibar \bar Z = -\frac \ii{4\bar Z^2} C_{\ibar\jbar\kbar} C^\jbar_{\ \ell m}\bar Z^\ell  \bar Z^m C^\kbar_{\ pq}\bar Z^p  \bar Z^q.
\end{equation}
By making use of the special geometry relation \cite{cv85,dwvds,bfgm}
\begin{equation}
C_{\ibar\jbar\kbar} C^\jbar_{\ (\ell m}C^\kbar_{\ pq)}= \frac 43 C_{(\ell m p}g_{q)\ibar}+
\bar E_{\ibar \ell m pq},\label{ssrel}
\end{equation}
where the tensor $\bar E_{\ibar \ell m pq}$ defined by this
relation is related to the covariant derivative of the Riemann
tensor and it exactly vanishes for all symmetric spaces \footnote{In
this case equation \eq{ssrel} is a consequence of the special
geometry relation $D_i C_{jk\ell}=0$.}, we may finally rewrite
\eq{extremum} as
\begin{equation}
2 \bar Z D_i Z =\frac \ii{6Z^2}D_i Z  C_{\jbar\kbar\lbar}Z^\jbar Z^\kbar Z^\lbar + \frac \ii{8Z^2} E_{i\jbar\kbar\lbar\mbar}Z^\jbar Z^\kbar Z^\lbar Z^\mbar. \end{equation}
Moreover, using also \eq{ridef2} we obtain
\begin{equation}
\left(|Z|^2 - \frac 13 |D_iZ|^2\right)D_i Z= \frac \ii {8Z} E_{i\jbar\kbar\lbar\mbar}Z^\jbar Z^\kbar Z^\lbar Z^\mbar.\label{finalcalc}
\end{equation}
For symmetric spaces eq. \eq{finalcalc} gives
\begin{equation}
|D_iZ|^2=3|Z|^2\label{nonbps}
\end{equation}
implying that for these black holes:
 $M^2_{\mbox{\small B-R}} =4|Z|^2_{(\partial_i
V_{\mbox{\small B-H}}=0)}$.

 This relation, for symmetric spaces, was obtained in
\cite{fk2006} and then all the solutions of this type have been
classified in \cite{bfgm}. In particular, solutions with
$C_{ijk}\equiv 0$ correspond to the special series of symmetric
special manifolds $\frac{SU(1,1+n)}{U(1)\times SU(1+n)}$ for which
only non-BPS solutions with $Z=0$ may exist.

 Solutions of the type in \eq{nonbps} have also been found for non-symmetric spaces based on cubic prepotentials in \cite{tt}.

 However, because of  \eq{finalcalc}, these cannot be the most general solutions.
For the generic case of non-symmetric special manifolds, we have instead
\begin{equation}
|D_iZ|^2=3|Z|^2 +\Delta\label{eqdelta}
\end{equation}
where
\begin{equation}
\Delta = - \frac 34\frac  {E_{i\jbar\kbar\lbar\mbar}Z^\jbar Z^\kbar Z^\lbar Z^\mbar}{N_3(Z^\kbar)}
\end{equation}
and the Bekenstein--Hawking entropy is
\begin{equation}
S_{\mbox{\small B-H}}=A/4= \pi \left(4|Z|^2 +\Delta\right).
\end{equation}
Note that, for these non-BPS black holes,   at the attractor point $\Delta $ is real and,
because of \eq{eqdelta}, it satisfies
$-\Delta < 3|Z|^2$.

 In all the cases, the attractive nature of the solution depends on the Hessian matrix, which however may have null directions.


\subsection{$N>2$  matter coupled attractors}\label{n=34attractors}

\subsubsection{The $N=3$ case}
The scalar manifold for this theory, as discussed in section \ref{sugras}, is the coset space
\begin{equation}
G/H=\frac{SU(3,n)}{SU(3)\times SU(n)\times U(1)}\label{n=3coset}
\end{equation}
and the relations among central and matter charges are (see \eq{dz1})
\begin{eqnarray}
  D(\omega) Z_{AB} &=&   Z_{I}  P^I_{AB }\,,\nonumber \\
D(\omega) Z_{I} &=& \frac{1}{2}   Z_{AB} \bar P^{AB}_I \,.
\end{eqnarray}
The extremum condition on the black-hole potential is then
\begin{eqnarray}
dV_{\mbox{\small B-H}}&=&\frac 12 D Z_{AB} \bar Z^{AB} +\frac 12  Z_{AB}D \bar Z^{AB}
+D Z_{I} \bar Z^{I} + Z_{I}D \bar Z^{I} \nonumber\\
&=&P_{AB }^I \bar Z^{AB}  Z_{I} +c.c.=0
\end{eqnarray}
and allows two different solutions with non-zero area. This is expected from section \ref{u-inv} because the isometry group of the symmetric space \eq{n=3coset} only has a quadratic invariant
\begin{equation}
I_2=\frac 12 |Z_{AB}|^2 -|Z_I|^2\,.
\end{equation}
Then,
\begin{itemize}
\item
either $Z_{AB} \neq 0$, $Z_I=0$, in this case we have a BPS
attractor and the black-hole potential becomes
\begin{equation}
V_{\mbox{\small B-H}}|_{\mbox{\small attr}}=I_2|_{\mbox{\small attr}}>0\,,
\end{equation}
\item or
$Z_{I} \neq 0$, $Z_{AB}=0$, which gives a non-BPS attractor solution
with black-hole potential
\begin{equation}
V_{\mbox{\small B-H}}|_{\mbox{\small attr}}= -I_2|_{\mbox{\small attr}}>0\,.
\end{equation}
\end{itemize}


 \subsubsection{The $N=4$ case}

In this case the scalar manifold  is the coset space
\begin{equation}
G/H=\frac{SU(1,1)}{U(1)} \times \frac{SO(6,n)}{SO(6)\times SO(n)}\label{n=4coset}
\end{equation}
and the relations among central and matter charges are (see \eq{dz1} and the discussion below)
\begin{eqnarray}
  D(\omega) Z_{AB} &=&   Z_{I}  P_{AB}^I +  \frac{1}{2} \bar Z^{CD} \epsilon_{ABCD}\, P\,, \nonumber \\
D(\omega) \bar Z_{I} &=& \frac{1}{2}  \bar Z^{AB}  P_{AB I} +
Z_I  \, P\,.
\end{eqnarray}
We recall that for this theory the vielbein $P_{AB I}$ satisfies the reality condition
$\bar P^{AB I} \equiv (P_{AB I})^\star = \frac 12 \epsilon^{ABCD}  P^I_{CD }$.

The extremum condition on the black-hole potential is then
\begin{eqnarray}
dV_{\mbox{\small B-H}}&=&\frac 12 D Z_{AB} \bar Z^{AB} +\frac 12  Z_{AB}D \bar Z^{AB}
+D Z_{I} \bar Z^{I} + Z_{I}D \bar Z^{I}=0 \nonumber\\
&=&P_{AB I} \left(\bar Z^{AB}  Z_{I} +\frac 12 \epsilon^{ABCD} Z_{CD} \bar  Z_{I}\right) + P \left( Z_{I}  Z_{I} +\frac 14 \epsilon_{ABCD} \bar Z^{AB}\bar Z^{CD} \right)
+\nonumber\\
&&+\bar P \left( \bar Z^{I} \bar  Z^{I} +\frac 14 \epsilon^{ABCD} \bar Z_{AB}\bar Z_{CD} \right)=0\label{extrn=4}\,.
\end{eqnarray}
Equation \eq{extrn=4} is satisfied for
\begin{equation}
\left\{\matrix{\bar Z^{AB}  Z^{I} +\frac 12 \epsilon^{ABCD} Z_{CD} \bar Z^{I}&=~0 \cr
Z^{I}  Z^{J}\delta_{IJ} +\frac 14 \epsilon_{ABCD} \bar Z^{AB}\bar Z^{CD} &=~0}
 \right.\,.
 \end{equation}
Therefore we have, in terms of the proper values $Z_1,Z_2$ of the central charge antisymmetric matrix $Z_{AB}$ (by means of a $U(1)\subset H$ transformation \cite{zum}, they may always be chosen real and positive) and of the complex matter charges $Z^I$
\begin{equation}
\left\{\matrix{\bar Z_1  Z^I + Z_{2}  \bar Z^I&=~0 \cr
Z^IZ^I+2 \bar Z_1\bar Z_2 &=~0}
 \right.\,. \label{n=4extremum}
 \end{equation}
\begin{itemize}
\item
 The BPS solution with finite area is found, as discussed in general in section \ref{sec:geopot}, for
 \begin{equation}
 Z_I =0  \,; \qquad Z_2=0 \quad (\mbox{for }Z_1>Z_2)
 \end{equation}
 and corresponds to the black-hole potential
 \begin{equation}
 V_{\mbox{\small B-H}}|_{\mbox{\small attr}}=(Z_1)^2\,.
 \end{equation}
 This solution partially breaks the symmetry of the moduli space, as
$$
\left\{\matrix{
SU(4)& \to & SU(2)\times SU(2) \times U(1)\cr
SO(n) &\to & SO(n)
}
 \right. \,.
 $$

\end{itemize}

 There are also two non-BPS solutions:
 \begin{itemize}
\item
 One is found by  choosing  $Z_I =(z,\vec 0)$
\begin{equation}
\left\{\matrix{
Z_1&=~ Z_2 =\rho\cr
z&=~\sqrt 2 \ii \rho}
 \right.
 \end{equation}
which gives, for the black-hole potential
\begin{equation}
V_{\mbox{\small B-H}}|_{\mbox{\small attr}} = (Z_1)^2 + (Z_2)^2 + |z|^2 = 4\rho^2\,.
\end{equation}
In this case the isotropy symmetry then becomes
$$
\left\{\matrix{
SU(4)& \to & USp(4)\cr
SO(n) &\to & SO(n-1)
}
 \right. \,.
 $$
\item
The other is obtained by choosing instead   $Z_I =(k_1,k_2,\vec 0)$ and $Z_{AB}=0$.
This solves \eq{n=4extremum} for
$k_1^2 +k_2^2=0$, that is for $k_2=\pm \ii k_1=\ii k$, giving
\begin{equation}
V_{\mbox{\small B-H}}|_{\mbox{\small attr}} = |k_1|^2 + |k_2|^2 = 2|k|^2 \,.
\end{equation}
For this case, then, the isotropy symmetry preserved is
$$
\left\{\matrix{
SU(4)& \to & SU(4)\cr
SO(n) &\to & SO(n-2)
}
 \right. \,.
 $$
\end{itemize}
\bigskip

 The analysis of this section is in accord with the discussion on U-invariants of section \ref{u-inv}.
 Indeed, the isometry group of the scalar manifold  \eq{n=4coset} admits the quartic invariant \eq{invar4}
 \begin{equation}
I_4 = S_1^2 - |S_2|^2
\end{equation}
where $S_1$ and $S_2$ are the $O(6,n)$ invariants introduced in \eq{invar41}, \eq{invar42} and
we have $S_{\mbox{\small B-H}}=\sqrt{|I_4|}$.

For the BPS case, $I_4>0$. For the non-BPS ones we have, in the first case $I_4=-|S_2|^2<0$, in the second case $I_4=S_1^2 >0$.

\bigskip

The case of the pure $N=4$ supergravity model anticipated as an
example in section \ref{sec:geopot} falls in this classification
and corresponds to the BPS solution (since in that case $Z_I \equiv
0$). It is however interesting to look at the $N=2$ reduction of
that model, where only 2 of the 6 vector fields survive, one as the
graviphoton and one inside a vector multiplet whose scalars span the
coset $\frac{SU(1,1)}{U(1)}$ (axion-dilaton system).
Correspondingly, the two proper-values of the $N=4$ central charge
play now two different roles: one, say $Z_1$, is the $N=2$ central
charge, while the other, $Z_2$, is  the matter charge. Equation
\eq{n=4extremum} has now two distinct solutions (corresponding to
the twice degenerate BPS solution in $N=4$): the BPS one, for
$Z_2=0$, $M_{ADM}=Z_1$, and a non-BPS one, for $Z_1=0$, $Z_2\neq 0$.
This is understood, in terms of invariants, from the fact that
$SU(1,1)$ does not have an independent quartic invariant, and in
fact, in this case, one finds that $I_4$ reduces to $I_4
=\left[(Z_1)^2 -(Z_2)^2\right]^2$.


\subsection{$N>4$  pure supergravity attractors}\label{n=568attractors}
We are going to discuss here the attractor solutions for the extended theories with $N>4$, where no matter multiplets may be coupled. We will include a discussion of their relation to $N=2$ BPS and non-BPS black holes, already presented in \cite{fgk06}.

\subsubsection{The $N=5$ case}
The moduli space of this model is
\begin{equation}
G/H=\frac{SU(1,5)}{U(5)} \label{n=5coset}\,,
\end{equation}
the theory contains 10 graviphotons and the relations among the central charges are
\begin{eqnarray}
  D(\omega) Z_{AB} &=&   +  \frac{1}{2} \bar Z^{CD} P_{ABCD}\, .
\end{eqnarray}
Correspondingly, the extremum condition on the black-hole potential
is
\begin{eqnarray}
dV_{\mbox{\small B-H}}&=&\frac 12 D Z_{AB} \bar Z^{AB} +\frac 12  Z_{AB}D \bar Z^{AB}
\nonumber\\
&=&\frac14 P_{AB CD} \bar Z^{AB} \bar Z^{CD} +c.c.=0 \label{extrn=5}\,.
\end{eqnarray}
This extremum condition allows only one solution with non-zero area, the BPS one.
Indeed, in terms of the proper values $Z_1, Z_2$ of $Z_{AB}$, equation \eq{extrn=5} becomes
\begin{equation}
Z_1 Z_2 + \bar Z_1 \bar Z_2 =0\,.
\end{equation}
However, by means of a $U(5)$ rotation $Z_1,Z_2$ may always be
chosen real and non-negative \cite{zum}, leaving as the only
solution with non-zero area $Z_1 > 0$, $Z_2=0$ (or viceversa). The
black-hole potential  on this solution is
\begin{equation}
V_{\mbox{\small B-H}}|_{\mbox{\small attr}}=|Z_1|^2 \qquad (\mbox{or } 1\leftrightarrow 2)
\end{equation}

This solution is $\frac 15$-BPS and breaks the symmetry of the moduli space:
$$U(5) \to SU(2)\times SU(3) \times U(1).$$

However, if we truncate this model   $N=5 \to N=2$, we have the following decomposition of the 10 vectors
$$\mathbf{10} \to \mathbf{1}+\bar \mathbf{3}+\mathbf{6}\,.$$
The singlet corresponds to the $N=2$ graviphoton, while  $\bar \mathbf{3}$ is the representation of the 3 vectors in the vector multiplets. The 6 extra vectors are projected out in the truncation. Correspondingly, the $N=5$ central charge $Z_{AB}$ reduces to:
\begin{equation}
Z_{AB} \to \pmatrix{Z_{ab}=Z \delta_{ab} &0\cr
0& Z_{IJ} =\epsilon_{IJK} \bar Z^K}\,, \quad a,b=1,2;~I,J,K=1,2,3
\end{equation}

 The two solutions $Z_1 >0,Z_2=0$ and $Z_1=0,Z_2>0$, which were BPS and degenerate in the $N=5$ theory, in the $N=2$ interpretation  correspond the first to a BPS solution (if we set $Z_1\equiv Z$) and the second to a non-BPS solution with $Z=0$, as for the quadratic series discussed in section \ref{n=2attractors}.

Let us inspect these results in terms of the discussion of  section \ref{u-inv}.
The $SU(5,1)$ invariant is (in terms of the $U(5)$ invariants introduced in section \ref{u-inv}):
\begin{equation}
I_4=   4 Tr(A^2) - (Tr A)^2
\end{equation}
that is, in terms of the proper-values of the central charge
\begin{equation}
I_4=  \left[ (Z_1)^2 -(Z_2)^2\right]^2 \label{invar5bis}\,.
\end{equation}
The solutions $Z_1\neq Z_2$ are separated by the solution $Z_1=Z_2$, which corresponds to a small black hole, with $I_4=0$. This is the solution which preserves the maximal amount of supersymmetry ($\frac 25$ unbroken), but it does not come from the attractor equations.


\subsubsection{The $N=6$ case}
The moduli space  is
\begin{equation}
G/H=\frac{SO^*(12)}{U(6)}\label{n=6coset}\,,
\end{equation}
and the theory  contains 16 graviphotons, 15 in the twice-antisymmetric representation of $U(6)$ plus a singlet. The attractor solutions for this theory have already been presented in \cite{bfgm}.

 The relations among the central charges are
\begin{eqnarray}
  D(\omega) Z_{AB} &=& \frac{1}{2} \bar Z^{CD} P_{ABCD}  + \frac 1{4!}\bar Z
  \epsilon_{ABCDEF} \bar P^{CDEF}\,,\nonumber \\
D(\omega) Z &=& \frac{1}{2!4!}  \bar Z^{AB}   \epsilon_{ABCDEF} \bar P^{CDEF}\,.
\end{eqnarray}

The  black-hole potential for this theory  is
\begin{equation}
V_{\mbox{\small B-H}}=\frac 12 Z_{AB}\bar Z^{AB} + Z\bar Z
\end{equation}
and the extremum condition is then
\begin{eqnarray}
dV_{\mbox{\small B-H}}&=&\frac 12 D Z_{AB} \bar Z^{AB} +\frac 12  Z_{AB}D \bar Z^{AB}
+D Z \bar Z + ZD \bar Z=0 \nonumber\\
&=&\frac 14 P_{ABCD}\left(\bar Z^{AB} \bar Z^{CD} +\frac 1{3!} \epsilon^{ABCDEF} Z_{EF}  Z\right) +c.c.=0\label{extrn=6}\,.
\end{eqnarray}
In terms of the proper-values $Z_1,Z_2,Z_3$ of $Z_{AB}$, which may always be chosen real and non negative by a $U(6)$ rotation, the condition to be satisfied on the extremum is:
\begin{equation}
Z_1 Z_2 + Z Z_3 =0\, \qquad (1\to 2\to 3\to 1 \mbox{ cyclically})\,.
\end{equation}
This equation admits one solution $\frac 16$-BPS with $Z=0$, and two independent non-BPS solutions, both with $Z\neq 0$.
\begin{itemize}
\item
The BPS solution is found for
\begin{equation}
Z=0\,\qquad Z_2=Z_3 =0 \,\qquad Z_1\neq 0\,,
\end{equation}
if we choose $Z_1 \geq Z_2\geq Z_3$. In this case the black-hole
potential becomes
\begin{equation}
V_{\mbox{\small B-H}}|_{\mbox{\small attr}}=|Z_1|^2 \qquad (\mbox{or } 1\leftrightarrow 2\leftrightarrow 3)
\end{equation}
and corresponds to $I_4>0$.

This solution breaks the symmetry
$$U(6) \to SU(2)\times U(4)$$
and corresponds to an $\frac{SO^*(12)}{SU(4,2)}$ orbit of the charge vector.
\item
One non-BPS solution is obtained
for
\begin{equation}
Z\neq 0\,\qquad Z_1=Z_2=Z_3 =0\,.
\end{equation}
It gives for the black-hole potential
\begin{equation}
V_{\mbox{\small B-H}}|_{\mbox{\small attr}}=|Z|^2
\end{equation}
and preserves all the $U(6)$ symmetry of the moduli space. This solution corresponds to the orbit
$\frac{SO^*(12)}{SU(6)}$.
Also for this solution the quartic invariant is positive $I_4>0$.
\item
The third
solution is found by setting
\begin{equation}
Z_1=Z_2=Z_3 =\rho \,,\qquad
 Z= -\rho \,.
\end{equation}
In this case the black-hole potential becomes
\begin{equation}
V_{\mbox{\small B-H}}|_{\mbox{\small attr}}=4 \rho^2 \,.
\end{equation}
This solution breaks the symmetry $U(6) \to USp(6)$, and corresponds
to the charge orbit $\frac{SO^*(12)}{SU^*(6)}$. The quartic
invariant for this solution is negative $I_4<0$.
\end{itemize}

It is interesting to note, as already observed in
\cite{adf96,fgk06,bfgm},  that the bosonic sector of the $N=6$ is
exactly the same as the one of the $N=2$ model coupled with 15
vector multiplets with scalar sector based on the same coset
\eq{n=6coset}. In the $N=2$ interpretation of this model, the
singlet  charge $Z$ plays the role of central charge, while the 15
charges $Z_{AB}$ are interpreted as matter charges.

The interpretation of the three attractor solutions is now
different: the first one, which was $\frac 16$-BPS in the  $N=6$
model, is now non-BPS and breaks  supersymmetry,  while the second
one in this model is $\frac 12$-BPS. The third solution, where all
the proper forms of the dressed charges are different from zero,  is
non-BPS in both interpretations.


\subsubsection{The $N=8$ case}
This model has been studied in detail in \cite{fk2006}.
Its scalar manifold is the coset
\begin{equation}
G/H=\frac{E_{7(7)}}{SU(8)}\label{n=8coset}\,.
\end{equation}

 The relations among the 28 central charges are
\begin{eqnarray}
  D(\omega) Z_{AB} &=& \frac{1}{2} \bar Z^{CD} P_{ABCD} \, ,
\end{eqnarray}
where the vielbein $P_{ABCD}$ satisfies the reality condition
\begin{equation}
\bar P^{ABCD}= \epsilon^{ABCDEFGH} P_{EFGH}\,.
\end{equation}

The extremum condition is then
\begin{eqnarray}
dV_{\mbox{\small B-H}}&=&\frac 12 D Z_{AB} \bar Z^{AB} +\frac 12  Z_{AB}D \bar Z^{AB}
=0 \nonumber\\
&=&\frac 14 P_{ABCD}\left(\bar Z^{AB} \bar Z^{CD} +\frac 1{4!} \epsilon^{ABCDEFGH} Z_{EF}  Z_{GH}\right) =0\label{extrn=8}\,.
\end{eqnarray}

In terms of the central charge proper-values $Z_1,\cdots Z_4$ the condition
for the  extremum may be written
\begin{equation}
\left\{\matrix{Z_1 Z_2 + \bar Z_3 \bar Z_4 &=&0\cr
Z_1 Z_3 + \bar Z_2 \bar Z_4 &=&0\cr
Z_2 Z_3 + \bar Z_1 \bar Z_4 &=&0}\right.
\end{equation}
and  admits two independent attractor solutions:
\begin{itemize}
\item
The BPS solution is found for
\begin{equation}
 Z_2=Z_3 =Z_4=0 \,\qquad Z_1\neq 0
\end{equation}
if we choose $Z_1 \geq Z_2\geq Z_3\geq Z_4$. In this case the black
hole potential becomes
\begin{equation}
V_{\mbox{\small B-H}}|_{\mbox{\small attr}}=|Z_1|^2 \qquad (\mbox{or } 1\leftrightarrow 2\leftrightarrow 3\leftrightarrow
4)
\end{equation}
and corresponds to $I_4>0$.
This solution breaks the symmetry
$$SU(8) \to SU(2)\times U(6)$$
and corresponds to an $\frac{E_7}{E_{6(2)}}$ orbit of the charge vector.
\item
The non-BPS solution is obtained
for
\begin{equation}
Z_1=Z_2=Z_3 =Z_4 = e^{\ii\frac \pi 4} \rho\,\qquad \rho \in
\mathbb{R}^+ \,.
\end{equation}
It gives for the black-hole potential
\begin{equation}
V_{\mbox{\small B-H}}|_{\mbox{\small attr}}=4 \rho^2 \,.
\end{equation}
This solution breaks the symmetry $SU(8) \to USp(8)$, and corresponds to the charge orbit $\frac{E_7}{E_{6(6)}}$. The quartic invariant for this solution is negative $I_4<0$.
\end{itemize}

\section{Conclusions}\label{conclusions}
This survey  has presented the main features of   the physics of black holes embedded in supersymmetric theories
of gravitation.  They have an  extremely rich structure and give an
interplay between space-time singularities in solutions of
Einstein-matter coupled equations and the solitonic, particle-like
structure of these configurations such as mass, spin and charge.

The present analysis may be extended to rotating black holes
and to geometries not necessarily asymptotically flat (such as, for
example, asymptotically anti de Sitter solutions).
Furthermore, the concept of entropy may be extended to theories
which include higher curvature and higher derivative matter terms \cite{w,cdm,dw,sen}.
This is important in order to make contact with superstring and
M-theory where these terms unavoidably appear.
In this context, a remarkable connection has been found between the
entropy functional and the topological string partition function, an
approach pioneered in \cite{osv}.

Black hole attractors fall in the class of possible superstring
vacua, which in a wide context have led to the study of the so-called
landscape \cite{landscape}.

It is a challenging problem to see which new directions towards a
fundamental theory of nature these investigations may suggest in the
future.


\section*{Acknowledgements}
 The present review is partly based on work and discussions with the following people: S. Bellucci, A. Ceresole, M. Duff, P. Fr\'e, E. Gimon, M. Gunaydin, R. Kallosh, M.A. Lled\'o, J. Maldacena, A. Marrani, A. Strominger.

Work supported in part by the European
Community's Human Potential Program under contract
MRTN-CT-2004-005104 `Constituents, fundamental forces and
symmetries of the universe', in which L.A., R.D'A. and M.T. are
associated to Torino University. The work of S.F. has been
supported in part by European Community's Human Potential Program
under contract MRTN-CT-2004-005104 `Constituents, fundamental
forces and symmetries of the universe' and the contract MRTN-
CT-2004-503369 `The quest for unification: Theory Confronts
Experiments', in association with INFN Frascati National
Laboratories and by D.O.E. grant DE-FG03-91ER40662, Task C.

\end{document}